\input epsf
\documentclass[12pt]{article}
\usepackage{latexsym,amsmath,amssymb,graphicx}
\renewcommand{\thefootnote}{\fnsymbol{footnote}}

\usepackage{amsmath}
\usepackage{amsfonts}

\numberwithin{equation}{section}

\def\doubleset#1#2{\bgroup%
\def\doit#1#2{%
\setbox\dblsetbox=\hbox{$\cstyle #1$}%
\raise#2\ht\dblsetbox\copy\dblsetbox%
\hskip-\wd\dblsetbox%
\raise-#2\ht\dblsetbox\box\dblsetbox}%
\mathchoice%
{\def\cstyle{\displaystyle}\doit#1#2}%
{\def\cstyle{\textstyle}\doit#1#2}%
{\def\cstyle{\scriptstyle}\doit#1#2}%
{\def\cstyle{\scriptscriptstyle}\doit#1#2}\egroup}
\def\underarrow#1{\vbox{\ialign{##\crcr$\hfil\displaystyle
 {#1}\hfil$\crcr\noalign{\kern1pt\nointerlineskip}$\longrightarrow$\crcr}}}

\newbox\dblsetbox

\newlength{\extraspace}
\newlength{\extraspaces}
\newcommand{\be}{\begin{equation}
\addtolength{\abovedisplayskip}{\extraspaces}
\addtolength{\belowdisplayskip}{\extraspaces}
\addtolength{\abovedisplayshortskip}{\extraspace}
\addtolength{\belowdisplayshortskip}{\extraspace}}
\newcommand{\ee}{\end{equation}}

\newcommand{\ba}{\begin{eqnarray}
\addtolength{\abovedisplayskip}{\extraspaces}
\addtolength{\belowdisplayskip}{\extraspaces}
\addtolength{\abovedisplayshortskip}{\extraspace}
\addtolength{\belowdisplayshortskip}{\extraspace}}
\newcommand{\ea}{\end{eqnarray}}

\newcommand{\bd}{\begin{displaymath}
\addtolength{\abovedisplayskip}{\extraspaces}
\addtolength{\belowdisplayskip}{\extraspaces}
\addtolength{\abovedisplayshortskip}{\extraspace}
\addtolength{\belowdisplayshortskip}{\extraspace}}
\newcommand{\ed}{\end{displaymath}}

\newcounter{saveeqn}

\newcommand{\newsection}[1]{
\vspace{12mm}
\pagebreak[3]
\addtocounter{section}{1}
\setcounter{equation}{0}
\setcounter{subsection}{0}
\noindent{\bf \thesection. #1}
\nopagebreak
\medskip
\nopagebreak}

\newcommand{\newsubsection}[1]{
\vspace{0.8cm}
\pagebreak[3]
\addtocounter{subsection}{1}
\noindent{\it \thesubsection. #1}
\nopagebreak
\vspace{2mm}
\nopagebreak}

\begin{document}
\addtolength{\baselineskip}{1.5mm}

\thispagestyle{empty}
\begin{flushright}
hep-th/   \\
\end{flushright}
\vbox{}
\vspace{3.0cm}

\begin{center}
\centerline{\LARGE{The Half-Twisted Orbifold Sigma Model And The}}
\medskip
\centerline{\LARGE{ Chiral de Rham Complex}}

\vspace{2.0cm}

{Meng-Chwan~Tan\footnote{E-mail: g0306155@nus.edu.sg}}
\\[0mm]
{\it Department of Physics\\
National University of Singapore \\
Singapore 119260}\\[8mm]
\end{center}

\vspace{2.0 cm}

\centerline{\bf Abstract}\bigskip \noindent

In this paper, we study the perturbative aspects of the half-twisted variant of Witten's topological A-model on a complex orbifold $X/G$, where $G$ is an isometry group of $X$. The objective is to furnish a purely physical interpretation of the mathematical theory of the Chiral de Rham complex on orbifolds recently constructed by Frenkel and Szczesny in \cite{Frenkel}. In turn, one can obtain a novel understanding of the holomorphic (twisted) $N=2$ superconformal structure underlying the untwisted and twisted sectors of the quantum sigma model, purely in terms of an obstruction (or a lack thereof) to a global definition of the relevant physical operators which correspond to $G$-invariant sections of the sheaf of Chiral de Rham complex on $X$. Explicit examples are provided to help illustrate this connection, and comparisons with their non-orbifold counterparts are also made in an aim to better understand the action of the $G$-orbifolding on the original half-twisted sigma model on $X$.

\newpage

\renewcommand{\thefootnote}{\arabic{footnote}}
\setcounter{footnote}{0}

\newsection{Introduction}

The theory of the Chiral de Rham complex or CDR for short, is a fairly well-developed subject in the mathematical literature. It aims  to provide a rigorous mathematical construction of conformal field theories in two-dimensions without    resorting to mathematically non-rigorous methods such as the path integral. The sheaf of CDR is a specialisation of a general sheaf of chiral differential operators, and it was first introduced and studied in two seminal papers by Malikov et al. \cite{MSV1, MSV2}. Since its inception, the sheaf of CDR has found interesting applications in various fields of geometry and representation theory, namely mirror symmetry \cite{Bo}, and the study of elliptic genera \cite{BL, BL1, BL2, grassi}.

Previous efforts to provide an explicit interpretation of the theory of CDR in terms of the physical model it is supposed to describe, has been undertaken in \cite{Ka, MC}. In \cite{Ka}, it is argued that on a Calabi-Yau manifold $X$, the mathematical theory of the CDR can be identified with the infinite-volume limit of a half-twisted variant of the topological A-model. On the other hand, a different and more general approach was taken in \cite{MC}, whereby a twisted version of the perturbative $(0,2)$ heterotic sigma model   is first considered and described using the general theory of chiral differential operators, where at the $(2,2)$ locus, the interpretation of the theory of CDR in terms of the half-twisted A-model on an arbitrary (not necessarily Calabi-Yau) smooth manifold is made manifest. Part of the results in \cite{MC}  hence serve as an alternative verification and generalisation of the specific findings established earlier in \cite{Ka}.

The sheaf of CDR, as constructed by Malikov et al. in \cite{MSV1, MSV2}, is defined over a smooth complex variety. However, this construction was recently extended by Frenkel and Szczesny in \cite{Frenkel} to allow for a definition over singular orbifolds. Moreover, further evidence of the physical relevance of this extended construction    comes from the fact that one can demonstrate a mathematical equivalence between the expressions of the orbifold elliptic genus and a genus one partition function \cite{Frenkel}.

In this paper, we will continue the program in \cite{MC} to study, in pertubation theory,  the half-twisted A-model on an orbifold. The main aim is to seek a purely physical interpretation of the above-described mathematical construction of the sheaf of CDR over orbifolds. In turn, we hope to obtain some novel insights into the physics via a reinterpretation of some established mathematical results.

\smallskip\noindent{\it A Brief Summary and Plan of the Paper}

A brief summary and plan of the paper is as follows. First, in Section 2, we will discuss some important and relevant features of the half-twisted A-model on a smooth manifold $X$. In particular, we will discuss its holomorphic chiral algebra and holomorphic $N=2$ (twisted) superconformal structure.

In Section 3, we will review the general construction of orbifold sigma models. We will quickly specialise our discussion to the half-twisted    model     on a general orbifold $X/G$, where the action of the finite group $G$ is an isometry of $X$. We will explain the additional features that one might expect to observe due to an orbifolding procedure of including twisted sectors and projecting onto $G$-invariant operators in defining the model on $X/G$ instead of $X$. We will also discuss the holomorphic $N=2$ (twisted) superconformal structure which also underlies the twisted sector.

In Section 4, we will introduce the notion of a sheaf of perturbative observables.  This will in turn  allow us to describe the physical operators of the chiral algebra as elements in the appropriate $G$-invariant Cech cohomology groups. Thereafter, we will furnish a local description of the sigma model on $X$ in terms of a free $bc$-$\beta\gamma$ system. We also describe how the local symmetries of the free $bc$-$\beta\gamma$ system can be used to `glue' the local descriptions together to furnish a global description of the sigma model on $X$ so that we can ultimately study the sigma model on $X/G$. It is at this juncture that the interpretation of the orbifold sigma model in terms of the theory of the CDR over an orbifold of \cite{Frenkel}  is made clear.

In Section 5, we consider an example on the non-Calabi-Yau orbifold $\mathbb {CP}^1 /{\mathbb Z_K}$. Via this example, we will be able to     obtain a novel understanding of the broken, holomorphic $N=2$ (twisted) superconformal structure underlying  the untwisted and twisted sectors of the quantum sigma model, purely in terms of an obstruction to a global definition of the stress tensors and their worldsheet superpartners which correspond to $\mathbb Z_K$-invariant local sections of the sheaves of CDR on $\mathbb {CP}^1$. By considering an example on a ${\mathbb Z}_K$ orbifold of the flat  $({\bf{S}^3 \times \bf{S}^1})$ manifold, one can also understand,   the restoration of the $N=2$ (twisted) superconformal structure in quantum perturbation theory,   purely as a vanishing obstruction to the above-mentioned  global definition of operators   in both sectors. We will also make various comparisons with the results obtained in \cite{MC} involving the half-twisted A-model on $\mathbb {CP}^1$ and $\bf{S}^3 \times \bf{S}^1$ respectively.

\smallskip\noindent{\it Beyond Perturbation Theory}

Instanton effects may change the picture radically, triggering a spontaneous breaking of supersymmetry, hence making the chiral algebra trivial as the elliptic genus vanishes. Therefore, beyond perturbation theory, the sigma model may no longer be described by the theory of CDR. This non-perturbative consideration is beyond the scope of the present paper, and will not be covered in the present study.

\newsection{The Half-Twisted  A-Model on a Smooth Manifold $X$}

\vspace{-0.6cm}\newsubsection{Some Salient Features of the Half-Twisted A-Model}

In this section, we will study the salient features of the half-twisted A-model on a smooth manifold. The reason for doing so is that many features of the non-orbifold theory do carry over to the orbifold sigma model. Hence, their discussion is relevant to the present paper.

To begin with, let us first recall the half-twisted variant of the A-model in perturbation theory. It governs maps $\Phi : \Sigma \to X$, with $\Sigma$ being the worldsheet Riemann surface. By picking local coordinates $z$, $\bar z$ on $\Sigma$, and $\phi^{i}$, $\phi^{\bar i}$ on the K\"ahler manifold $X$, the map $\Phi$ can then be described locally via the functions $\phi^{i}(z, \bar z)$ and $\phi^{\bar i}(z, \bar z)$. Let $K$ and ${\overline K}$ be the canonical and anti-canonical bundles of $\Sigma$ (the bundles of one-forms of types $(1,0)$ and $(0,1)$ respectively), whereby the spinor bundles of $\Sigma$ with opposite chiralities are given by $K^{1/2}$ and ${\overline K}^{1/2}$. Let $TX$ and $\overline {TX}$ be the holomorphic and anti-holomorphic tangent bundle of $X$. The half-twisted variant as defined in \cite{n=2}, will have the same classical Lagrangian as that of the original A-model in \cite{mirror manifolds}. (The only difference is that the cohomology of operators and states is taken with respect to a $\it{single}$ right-moving supercharge only instead of a linear combination of a left- and right-moving supercharge. This will be clear shortly).  It  is thus given by\footnote{The action just differs from the A-model action in \cite{mirror manifolds} by a term $\int_{\Sigma} \Phi^*(K)$, where $K$ is the K\"ahler $(1,1)$-form on $X$. This term is irrelevant in perturbation theory where one considers only trivial maps $\Phi$ of degree zero.}
\begin{eqnarray}
S & = & \int_{\Sigma} |d^2z| \left( g_{i{\bar j}} \partial_z \phi^{\bar j} \partial_{\bar z}\phi^i + g_{i{\bar j}} \psi_{\bar z}^i D_z \psi^{\bar j} + g_{i \bar j} \psi^{\bar j}_z D_{\bar z} \psi^i  - R_{i {\bar k}  j {\bar l}} {\psi}^{i}_{\bar z} {\psi}^{\bar k}_z  \psi^j \psi^{\bar l}  \right),
\label{S}
\end{eqnarray}
where $|d^2z| = i dz \wedge d\bar z$ and $i,j,k, l = 1, 2, \dots, {\textrm{dim}_{\mathbb C} X}$. $R_{i {\bar k}  j {\bar l}}$ is the curvature tensor with respect to the Levi-Civita connection $\Gamma^i{}_{lj} = g^{i \bar k}\partial_lg_{j \bar k}$, and the covariant derivatives with respect to the connection induced on the worldsheet are given by
\be
D_z\psi^{\bar j} = \partial_z \psi^{\bar j} + \Gamma^{\bar j}{}_{\bar i \bar k} \partial_z\phi^{\bar i}\psi^{\bar k}, \qquad  D_{\bar z}\psi^{i} = \partial_{\bar z} \psi^{i} + \Gamma^{i}{}_{j k} \partial_{\bar z} \phi^{j}\psi^{k}.
\ee
The various fermi fields transform as smooth sections of the following bundles:
\begin{eqnarray}
\psi^i  \in  \Gamma \left(\Phi^*{TX}  \right), & \qquad & \psi^{\bar i}_{z}  \in  \Gamma \left( K \otimes \Phi^*{\overline{TX}}\right), \nonumber \\
\psi^i_{\bar z} \in  \Gamma \left({\overline K} \otimes \Phi^*{TX} \right), & \qquad &  \psi^{\bar i} \in  \Gamma \left(\Phi^*{\overline{TX}}\right), \\
\nonumber
\end{eqnarray}
Notice that we have included additional indices in the above fermi fields so as to reflect their geometrical characteristics on $\Sigma$; fields without a $z$ or $\bar z$ index transform as worldsheet scalars, while fields with a $z$ or $\bar z$ index transform as $(1,0)$ or $(0,1)$ forms on the worldsheet respectively. In addition, as reflected by the $i$, and $\bar i$ indices, all fields continue to be valued in the pull-back of the corresponding bundles on $X$.

Let us next discuss the classical symmetries of the action $S$. Firstly, note that $S$ has a left and right-moving ghost number symmetry whereby the left-moving fermionic fields transform as $\psi^i \to e^{i\alpha}\psi^i$ and $\psi^{\bar i}_{z} \to e^{-i \alpha} \psi^{\bar i}_{z}$, and the right-moving fermionic fields transform as $\psi^{\bar i} \to e^{i \alpha}\psi^{\bar i}$ and $\psi^i_{\bar z} \to e^{-i \alpha}\psi^i_{\bar z}$, where $\alpha$ is real. In other words, the fields $\psi^i$, $\psi^{\bar i}_{ z}$, $\psi^{\bar i}$ and $\psi^i_{\bar z}$ can be assigned the $(g_L, g_R)$ left-right ghost numbers $(1,0)$, $(-1,0)$, $(0,1)$ and $(0,-1)$ respectively. The infinitesimal version of this symmetry transformation of the left-moving fermi fields read (after absorbing some trivial constants)
\be
\delta\psi^i = \psi^i, \quad \delta \psi^{\bar i}_{z} = - \psi^{\bar i}_{z},
\label{ghostL}
\ee
while those of the right-moving fermi fields read
\be
\delta \psi^{\bar i} = \psi^{\bar i}, \quad \delta \psi^i_{\bar z} = - \psi^i_{\bar z}.
\label{ghostR}
\ee
The conserved holomorphic (i.e.\,left-moving) current associated with the transformation (\ref{ghostL}) will then be given by
\be
J(z) = g_{i\bar j} \psi^{\bar j}_z \psi^i.
\label{J}
\ee
$J(z)$ is clearly a dimension one bosonic current. (There is also an anti-holomorphic conserved current associated with the right-moving ghost symmetry. However, it is irrelevant to our discussion). Secondly, note that $S$ is also invariant under the following field transformations:
\begin{eqnarray}
\delta \phi^i = \psi^i, &\quad & \delta\phi^{\bar i} = 0, \nonumber \\
\label{susyq}
\delta\psi^{\bar i}_{z} = - \partial_z \phi^{\bar i}, &  \quad & \delta \psi^i_{\bar z} = - \Gamma^i{}_{j k} \psi^j \psi^k_{\bar z}, \\
\delta \psi^i = 0, & \quad & \delta \psi^{\bar i} = 0. \nonumber
\end{eqnarray}
The conserved, dimension one fermionic current in this case will be given by
\be
Q(z) = g_{i \bar j} \psi^i \partial_z \phi^{\bar j}.
\label{Q}
\ee
For later convenience, let us label the charge corresponding to the current $Q(z)$ as $Q_L$.

The third set of field transformations that leave $S$ invariant are given by
\begin{eqnarray}
\delta \phi^{\bar i} = \psi^{\bar i}, &\quad & \delta\phi^{i} = 0, \nonumber \\
\label{susyqr}
\delta\psi^{i}_{\bar z} = - \partial_{\bar z} \phi^{i}, &  \quad & \delta \psi^{\bar i}_{z} = - \Gamma^{\bar i}{}_{\bar j \bar k} \psi^{\bar j} \psi^{\bar k}_{z}, \\
\delta \psi^i = 0, & \quad & \delta \psi^{\bar i} = 0. \nonumber
\end{eqnarray}
The corresponding current of the above symmetry is given by $Q_R(\bar z) = g_{i \bar j}\psi^{\bar j}\partial_{\bar z} \phi^i$. Let us also label the conserved charge of $Q_R(\bar z)$ as $Q_R$.

In Witten's topological A-model, the BRST-charge operator that defines the BRST cohomology is given by $Q_{BRST} = Q_L + Q_R$, where $Q_L$ and $Q_R$ are the above-mentioned left and right-moving (scalar) supercharges which generate the symmetry transformations in (\ref{susyq}) and (\ref{susyqr}) respectively. However, the half-twisted A-model is a greatly enriched variant in which one ignores $Q_L$ and considers $ Q_R$ as the BRST operator \cite{n=2}. Since the corresponding cohomology is now defined with respect to a single, right-moving, scalar  supercharge $Q_R$, its classes need not be restricted to dimension $(0,0)$ operators (which correspond to ground states). In fact, the physical operators will have dimension $(n,0)$, where $n \geq 0$. Let us verify this important statement.

From (\ref{S}), we find that the anti-holomorphic stress tensor takes the form $ T_{\bar z \bar z} =g_{i \bar j}  \partial_{\bar z} \phi^i \partial_{\bar z} \phi^{\bar j} + g_{i \bar j}  \psi_{\bar z}^i \left ( \partial_{\bar z} \psi^{\bar j} + \Gamma^{\bar j}_{\bar l \bar k}\partial_{\bar z} \phi^{\bar l} \psi^{\bar k} \right)$. One can go on to show that $ T_{\bar z \bar z} = \{ Q_R , - g_{i \bar j} \psi_{\bar z}^i \partial_{\bar z} \phi^{\bar j} \}$, that is, $T_{\bar z \bar z}$ is trivial in $Q_R$-cohomology. Now, we say that a local operator $\cal O$ inserted at the origin has
dimension $(n,m)$ if under a rescaling $z\to \lambda z$, $\bar
z\to \bar\lambda z$,
it transforms as $\partial^{n+m}/\partial z^n\partial\bar z^m$,
that is, as $\lambda^{-n}\bar\lambda{}^{-m}$. Classical local
operators have dimensions $(n,m)$ where $n$ and $m$ are
non-negative integers.\footnote{Anomalous
dimensions under RG flow may shift the values of $n$ and $m$ quantum mechanically, but the spin given by
$(n-m)$, being an intrinsic property, remains unchanged.} However, only local operators with $m = 0$ survive in $Q_R$-cohomology. The reason for the last statement is that the rescaling of $\bar z$ is generated by $\bar L_0=\oint d\bar z\, \bar z T_{\bar
z\bar z}$.  As we noted above, $T_{\bar z\,\bar z}$
is of the form $\{{Q}_R,\dots\}$, so $\bar L_0=\{{Q}_R,V_0\} $ for some $V_0$. If $\cal O$ is to be admissible as a local physical operator, it must at least be true that $\{{Q}_R, {\cal O}\}=0$. Consequently, $[\bar L_0,{\cal
O}]=\{{Q}_R,[V_0,{\cal O}]\}$.  Since the eigenvalue of $\bar L_0$ on $\cal O$ is $m$, we have $[\bar L_0,{\cal O}]=m{\cal O}$. Therefore, if $m\not= 0$, it follows that ${\cal O}$ is $Q_R$-exact and thus trivial in $Q_R$-cohomology. On the other hand, the holomorphic stress tensor is given by $T_{zz} =   g_{i \bar j} \partial_z \phi^i \partial_z \phi^{\bar j} + g_{i \bar j} \psi^{\bar j}_z D_z \psi^i$, and one can verify that it can be written as  $T_{zz} = \{ Q_L , -g_{i \bar j} \psi^{\bar j}_z \partial_z \phi^i \}$,  that is, it is $Q_L$-exact. Since we are only interested in $Q_R$-closed modulo $Q_R$-exact operators, there is no restriction on the value that $n$ can take. These arguments persist in the quantum theory, since a vanishing cohomology in the classical theory continues to vanish when quantum effects are small enough in the perturbative limit.

Hence, in contrast to the A-model, the BRST spectrum of physical operators and states in the half-twisted variant is infinite-dimensional. A specialisation of its genus one partition function, also known as the elliptic genus of $X$, is given by the index of the $Q_R$ operator. Indeed, the half-twisted  model is not a topological field theory, rather, it is a 2d conformal field theory - the full stress tensor derived from its action is exact with respect to the  combination $Q_L +Q_R$, but not $Q_R$ alone.

In fact, more can be said about the observables of the half-twisted model. By a similar argument, we can show that $\cal O$,  as an element of the $ Q_R$-cohomology, varies homolomorphically with $z$. Indeed, because the momentum operator (which acts on $\cal O$ as $\partial_{\bar z}$) is given by $\bar L_{-1}$, the term $\partial_{\bar z} \cal O$ will be given by the commutator $[ \bar L_{-1}, \cal O]$. Since $\bar L_{-1} = \oint d\bar z\,T_{\bar z\bar z}$, we will  have $\bar L_{-1}=\{{Q_R},V_{-1}\}$ for some $V_{-1}$. Hence, because $\cal O$ is physical such that ${\{Q_R, \cal O\}} =  0$, it will be true that $\partial_{\bar z}{\cal O}=\{{Q_R},[V_{-1},{\cal O}]\}$ and thus vanishes in $Q_R$-cohomology. As before, since a vanishing cohomology in the classical theory continues to vanish when quantum effects are small enough in perturbation theory, this observation will continue to hold at the quantum level. Moreover, since the holomorphic stress tensor can be verified to be  $Q_R$-closed but not $Q_R$-exact (even at the quantum level), the space of local operators will be invariant under holomorphic reparameterisations of the coordinates on the worldsheet.

One can also make some observations about the correlation functions of these local operators. Firstly, note that the $\partial_{\bar z}$ operator on $\Sigma$ is given by ${\bar L}_{-1} = \oint d{\bar z}\ T_{\bar z \bar z}$. Therefore, ${\partial_{\bar z}\left <{\cal O}_1(z_1) {\cal O}_2(z_2) \dots {\cal O}_s(z_s)  \right >}$ will be given by $ \oint d{\bar z} \left <T_{\bar z \bar z} \ {\cal O}_1(z_1) {\cal O}_2(z_2) \dots {\cal O}_s(z_s) \right >$, which vanishes because $T_{\bar z \bar z} \sim 0$ in $Q_R$-cohomology. Thus, the correlation functions are always holomorphic in $z$. Secondly, note that the trace of the stress tensor is also trivial in $Q_R$-cohomology, that is, we can express $T_{z \bar z} = \{Q_R, G_{z \bar z}\}$ for some $G_{z \bar z}$. Hence, the variation of the correlation functions due to a change in the scale of $\Sigma$ will be given by $\left <{{\cal O}_1(z_1)} {{\cal O}_2(z_2)} \dots {{\cal O}_s(z_s)} T_{z \bar z} \right >$, which also vanishes because $T_{z \bar z} \sim 0$ in $Q_R$-cohomology. In other words, the correlation functions of local physical operators will continue to be invariant under arbitrary scalings of $\Sigma$. Thus, the correlation functions are always independent of the K\"ahler structure on $\Sigma$ but vary holomorphically with its complex structure (as is familiar for chiral algebras). Since the correlation functions are holomorphic in the parameters of the theory, they are protected from perturbative corrections. Note that all the observations made so far   apply to the observables of the orbifold theory as well.

\newsubsection{A Holomorphic Chiral Algebra}

Let ${\cal O} (z)$ and $\widetilde {\cal O} (z')$ be two $Q_R$-closed operators such that their product is $Q_R$-closed as well. Now, consider their operator product expansion or OPE:
\be
{\cal O}(z)  {\widetilde {\cal O}}(z') \sim \sum_k f_k (z-z') {\cal O}_k (z'),
\label{OPE}
\ee
in which the explicit form of the coefficients $f_k$ must be such that the scaling dimensions and $(g_L, g_R)$ ghost numbers of the operators agree on both sides of the OPE. In general, $f_k$ is not holomorphic in $z$. However, if we work modulo $Q_R$-exact operators in passing to the $Q_R$-cohomology, the $f_k$'s which are non-holomorphic and are thus not annihilated by $\partial / \partial {\bar z}$, drop out from the OPE because they multiply operators ${\cal O}_k$ which are $Q_R$-exact. This is true because $\partial / \partial{\bar z}$ acts on the LHS of (\ref{OPE}) to give terms which are cohomologically trivial.\footnote{Since $\{Q_R,{\cal O}\}=0$, we have $\partial_{\bar z}{\cal O}=\{Q_R, V(z)\}$ for some $V(z)$, as argued before. Hence $\partial_{\bar z}{\cal O}(z)\cdot {\widetilde {\cal O}}(z')=\{Q_R,V(z){\widetilde {\cal O}}(z')\}$.} In other words, we can take the $f_k$'s to be holomorphic coefficients in studying the $Q_R$-cohomology, and the OPE of (\ref{OPE}) has a holomorphic structure. Hence, we have established that the $Q_R$-cohomology of holomorphic local operators has a natural structure of a holomorphic chiral algebra which we shall denote as $\cal A$; in addition to having holomorphic expansion coefficients $f_k$, the OPEs of the local operators in the  chiral algebra also obey the usual relations of holomorphy, associativity, and invariance under scalings and arbitrary holomorphic reparameterisations of $z$.

\newsubsection{The Moduli of the Chiral Algebra}

We shall now discuss the moduli of the chiral algebra $\cal A$. Note that the chiral algebra does depend on the complex structure of $X$ because it enters in the definition of the fields and the fermionic symmetry transformation generated by $Q_R$. In addition, the moduli also depends on a certain type of cohomology class. We shall now determine what this cohomology class is. To this end, we shall consider adding to $S$, a term which will represent a modulus of $\cal A$. As was shown in \cite{CDO, MC}, this term results in a non-K\"ahler deformation of the target space $X$. Thus, $X$ will be a complex, hermitian manifold in all our following discussions.

To proceed, let $T = {1\over 2}T_{ij} d\phi^i \wedge d\phi^j$ be any two-form on $X$ that is of type $(2,0).$\footnote{As noted in \cite{MC}, the restriction of $T$ to be a gauge field of type $(2,0)$, will enable us to associate the moduli of the chiral algebra with the moduli of sheaves of vertex superalgebras of which the CDR is a special case.} The term  that deforms $S$ will then be given by
\be
{S_T=\int_{\Sigma} |d^2z| \{Q_R, T_{ij}\psi_{\bar z}^i\partial_z\phi^j\}}.
\label{ST}
\ee
By construction, $S_T$ is $Q_R$-invariant. Moreover, since it has vanishing $(g_L, g_R)$ ghost numbers, it is also invariant under the global $U(1)_L \times U(1)_R$ ghost symmetry. Hence,  as required, the addition of $S_T$ preserves  the classical symmetries of the theory. Explicitly, we then have
\be
S_T = \int_{\Sigma} |d^2 z| \left( T_{ij, {\bar k}}\psi^{\bar k} \psi_{\bar z}^i \partial_z \phi^j - T_{ij} \partial_{\bar z} \phi^i \partial_z \phi^j  \right),
\label{STex}
\ee
where $T_{ij,{\bar k}} = \partial T_{ij}/ \partial \phi^{\bar k}$. Note that since $|d^2 z| = i dz \wedge d{\bar z}$, we can write the second term on the RHS of (\ref{STex}) as
\be
S^{(2)}_T = {{i \over 2} \int_{\Sigma} T_{ij} d\phi^i \wedge d\phi^j }= {i \int_{\Sigma} \Phi^*(T)}.
\label{T}
\ee
Recall that in perturbation theory, we are considering degree-zero maps $\Phi$ with no multiplicity. Hence, for $S^{(2)}_T$ to be non-vanishing, $T$ must $\it{not}$ be closed, i.e. $dT \neq 0$. In other words, one must have a non-zero flux ${\cal H} = {dT}$.  As $T$ is of type $(2,0)$, $\cal H$ will be a three-form of type $(3,0) \oplus (2,1)$.

Notice here that the first term on the RHS of (\ref{STex}) is expressed in terms of $\cal H$, since $T_{ij,{\bar k}}$  is simply the $(2,1)$ part of $\cal H$. In fact, $S^{(2)}_T$ can also be written in terms of $\cal H$ as follows. Suppose that $C$ is a three-manifold whose boundary is $\Sigma$ and over which the map $\Phi : \Sigma \to X$ extends. Then, if $T$ is globally-defined as a $(2,0)$-form, the relation ${\cal H} =dT$ implies, via Stoke's theorem, that
\be
S^{(2)}_T = i \int_{C} \Phi^*(\cal H).
\label{H}
\ee
Hence, we see that $S_T$ can be expressed solely in terms of the three-form flux $\cal H$ (modulo terms that do not affect perturbation theory). The relevant thing to note for the present paper is that $\cal H$ represents a class in the Cech cohomology group  $H^1(X, \Omega^{2,cl}_X)$, where $\Omega^{2,cl}_X$ is the sheaf of $\partial$-closed $(2,0)$-forms on $X$. This has been shown in \cite{CDO} and reviewed in \cite{MC}. Thus, the modulus of the chiral algebra is represented by a class in $H^1(X, \Omega^{2,cl}_X)$.

One last thing to note at this point is that we do not actually want to limit ourselves to the case where $T$ is globally-defined; as is clear from (\ref{ST}), if $T$ were to be globally-defined, $S_T$ and therefore the modulus of the chiral algebra would vanish in $Q_R$-cohomology. Fortunately, the RHS of (\ref{H}) makes sense as long as $\cal H$ is globally-defined, with the extra condition that $\cal H$ be closed, since $C$ cannot be the boundary of a four-manifold.\footnote{From homology theory, the boundary of a boundary vanishes. Hence, since $\Sigma$ exists as the boundary of $C$, the three-manifold $C$ itself cannot be a boundary of a higher-dimensional four-manifold.} Therefore, it suffices for $T$ to be locally-defined such that ${\cal H} =dT$ is true only $\it{locally}$. Hence, $T$ must be interpreted a a two-form gauge field in string theory (or a $\it{non}$-$\it{trivial}$ connection on gerbes in mathematical theories). This has been emphasised in a similar context in \cite{CDO, MC}.

\newsubsection{A Holomorphic (Twisted) $N=2$ Superconformal Algebra}

Let us write the conserved, dimension two holomorphic stress tensor  associated with the symmetry under holomorphic reparameterisations of the coordinates on the worldsheet  as $T(z) = - T_{zz}$ . Recall that it is given by
\be
T(z) = - g_{i \bar j} \partial_z \phi^i \partial_z \phi^{\bar j} - g_{i \bar j} \psi^{\bar j}_z D_z \psi^i.
\label{T}
\ee
Also recall that one can write $T(z) = \{ Q_L, G(z) \} =\delta G(z)$, the variation of $G(z)$ under the field transformations  (\ref{susyq}), where
\be
G(z) = g_{i \bar j} \psi^{\bar j}_z \partial_z \phi^i.
\label{G}
\ee
Hence, $G(z)$ is a conserved, dimension two fermionic current. Notice that the conserved currents and tensors $J(z)$, $Q(z)$, $T(z)$, $G(z)$ possess only holomorphic scaling dimensions. Thus, their respective spins will also be given by their dimensions.

One can verify that  $J(z)$, $Q(z)$, $T(z)$ and   $G(z)$ are all invariant under the field transformations of (\ref{susyqr}). In fact, we find that $J(z)$, $Q(z)$, $T(z)$ and $G(z)$ are all $Q_R$-closed operators in the $Q_R$-cohomology of the half-twisted model, at least at the classical level. Also note that if $\cal O$ and $\cal O'$ are $Q_R$-closed operators in the $Q_R$-cohomology, i.e., ${\{Q_R, \cal O\}}={\{Q_R, {\cal O}'\}}= 0$, then $\{Q_R, {\cal O}{\cal O}'\} =0$. Moreover, if $\{Q_R, {\cal O}\}=0$, then ${\cal O}\{Q_R, W\}= \{Q_R, {\cal O}W\}$  for any operator $W$.  These two statements mean that the cohomology classes of operators that (anti)commute with $Q_R$ form a closed (and well-defined) algebra under operator products. One can indeed show that $J(z)$, $Q(z)$, $T(z)$ and $G(z)$ form a complete multiplet which generates a closed, holomorphic, ($\it{twisted}$) $N=2$ superconformal algebra with the following OPE relations  \cite{RD}:
$$
T(z)T(w) \sim \frac{2T(w)}{(z-w)^2}+\frac{\partial T(w)}{z-w}
\eqno{(2.17a)}
$$
$$
J(z)J(w) \sim \frac{d}{(z-w)^2};\
T(z)J(w) \sim -\frac{d}{(z-w)^3}+\frac{J(w)}{(z-w)^2}+\frac{\partial J(w)}{z-w}
\eqno{(2.17b)}
$$
$$
G(z)G(w) \sim 0;\ T(z)G(w)\sim \frac{2G(w)}{(z-w)^2}+\frac{\partial G(w)}{z-w};\
J(z)G(w)\sim-\frac{G(w)}{z-w}
\eqno{(2.17c)}
$$
$$
Q(z)Q(w)\sim 0; \ T(z)Q(w) \sim\frac{Q(w)}{(z-w)^2}+\frac{\partial Q(w)}{z-w};\
J(z)Q(w) \sim\frac{Q(w)}{z-w}
\eqno{(2.17d)}
$$
$$
Q(z)G(w)\sim \frac{d}{(z-w)^3}+\frac{J(w)}{(z-w)^2}+\frac{T(w)}{z-w},
\eqno{(2.17e)}
$$
where $d = \textrm{dim}_{\mathbb C} X$. This structure is also known as a structure of a topological vertex algebra of rank $d$ in the mathematical literature \cite{MSV1}. Thus, we see that $G(z)$ is a (worldsheet) superpartner of  $T(z)$ under the supersymmetry generated by the charge $Q_L$ of the supercurrent $Q(z)$. In addition, we also find from the OPEs that $[Q_L, J(z)] = - Q(z)$, i.e., $J(z)$ is a (worldsheet) superpartner of $Q(z)$.  These observations will be relevant to our discussion momentarily. Also notice that the central charge in the stress tensor OPE (2.13a) is zero. This means that the Weyl anomaly vanishes and that the trace of the stress tensor is trivial in $Q_R$-cohomology at the quantum level. This simply reflects the invariance of the correlation functions under scalings of the worldsheet as noted earlier.

The classical, holomorphic, OPE algebra of the half-twisted model in (2.13) may or may not persist in the quantum theory. In fact, in a `massive' model where the first Chern class $c_1(X)$ is non-vanishing, the global $U(1)$ symmetry associated with $J(z)$ will be broken. Likewise for the symmetry associated with its superpartner $Q(z)$. Hence, $J(z)$ and $Q(z)$ will cease to remain in the $Q_R$-cohomology at the quantum level. However, the symmetries associated with $T(z)$ and $G(z)$ are exact in quantum perturbation theory, and these operators will remain in the $Q_R$-cohomology, regardless of the value of $c_1(X)$.  Hence, for $c_1(X) \neq 0$, we have in some sense a reduction  from an $N=2$ to an $N=1$ algebra.

\newsection{The Half-Twisted A-Model on an Orbifold $X/G$}

\vspace{-0.6cm}\newsubsection{Orbifolding the Half-Twisted A-Model}

Orbifolds are among the simplest class of solutions in string theory \cite{Dixon}. It is a possibly singular space, defined by equating points on an underlying manifold related by the action of its isometry group. However, despite the singularity of the geometry that arises due to the presence of fixed points, the corresponding 2d CFT is non-singular; roughly speaking, orbifolding just amounts to gauging the worldsheet theory by the isometry group of the target space. This leads to a set of standard procedures that one can employ to consistently define an orbifold theory.

In this section, we will apply the well-known orbifolding procedure to the half-twisted model discussed in section 2. In order to define the half-twisted model on the orbifold $X/G$, where $G$ is a finite group isometry of $X$, one starts with the original theory on the smooth manifold $X$, add twisted sectors, and project onto $G$-invariant operators and states in both the untwisted and twisted sectors. There will be a modification of the fermion number of the vacuum  as well.  Let us look at this procedure in greater detail.

\vspace{0.4cm}\smallskip\noindent{\it The Action of a Finite Isometry Group $G$ of $X$}

In order to for us to elaborate on the twisted sectors and ascertain which operators will eventually survive the $G$-projection, we will first need to specify the action of the finite isometry group $G$ on the various fields of the sigma model. For the purpose of making contact with the results of \cite{Frenkel}, we shall specialise to the case where $G$ is an abelian  cyclic group of order $K$, that is, $G = \mathbb Z_K$. Since $\mathbb Z_K$ is a subset of the rotation group, let us first specify the generator of rotations - it takes the general form
\be
{\cal R} = exp \ [2 \pi i  \sum_{j=1}^{\mathrm{dim}_{\mathbb C}X } (\theta_j J_{2j-1, 2j})],
\label{rot}
\ee
where $J_{2j-1, 2j}$ is the angular momentum generator which generates rotations in the $(2j -1,2j)$ plane, and $\theta_j$ is the corresponding rotation parameter.  Recall that on $X$, the $\phi^j$ and $\phi^{\bar j}$ fields transform as holomorphic and anti-holomorphic coordinates. They are given by complex linear combinations $\phi^j = 2^{-1/2}(\phi^{2j -1} + i  \phi^{2j})$ and $\phi^{\bar j} = 2^{-1/2}(\phi^{2j -1} - i  \phi^{2j})$, where $j, \bar j = 1, 2, \dots, \textrm{dim}_{\mathbb C}X$. Recall also that $\psi^j$ and $\psi^{\bar j}$ transform as (pullbacks of)  holomorphic and anti-holomorphic tangent bundles, where they are also given by the complex linear combinations $\psi^j = 2^{-1/2} (\psi^{2j -1} + i  \psi^{2j})$ and $\psi^{\bar j} = 2^{-1/2}(\psi^{2j -1} - i  \psi^{2j})$. Last but not least, note that  $g_{j \bar i}\psi^{\bar i}_z = \psi_{z j} $ and $ g_{\bar j  i}\psi^{i}_{\bar z}= \psi_{\bar z \bar j}$ transform as  holomorphic and anti-holomorphic one-forms on $X$ respectively. Thus, in order for the action of $G$ to commute with the worldsheet supersymmetries, and since scalars should be preserved under a rotation generated by $\cal R$,  the action of $G= \mathbb Z_K$ on the various fields must be given as follows:
\begin{eqnarray}
\label{1}
\phi^j  \to e^{2\pi i \theta_j} \phi^j, & \qquad & \phi_j \to e^{-2\pi i \theta_j} \phi_j,\\
\label{2}
\psi^j \to e^{2\pi i \theta_j} \psi^j, & \qquad  & \psi_{z j} \to e^{-2\pi i \theta_j} \psi_{z j},\\
\label{3}
\psi^{\bar j} \to e^{-2\pi i \theta_j} \psi^{\bar j}, & \qquad & \psi_{\bar z \bar j} \to e^{2\pi i \theta_j} \psi_{\bar z \bar j},
\end{eqnarray}
where $\theta_j = {m_j / K}$, and $m_j = 0, 1, 2 \dots K-1$. We have also written $g_{j \bar i}\phi^{\bar i}$ as $\phi_j$.

\vspace{0.4cm}\smallskip\noindent{\it Twisted Sectors}

Modular invariance of the partition function also requires the addition of twisted sectors. Moreover, from a string theoretic perspective, interactions between untwisted strings can produce twisted strings. The corresponding fields in the twisted sectors are defined to have non-trivial monodromy on the worldsheet, that is, for each $g \in \mathbb Z_K$, where $g^K =1$, we have
\be
\varphi(\sigma + 2\pi)  = g \cdot \varphi(\sigma),
\label{twist}
\ee
where $\sigma$ is the usual worldsheet spatial coordinate and $\varphi$ represents  a generic worldsheet field. Note that  $\sigma \to \sigma + 2 \pi$ is simply effected by $z \to e^{2\pi i} z$. The relation in (\ref{twist}) is natural because $\varphi(\sigma + 2 \pi) \sim \varphi (\sigma)$ for closed strings (which we are considering in this paper), and $G$ is an isometry of the target space and thus, the worldsheet theory.

Let us denote the fields in the $g$-twisted sector as $\phi^{g}$ and $\psi^{ g}$. Then, these fields will obey the following twist conditions\footnote{Our twisting convention differs from that found in the physics literature by a `-' sign in $m_j$. This is done so that we can identify with the results of \cite{Frenkel}. This change in convention is inconsequential as $g$ in (\ref{twist}) can be any element of $\mathbb Z_K$, and $e^{- 2 \pi i m_j / K} \sim e^{2 \pi i (K - m_j) /K} \in  \mathbb Z_K$. What is important is that the fields which transform as sections of the contagent and tangent bundles should have opposite twists.}
\begin{eqnarray}
\label{twistphi}
\phi^{j,g} (\sigma + 2 \pi) = e^{- 2 \pi i \theta_j} \phi^{j,g} (\sigma), & \qquad & \phi^g_j (\sigma + 2\pi) = e^{ 2 \pi i \theta_j} \phi^g_j (\sigma), \\
\label{twistpsi}
\psi^{j,g} (\sigma + 2 \pi) =  e^{- 2\pi i \theta_j} \psi^{j,g} (\sigma), & \qquad  & \psi^g_{z j}(\sigma + 2 \pi) = e^{ 2\pi i \theta_j} \psi^g_{z j}(\sigma),\\
\label{twist2psi}
\psi^{\bar j, g}(\sigma + 2 \pi) = e^{2\pi i \theta_j} \psi^{\bar j, g} (\sigma), & \qquad & \psi^g_{\bar z \bar j}(\sigma + 2 \pi) = e^{-2\pi i \theta_j} \psi^g_{\bar z \bar j}(\sigma),
\end{eqnarray}
where we consider the fermionic fields to come from the R-sector only.

Note that the mode numbers that appear in the Laurent expansion of the twisted fields must be shifted accordingly so that the fields will exhibit the required monodromies stated in (\ref{twistphi})-(\ref{twist2psi}). Therefore, we can expand the twisted fields (when $X$ is flat) as
\begin{eqnarray}
\label{phiexpand}
\partial_z \phi^{j,g} (z) = \sum_{m \in \, \theta_j + {\mathbb Z}} {\alpha^{j,g}_m \over z^{m+1}}, & \qquad & \partial_z \phi^g_j (z) = \sum_{m \in \, -\theta_j + {\mathbb Z}} {\alpha^g_{jm} \over z^{m+1}}, \\
\label{phiexpand2}
{\bar \partial}_{\bar z} \phi^{j,g} (\bar z) = \sum_{m \in \, - \theta_j + {\mathbb Z}} {{\tilde \alpha}^{j,g}_m \over {\bar z}^{m+1}}, & \qquad & {\bar \partial}_{\bar z} \phi^g_j (\bar z) = \sum_{m \in \, \theta_j + {\mathbb Z}} {{\tilde \alpha}^g_{jm} \over {\bar z}^{m+1}}, \\
\label{psiexpand}
\psi^{j,g} (z) = \sum_{m \in \, \theta_j + {\mathbb Z}} {\psi^{j,g}_m \over z^{m}},  & \qquad  & \psi^g_{z j}(z) = \sum_{m \in \, - \theta_j + {\mathbb Z}} {\psi^{g}_{j m} \over z^{m+1}},\\
\label{psiexpand2}
\psi^{\bar j, g}(\bar z) = \sum_{m \in \, \theta_j + {\mathbb Z}} {\psi^{\bar j,g}_m \over {\bar z}^{m}}, & \qquad & \psi^g_{\bar z \bar j}(\bar z) = \sum_{m \in \, -\theta_j + {\mathbb Z}} {\psi^{g}_{\bar j m} \over {\bar z}^{m+1}}.
\end{eqnarray}

\vspace{0.4cm}\smallskip\noindent{\it G-Invariance and the Hilbert Space of the Orbifold Theory}

Since orbifolding amounts to a form of gauging the worldsheet theory by the isometry group, one naturally has to project onto $G$-invariant operators and states. One can immediately see that  the admissible operators are those which are invariant under the action of $G = \mathbb Z_K$ specified in (\ref{1})-(\ref{3}).

The projection of the operators and states of the untwisted sector onto the  $G$-invariant subspace is straightforward. However, the projection of the twisted sectors onto the $G$-invariant subspace is a little less straightforward if the group that we are gauging by is non-abelian. Even though we are not  considering $G$ to be non-abelian in our paper, it will be useful to look at the generalised case so as to make contact with the definitions found in the mathematical literature.

Suppose we consider a $g$-twisted state in $H_g$, the $g$-twisted sector of the Hilbert space. Thus, if $\varphi$ is the operator corresponding to this state, we will have $\varphi(2 \pi ) = g  \varphi(0)$. If we act on this state by some other group element $h$, we are taken to the state whose corresponding operator is given by $h  \varphi (2 \pi) = h  g  \varphi(0) = hgh^{-1} (h \varphi(0))$. In other words, if $g$ and $h$ do not commute, the action of $h$ is to take the original state to another state in $H_{hgh^{-1}}$. Hence, under the action of the group, sectors within a given conjugacy class mix. (Two elements $g_1$ and $g_2$ are conjugate if $g_2 = h g_1 h^{-1}$ for some $h$ in the group.) This means that in order for one to project onto group invariant states, we will have to take a state in the sector $H_g$, and project onto the invariant subspace of $C(g)$, the centraliser of $g$ (the subgroup of elements commuting with $g$, which therefore take the state in $H_g$ to itself.). The total Hilbert space is then obtained by taking the sum of the corresponding states from the sectors in the same conjugacy class, while projecting each of these states onto the invariant subspace of its corresponding centraliser. Thus in general, the independent twisted sectors are labelled not by elements of $G$, but by the conjugacy classes of $G$. In other words, we can write the Hilbert space of the orbifold theory as a direct sum of twisted sectors, one for each conjugacy class $[g]$ in the group $G$:
\be
{\cal H} = \bigoplus_{[g]} {\cal H}_{[g]},
\label{hilbert space}
\ee
 where
\be
{{\cal H}_{[g]}} = \bigoplus_{g \in [g]} {{\cal H}^{C(g)}_g} = \bigoplus^{|[g]|}_{i=1} {{\cal H}^{C(g)}_{h_i g {h_i}^{-1}}}
\label{twisted sectors}
\ee
for an appropriate set $\{h_i \}$. The $C(g)$ superscript just denotes the $C(g)$-invariant subspace of the Hilbert space of states in the respective sectors. In addition, note that for an abelian orbifold, we have
\be
{\cal H}^{C(g)}_g \simeq {\cal H}^{C(g')}_{g'},
\label{isomorphism}
\ee
where $g$ and $g'$ belong in the same conjugacy class. This is true because $g' = g$ in such a case, and the conjugacy class of elements is the group itself. Notice also that for an abelian group, $C(g) = G$,  i.e., one still takes $G$-invariant operators and states in an abelian orbifold theory.  In general, $C(1) = G$, so the sum in  (\ref{hilbert space}) includes the $G$-invariant untwisted sector as required.

\vspace{0.4cm}\smallskip\noindent{\it Fermion Number Shift}

In the twisted sectors, there is a shift in the fermion number of the vacuum \cite{Jackiw, Wen}. To understand the implications of this statement, let us first explain the origin of the twisted boundary conditions for the fermions. Let the fixed-point set of $X$ be written as $X_g$. By definition of a fixed-point set, the group action $g$ must act trivially on the tangent bundle of $X_g$, but non-trivially on its normal bundle. Since the fermi fields transform as (pullbacks of) tangent bundles on $X$, i.e., they have tangent space indices, it will mean that $g$ will act non-trivially on the fermi fields whose indices correspond to the normal directions to $X_g$. Thus, the fermionic vacuum corresponds to a sector with generalised boundary conditions on the ends of the strings. It can be shown \cite{Wen} that for chiral fermions in one spatial dimension, when the chiral fermion number is properly regularised to account for an infinite spectrum of energies, the general boundary condition $\psi(\sigma + 2 \pi) = e^{-2\pi i f} \psi (\sigma)$ leads to the result $F =f$ for the twisted fermion vacuum, where $F$ is the fermion number.

The above argument can be extended to the multi-fermion case. By choosing a basis for the tangent space such that the matrix of $g$ is diagonalised, one can see that we actually have a seperate shift for each of the chiral fermions. If the eigenvalues of $g$ are given by $e^{-2\pi i \theta_l}$, $l = 1, \dots, r$, where $r = \textrm{codim}_{\mathbb C} X_g$, the chiral fermion number of the vacuum is shifted by
\be
F_g = \sum_{l=1}^{r} \theta_l,
\label{shift}
\ee
where $\theta_l = m_l /K$ and $0< \theta_l <1$. Note that the present discussion applies to anti-chiral fermions as well. A relevant point to note however, is that since the chiral fermion number of the sigma model is in one-to-one correspondence with the grading by left-moving ghost number $g_L$, the value of $g_L$ will be shifted by $F_g$ in the $g$-twisted sector. Likewise, by considering the anti-chiral fermion number, we find that the value of $g_R$ will also be shifted by $F_g$ in the $g$-twisted sector. These observations will be important shortly.

\newsubsection{The Holomorphic Chiral Algebra}

We shall now generalise the arguments in section 2.2 to the orbifold case. Let us first consider the untwisted sector. Let us take ${\cal O}_G(z)$ and $\tilde {\cal O}_G(z)$ to be two untwisted, $Q_R$-closed operators of the half-twisted sigma model on $X$ which survive a $G$-projection, i.e., they correspond to observables in the untwisted sector of the model on $X/G$. Taking a $G$-projection just picks out the subset of operators which are $G$-invariant. This means that the analysis of the structure of the algebra furnished by the OPE's of the untwisted local operators ${\cal O}_G(z)$ and $\tilde {\cal O}_G(z)$, is exactly the same as that found in section 2.2. Thus, we can conclude that the local operators in the $Q_R$-cohomology of the untwisted sector of the half-twisted sigma model on $X/G$, supports a natural structure of a holomorphic chiral algebra as defined in section 2.2. We shall denote this chiral algebra as ${\cal A}_G$.

Let us now consider the twisted sector. Let us take ${\cal O}^g_G(z)$ and $\tilde {\cal O}^g_G(z)$ to be two $g$-twisted, $Q_R$-closed operators of the half-twisted sigma model on $X$ which survive a $G$-projection, i.e., they correspond to twisted sector observables of the model on $X/G$.  As before, the $G$-projection just restricts to the subset of operators which are $G$-invariant. Also, notice that the analysis of section 2.2 is based on worldsheet supersymmetries.\footnote{The analysis is based on the $Q_R$-invariance of the operators and the $Q_R$-exactness of the antiholomorphic stress tensor $T_{\bar z \bar z}$, which certainly involves the variation of the fields generated by the worldsheet supercharge $Q_R$.} Since the twisting by $g$ commutes with the worldsheet supersymmetries, the analysis of the structure of the algebra furnished by the    OPE's  of the twisted local operators ${\cal O}^g_G(z)$ and $\tilde {\cal O}^g_G(z)$, will be the same as that in the untwisted case. Hence, we can conclude that the local operators in the $Q_R$-cohomology of the twisted sector of the half-twisted sigma model on $X/G$, also supports a natural structure of a holomorphic chiral algebra as defined in section 2.2. We shall denote this chiral algebra as ${\cal A}^g_G$.

\newsubsection{The Holomorphic (Twisted) $N=2$ Superconformal Structure}

\smallskip\noindent{\it The Untwisted Sector}

From the expressions of $J(z)$, $Q(z)$, $T(z)$ and $G(z)$ in (\ref{J}), (\ref{Q}), (\ref{T}) and (\ref{G}) respectively, one finds that they are all invariant under the $G$-action specified in (\ref{1})-(\ref{3}). Thus, the $Q_R$-closed operators $J(z)$, $Q(z)$, $T(z)$ and $G(z)$  qualify as admissible operators in the untwisted sector of the orbifold theory.  As in section 2.2, one can verify that these operators form a complete multiplet which generates a closed, holomorphic, ($\it{twisted}$) $N=2$ superconformal algebra at the classical level. The corresponding OPE relations are as given in (2.17).

The observations made about the non-orbifolded theory in section 2.4 are also valid in the orbifold theory at hand. In particular, since the central charge in the stress tensor OPE is zero, it wll mean that the Weyl anomaly vanishes, and that the trace of the stress tensor is trivial in $Q_R$-cohomology at the quantum level. Thus, the correlation functions of observables from the untwisted sector are invariant under scalings of the worldsheet.  In addition, as with the non-orbifolded theory, the holomorphic (twisted) $N=2$ superconformal algebra of the untwisted sector will only persist in the quantum theory if there are no massive excitations, that is, if $X/G$ is a Calabi-Yau orbifold. Else, the symmetry associated with $J(z)$ and $Q(z)$ will be broken in the quantum theory, i.e., $J(z)$ and $Q(z)$ will cease to remain in the $Q_R$-cohomology of the orbifold sigma model at the quantum level. On the other hand, the symmetries associated with $T(z)$ and $G(z)$ are exact in quantum perturbation theory, and these operators will remain in the $Q_R$-cohomology of the quantum theory, regardless of whether $X/G$ is Calabi-Yau or not. (Note that $X/G$ will only be Calabi-Yau if $X$ is Calabi-Yau,  and if $G$ preserves the holomorphic $n$-form on $X$, where $n = \textrm{dim}_{\mathbb C} X$. We shall restrict our discussions    to such examples of $G$ in this paper.) Hence, we have in some sense, a reduction from an $N=2$ to $N=1$ algebra when $X/G$ is not Calabi-Yau.  We will examine this reduction more closely from a different point of view when we consider examples in section 5, where we describe the half-twisted orbifold model in terms of sheaves of CDR. We will then be able to obtain a purely mathematical interpretation of the above physical observations.

\vspace{0.2cm}\smallskip\noindent{\it The Twisted Sector}

Let us now consider an analogous set of $Q_R$-closed operators made up of twisted fields instead. To be specific, let us define
\begin{eqnarray}
\label{Jtwist}
J^g(z) & = &  \psi^g_{z i} \psi^{i,g}(z) + {F_g z^{-1}} ,\\
\label{Qtwist}
Q^g(z) & = & \psi^{i, g} \partial_z \phi^{g}_i (z), \\
\label{Ttwist}
T^g(z) & = & - \partial_z \phi^{i, g} \partial_z \phi^{g}_i (z) -  \psi^{g}_{z i} D_z \psi^{i,g}(z),\\
\label{Gtwist}
G^g(z) & = & \psi^{g}_{z i} \partial_z \phi^{i,g}(z).
\end{eqnarray}
Apart from the additional $F_g {z}^{-1}$ term in $J^g(z)$, the operators $J^g(z)$, $Q^g(z)$, $T^g(z)$ and $G^g(z)$ can be obtained by replacing the untwisted fields in $J(z)$, $Q(z)$, $T(z)$ and $G(z)$, with $g$-twisted ones (obeying the twist conditions in (\ref{twistphi})-(\ref{twist2psi})). The addition of the ${F_g z^{-1}}$ term is to account for the shift in the value of $g_L$ in the $g$-twisted sector as discussed earlier.\footnote{Recall that the charge of $J^g(z)$ is given by $g_L$, i.e., $g_L = \oint {dz \over {2\pi i}}J^g(z)$ in the $g$-twisted sector. Hence, the additional term results in a shift of $\oint {dz \over {2\pi i}} {F_g \over z} = F_g$ in $g_L$ as required.}  (Note that in writing (\ref{Jtwist})-(\ref{Gtwist}), we have used the fact that the metric $g_{i \bar j}$ is a function in the $\phi^j$ and $\phi^{\bar j}$ fields, and thus, $\partial_z g_{i \bar j} = 0$.)

Notice that $J^g(z)$, $Q^g(z)$, $T^g(z)$ and $G^g(z)$ are also invariant under the action of $G$ specified in (\ref{1})-(\ref{3}). Hence, they are admissible as operators in the twisted sector of the orbifold sigma model. Moreover, since the OPE's between the twisted bosonic and fermionic fields take the same form as the OPE's between their untwisted counterparts, one can verify that $J^g(z)$, $Q^g(z)$, $T^g(z)$ and $G^g(z)$ also satisfy the OPEs of  a holomorphic, (twisted) $N=2$ superconformal algebra:
$$
T^g(z)T^g(w) \sim \frac{2T^g(w)}{(z-w)^2}+\frac{\partial T^g(w)}{z-w}
\eqno{(3.21a)}
$$
$$
J^g(z)J^g(w) \sim \frac{d}{(z-w)^2};\
T^g(z)J^g(w) \sim -\frac{d}{(z-w)^3}+\frac{J^g(w)}{(z-w)^2}+\frac{\partial J^g(w)}{z-w}
\eqno{(3.21b)}
$$
$$
G^g(z)G^g(w) \sim 0;\ T^g(z)G^g(w)\sim \frac{2G^g(w)}{(z-w)^2}+\frac{\partial G^g(w)}{z-w};\
J^g(z)G^g(w)\sim-\frac{G^g(w)}{z-w}
\eqno{(3.21c)}
$$
$$
Q^g(z)Q^g(w)\sim 0; \ T^g(z)Q^g(w) \sim\frac{Q^g(w)}{(z-w)^2}+\frac{\partial Q^g(w)}{z-w};\
J^g(z)Q^g(w) \sim\frac{Q^g(w)}{z-w}
\eqno{(3.21d)}
$$
$$
Q^g(z)G^g(w)\sim \frac{d}{(z-w)^3}+\frac{J^g(w)}{(z-w)^2}+\frac{T^g(w)}{z-w},
\eqno{(3.21e)}
$$
where $d = \textrm{dim}_{\mathbb C}X$. Since the twist commutes with the worldsheet supersymmetries, we can make the same observations about the twisted sectors as we did for the untwisted sector. Firstly, the correlation functions of observables from the twisted sectors are invariant under scalings of the worldsheet.  Secondly, the holomorphic (twisted) $N=2$ superconformal structure of the twisted sectors will only persist in the quantum theory if there are no massive excitations, that is, if $X/G$ is a Calabi-Yau orbifold. Else, the symmetry associated with $J^g(z)$ and $Q^g(z)$ will be broken at the quantum level, i.e., $J^g(z)$ and $Q^g(z)$ will cease to remain in the $Q_R$-cohomology of the orbifold sigma model in the quantum theory. Thirdly, the symmetries associated with $T^g(z)$ and $G^g(z)$ are exact in quantum perturbation theory, and these operators will remain in the $Q_R$-cohomology at the quantum level, regardless of whether $X/G$ is Calabi-Yau or not. Thus, we also have a reduction from an $N=2$ to $N=1$ structure in the twisted sectors when $X/G$ is not Calabi-Yau. Likewise, we will be able to obtain a purely mathematical interpretation of the above physical observations when we consider some examples in section 5, where we describe the twisted sectors of the orbifold sigma model in terms of a twisted variant of the sheaf of CDR.

\newsection{Sheaf of Perturbative Observables}

\vspace{-0.4cm}
\newsubsection{General Description and Considerations}

From the orbifolding procedure outlined in section 3.1, we learn that the observables or more precisely, the local operators in the $Q_R$-cohomology of the half-twisted model on $X/G$, can be obtained by first considering the untwisted $\it{and}$ twisted local operators in the $Q_R$-cohomology of the half-twisted model on $X$, and then projecting onto those which are $G$-invariant only. Let us describe these observables in greater detail.

\smallskip\noindent{\it The Untwisted Sector}

We will start by describing the untwisted local operators in the $Q_R$-cohomology of the half-twisted model on $X$. In general, an untwisted local operator is an operator $\cal F$ that is a function of the untwisted fields $\phi^i$, $\phi^{\bar i}$, $\psi_{\bar z \bar i}$, $\psi^{\bar i}$,  $\psi_{zi}$, $\psi^i$, and their derivatives with respect to $z$ and $\bar z$.\footnote{Note here that since we are interested in local operators which define a holomorphic chiral algebra on the Riemann surface $\Sigma$, we will work locally on a flat $\Sigma$ with local parameter $z$. Hence, we need not include in our operators the dependence on the scalar curvature of $\Sigma$.} However, as we saw in section 2.1, the $Q_R$-cohomology vanishes for operators of dimension $(n,m)$ with $m \neq 0$. Since $\psi_{\bar z \bar i}$ and the derivative $\partial_{\bar z}$ both have $m=1$ (and recall from section 2.1 that a physical operator cannot have negative $m$ or $n$), $Q_R$-cohomology classes can be constructed from just $\phi^i$, $\phi^{\bar i}$, $\psi^{\bar i}$, $\psi_{z i}$, $\psi^i$ and their derivatives with respect to $z$. Note that the equation of motion for $\psi^{\bar i}$ is   $D_z \psi^{\bar i}= -{R^{\bar i}{}_{\bar k}{} _j {}_{\bar l}(\phi) \psi^{\bar k}_{z} \psi^j \psi^{\bar l}}$. Thus, we can ignore the $z$-derivatives of $\psi^{\bar i}$, since it can be expressed in terms of the other fields and their corresponding derivatives. Therefore, a chiral (i.e., $Q_R$-invariant) operator which represents a $Q_R$-cohomology class is given by
\be
{\cal F}(\phi^i,\partial_z\phi^i,\partial_z^2\phi^i,\dots;
\phi^{\bar i},\partial_z\phi^{\bar i},\partial_z^2\phi^{\bar i},\dots; \psi_{zi}, \partial_z\psi_{zi}, \partial_z^2\psi_{zi} \dots; \psi^i, \partial_z\psi^i, \partial_z^2\psi^i \dots ; \psi^{\bar i}),
\label{F}
\ee
where we have tried to indicate that $\cal F$ might depend on $z$
derivatives of $\phi^{i}$, $\phi^{\bar i}$, $\psi_{zi}$ and $\psi ^i$ of arbitrarily high order, though not on derivatives of $\psi^{\bar i}$. If the scaling dimension of $\cal F$ is bounded, it will mean that $\cal F$ depends only on the derivatives of fields up to some finite order, is a polynomial of bounded degree in those, and/or is a bounded polynomial in $\psi_{zi}$. Notice that $\cal F$ will always be a polynomial of finite degree in $\psi^i$, $\psi_{zi}$ and $\psi ^{\bar i}$, simply because $\psi^i$, $\psi_{zi}$ and $\psi^{\bar i}$ are fermionic and can only have a finite number of components before they vanish due to their anti-commutativity. However, the dependence of $\cal F$ on $\phi^i$, $\phi^{\bar i}$ (as opposed to their derivatives) need not have any simple form. Nevertheless, we can make the following observation - from the $(g_L, g_R)$ ghost numbers of the fields in section 2.1, we see that if $\cal F$ is homogeneous of degree $k$ in $\psi^{\bar i}$,     then it has ghost numbers $(g_L, g_R) =( p, k)$, where $p$ is determined by the net number of  $\psi^i$ over $\psi_{zi}$ fields (and/or of their corresponding derivatives) in $\cal F$.

\hspace {-0.2cm}A general $g_R= k$ operator ${\cal F} (\phi^i,\partial_z\phi^i,\dots; \phi^{\bar i},\partial_z\phi^{\bar i},\dots; \psi_{zi}, \partial_z \psi_{zi}, \dots; \psi^i, \partial_z \psi^i, \dots ; \psi^{\bar i})$ can be interpreted as a $(0,k)$-form on $X$ with values in a
certain tensor product bundle. In order to illustrate the general idea behind this interpretation, we will make things explicit for
operators of dimension $(0,0)$ and $(1,0)$. Similiar arguments will likewise apply for operators of higher dimension. For dimension $(0,0)$, the most general operator takes the form ${\cal F}(\phi^i,\phi^{\bar i}; \psi^j ; \psi^{\bar j})= f_{\bar j_1,\dots,\bar
j_k; j_1, \dots, j_q}(\phi^i, \phi^{\bar i}) \psi^{\bar j_i}\dots \psi^{\bar j_k} \psi^{j_1} \dots \psi^{j_q}$; thus, $\cal F$ may depend on $\phi^i$, $\phi^{\bar i}$ and $\psi^j$, but not on their derivatives, and is $k^{th}$ order in $\psi^{\bar j}$. Mapping $\psi^{j}$ to $d\phi^{j}$ and $\psi^{\bar j}$ to $d\phi^{\bar j}$, such an operator corresponds to an ordinary $(0,k)$-form $f_{\bar j_1,\dots,\bar j_k}(\phi^i, \phi^{\bar i})d\phi^{\bar j_1}\dots d\phi^{\bar j_k}$ on $X$ with values in the bundle $\Lambda^q {T^*X}$. Alternatively, it can be interpreted as an ordinary $(q,k)$-form $f_{\bar j_1,\dots,\bar j_k; j_1, \dots, j_q}(\phi^i, \phi^{\bar i}) d\phi^{\bar j_1}\dots d\phi^{\bar j_k}  d\phi^{j_1}\dots d\phi^{j_q}$ on $X$.\footnote{Note that $q, k \leq \textrm{dim}_{\mathbb C} X$ due to the anti-commutativity of $\psi^j$ and $\psi^{\bar j}$ as required of the wedge product of one-forms $d\phi^i$ and $d\phi^{\bar i}$.} For dimension $(1,0)$, there are four general cases. In the first case, we have an operator ${\cal F}(\phi^l,\partial_z\phi^j, \phi^{\bar l};\psi^j; \psi^{\bar j})=f^{\bar i}{}_{\bar j_1,\dots,\bar j_k ; j_1, \dots, j_q}(\phi^l,\phi^{\bar l}) g_{\bar i j}\partial_z\phi^j \psi^{\bar j_1}\dots\psi^{\bar j_k}\psi^{j_1} \dots \psi^{j_q}$ that is linear in $\partial_z\phi^j$ and
does not depend on any other derivatives. It is a $(0,k)$-form on $X$ with values in the tensor product bundle of $\overline{TX}$ with $\Lambda^q T^*X$; alternatively, it is a $(q,k)$-form on $X$ with values in the bundle $\overline {TX}$. Next, we can have an operator ${\cal F}(\phi^l, \phi^{\bar l}, \partial_z \phi^{\bar s};\psi^j ; \psi^{\bar j})=f^i{}_{\bar j_1,\dots,\bar j_k; j_1, \dots, j_q}(\phi^l, \phi^{\bar l}) g_{i \bar s}\partial_z \phi^{\bar s}\psi^{\bar j_i}\dots \psi^{\bar j_k} \psi^{j_1} \dots \psi^{j_q}$ that is linear in $\partial_z \phi^{\bar s}$ and does not depend on any other derivatives. It is a $(0,k)$-form on $X$ with values in the tensor product bundle of $TX$ with $\Lambda^q T^*X$; alternatively, it is a $(q,k)$-form on $X$ with values in the bundle $TX$. In the third case, we have an operator ${\cal F}(\phi^l, \phi^{\bar l};\psi^j, \partial_z\psi^i; \psi^{\bar j})=  f_{i \bar j_1,\dots,\bar j_k ; j_1, \dots, j_q}(\phi^l,\phi^{\bar l}) \partial_z\psi^{i} \psi^{\bar j_1}\dots\psi^{\bar j_k} \psi^{j_1} \dots \psi^{j_q}$ that is linear in $\partial_z\psi^{i}$ and does not depend on any other derivatives. Such an operator corresponds to a $(0,k)$-form on $X$ with values in the (antisymmetric) tensor product bundle of ${\cal V}^*$ with $\Lambda^q T^*X$, where the local holomorphic sections of the vector bundle ${\cal V}$ are spanned  by $\partial_z \psi^i$; alternatively, it is a $(q,k)$-form with values in the bundle $\cal V^*$.  In the last case, we have an operator ${\cal F} (\phi^l, \phi^{\bar l}; \psi_{zi}, \psi^j ; \psi^{\bar j})=f^{i}{}_{\bar j_1,\dots,\bar j_k; j_1, \dots, j_q}(\phi^l, \phi^{\bar l}) \psi_{zi}\psi^{\bar j_i}\dots \psi^{\bar j_k} \psi^{j_1} \dots \psi^{j_q}$; here, $\cal F$ may depend on $\phi^i$, $\phi^{\bar i}$, $\psi_{zi}$ and $\psi^i$, but not on their derivatives. Such an operator corresponds to a $(0,k)$-form on $X$ with values in the (antisymmetric) tensor product bundle of $TX$ with $\Lambda^qT^*X$. In a similiar fashion, for any integer $n>0$, the operators of dimension $(n,0)$ and ghost number $g_R = k$ can be interpreted as $(0,k$)-forms with values in a certain tensor product bundle over $X$. This structure persists in quantum perturbation theory, but there  may be perturbative corrections to the complex structure of the bundle.

Having described the untwisted local operators of the half-twisted model on $X$, one just needs to single out those operators which are $Q_R$-closed and $G$-invariant. In other words, the operators in the untwisted sector of the $Q_R$-cohomology of the half-twisted orbifold model on $X/G$, where $G=\mathbb Z_K$, are given by the operators $\cal F$ on $X$ which are $Q_R$-closed  and invariant under the transformations  (\ref{1})-(\ref{3}). We shall henceforth call them ${\cal F}_G$.

\smallskip\noindent{\it The Twisted Sector}

The local operators corresponding to the observables in the twisted sector  can be obtained by considering its composition in terms of the twisted fields $\phi^{i,g}$, $\phi^{\bar i, g}$, $\psi^{i,g}$, $\psi^{\bar i ,g}$, $\psi^g_{zi}$,  $\psi^g_{\bar z \bar i}$ and their respective derivatives with respect to $z$ and $\bar z$. However, as mentioned above, the $Q_R$-cohomology vanishes for operators of dimension $(n,m)$ with $m \neq 0$.\footnote{The analysis of the untwisted case carries over the twisted case at hand because the twist commutes with the worldsheet supersymmetries.} Since $\psi^g_{\bar z \bar i}$ and the derivative $\partial_{\bar z}$ both have $m=1$, $Q_R$-cohomology classes in the twisted sector can be constructed from just $\phi^{i,g}$, $\phi^{\bar i ,g}$, $\psi^{\bar i , g}$, $\psi^g_{z i}$, $\psi^{i,g}$ and their derivatives with respect to $z$. As in the untwisted case explained above, we can ignore the $z$-derivatives of $\psi^{\bar i ,g}$ since it can be expressed in terms of the other twisted fields and their corresponding derivatives. Therefore, a local operator which represents a $Q_R$-cohomology class in the twisted sector is given by
\be
{\cal F}^g(\phi^{i,g},\partial_z\phi^{i,g} ,\partial_z^2\phi^{i,g},\dots;
\phi^{\bar i, g},\partial_z\phi^{\bar i, g},\partial_z^2\phi^{\bar i, g},\dots; \psi^g_{zi}, \partial_z\psi^g_{zi}, \partial_z^2\psi^g_{zi} \dots; \psi^{i,g}, \partial_z\psi^{i,g}, \partial_z^2\psi^{i,g} \dots ; \psi^{\bar i,g}),
\label{Fg}
\ee
where we have tried to indicate that ${\cal F}^g$ might depend on $z$ derivatives of $\phi^{i,g}$, $\phi^{\bar i,g}$, $\psi^g_{zi}$ and $\psi ^{i,g}$ of arbitrarily high order, though not on derivatives of $\psi^{\bar i,g}$. Likewise, if the scaling dimension of ${\cal F}^g$ is bounded, it will mean that ${\cal F}^g$ depends only on the derivatives of fields up to some finite order, is a polynomial of bounded degree in those, and/or is a bounded polynomial in $\psi^g_{zi}$. Notice that ${\cal F}^g$ will always be a polynomial of finite degree in $\psi^{i,g}$, $\psi^g_{zi}$ and $\psi ^{\bar i , g}$, simply because $\psi^{i,g}$, $\psi^g_{zi}$ and $\psi^{\bar i,g}$ are fermionic and can only have a finite number of components before they vanish due to their anti-commutativity. However, the dependence of ${\cal F}^g$ on $\phi^{i,g}$, $\phi^{\bar i , g}$ (as opposed to their derivatives) need not have any simple form. Nevertheless, following the same arguments above involving the untwisted local operators, one can in general interpret a twisted local operator ${\cal F}^g$ with $k$ number of $\psi^{\bar i, g}$ fields, as a $(0,k)$-form on $X$ with values in a certain tensor product bundle. As mentioned before, this structure persists in quantum perturbation theory, but there  may be perturbative corrections to the complex structure of the bundle. In addition, note that due to the shift in the fermion number of the vacuum of the twisted sector, we can make the following observation - we see that if ${\cal F}^g$ is homogeneous of degree $k$ in $\psi^{\bar i}$, then it has (possibly fractional) ghost numbers $(g^g_L , g^g_R) =( p + F_g, k + F_g)$, where $p$ is determined by the net number of  $\psi^{i,g}$ over $\psi^g_{zi}$ fields (and/or of their corresponding derivatives) in ${\cal F}^g$.

Having described the twisted local operators of the half-twisted model on $X$, one just needs to single out those operators which are $Q_R$-closed and $G$-invariant. In other words, the operators in the twisted sector of the $Q_R$-cohomology of the half-twisted orbifold model on $X/G$, where $G=\mathbb Z_K$, are given by the operators ${\cal F}^g$ on $X$ which are $Q_R$-closed and invariant under the transformations  (\ref{1})-(\ref{3}). We shall henceforth call them ${\cal F}^g_G$.

\vspace{0.0cm}\smallskip\noindent{\it The Action of $Q_R$}

Let us now describe the action of $Q_R$ on such operators. Note that the following arguments involving the operators in the twisted and untwisted sectors are identical.\footnote{The follow-on arguments involve the supersymmetry transformations on the worldsheet, and since the twist commutes with the worldsheet supersymmetries, the discussion involving the twisted fields is the same.} Hence, for brevity, we shall restrict our discussion to the operators in the untwisted sector only.

At the classical level, if we interpret $\psi^{\bar i}$ as $d\phi^{\bar i}$, then $Q_R$ acts on functions of $\phi^i$ and $\phi^{\bar i}$, and  is simply the $\bar\partial $ operator on $X$. This follows from the transformation laws $\delta\phi^{\bar i}=\psi^{\bar i}$, ${\delta\phi^i} = 0$, ${\delta \psi^{\bar i}}=0$. If $X$ is flat, the interpretation of $Q_R$ as the $\bar\partial$ operator will remain valid when $Q_R$ acts on a more general operator ${\cal F}(\phi^i,\partial_z\phi^i,\dots;\phi^{\bar i},\partial_z \phi^{\bar i},\dots; \psi_{zi}, \dots ; \psi^i, \dots; \psi^{\bar i})$ that does depend on the derivatives of $\phi^i$ and $\phi^{\bar i}$. The reason for this is that if $R_{i \bar j k \bar l} = 0$, we will have the equation of motion $D_z \psi^{\bar i}=0$, which then means that one can neglect the action of $Q_R$ on derivatives $\partial_z^m\phi^{\bar i}$ with $m>0$. Moreover, since $\delta \psi_{zi} = 0$ for a flat metric, one can also ignore the action of $Q_R$ on the $\psi_{zi}$ fields and their derivatives $\partial_z^m \psi_{zi}$ with $m>0$. However, $X$ is not flat in general, and $Q_R$ need not always act as the $\bar \partial$ operator on a general operator at the classical level.

Perturbatively at the quantum level, there will be corrections to the action of $Q_R$. Let us now attempt to better understand the nature of such perturbative corrections. To this end, let $Q^{cl}_R$  denote the classical approximation to $Q_R$. Note that since sigma model perturbation theory is local on $X$, and it depends on an expansion of fields such as the metric tensor of $X$ in a Taylor series up to some given order, the perturbative corrections to $Q^{cl}_R$ will also be local on $X$, where order by order, they consist  of differential operators whose possible degree grows with the order of perturbation theory. In fact, the perturbative corrections to $Q^{cl}_R$ must represent $Q^{cl}_R$-cohomology classes. To see this, let us perturb the classical expression so that $Q_R = Q^{cl}_R + \epsilon Q' + O(\epsilon^2) $, where $\epsilon$ is a parameter that controls the magnitude of the perturbative quantum corrections at each order of the expansion. To ensure that we continue to have $Q_R^2 = 0$, we require that $\{Q^{cl}_R, Q' \} = 0$. In addition, if $Q'=\{Q^{cl}_R,\Lambda\}$ for some $\Lambda$, then via the conjugation of $Q_R$ with $\exp(-\epsilon\Lambda)$ (which results in a trivial change of basis in the space of $Q_R$-closed local operators), the correction by $Q'$ can be removed. Hence, $Q'$ represents a $Q^{cl}_R$-cohomology class. Since $Q'$ is to be generated in sigma model perturbation theory, it must be constructed locally from the fields appearing in the sigma model action.

It will be useful later for us to discuss the case when $X$ is flat now. In this case, $Q^{cl}_R$ will act as the $\bar \partial$ operator as argued above. In other words,  perturbative corrections to $Q^{cl}_R$ will come from representatives of $\bar\partial$-cohomology classes on $X$.  An example of a representative of a $\bar\partial$-cohomology class on $X$ which may contribute as a perturbative correction to the classical expression $Q_R = Q^{cl}_R$ would be an element of $H^1(X, \Omega^{2, cl}_X)$. It is also constructed locally from fields appearing in the action $S$, and is used to deform the action. In fact, its interpretation as a perturbative correction $Q'$ can be shown to be consistent with its interpretation as the moduli of the chiral algebra in this case. To see this, notice that its interpretation as $Q'$ means that it will parameterise a family of $Q_R = Q^{cl}_R + \epsilon Q'$ operators at the quantum level. Since the chiral algebra of local operators is defined to be closed with respect to the $Q_R$ operator, it will vary with the $Q_R$ operator  and consequently with $H^1(X, \Omega^{2,cl}_X)$, that is, one can associate the moduli of the chiral algebra with $H^1(X, \Omega^{2,cl}_X)$. Apparently, this is the only one-dimensional $\bar \partial$-cohomology class on $X$ that can be constructed locally from fields appearing in the action, and it may be that it completely determines the perturbative corrections to $Q_R = Q^{cl}_R$. This observation  will be important in section 4.3, when we discuss the $Q_R$-cohomology of local operators  (on a small open set $U \subset X$) furnished by a sheaf of CDR associated with a free $bc$-$\beta\gamma$ system.

The fact that $Q_R$ does not even act as the $\bar \partial$ operator at the classical level seems to suggest that one needs a more general framework than just ordinary Dolbeault or $\bar \partial$-cohomology to describe the $Q_R$-cohomology of the half-twisted orbifold model. Indeed, as we will show shortly in section 4.2, the appropriate description of the $Q_R$-cohomology of local operators spanning the chiral algebra will be given in terms of the more abstract notion of Cech cohomology.

\vspace{0.0cm}\smallskip\noindent{\it Support of Twisted Sector Observables on Fixed-Point Set of $X/G$}

From the fixed-point theorem \cite{mirror manifolds}, and the variation of the fermionic fields in (\ref{susyqr}), we see that for the half-twisted model on $X$,  the field configurations will  be governed by single-valued holomorphic maps characterised by $\partial_{\bar z} \phi^i = 0$. This continues to be true for the untwisted sector of the half-twisted orbifold model on $X/G$ because the various untwisted fields have trivial monodromy around points on the worldsheet. However, the same cannot be said about the twisted sectors. In fact, the notion of a single-valued holomorphic map cannot be defined in this case. This leads to some important consequences for the twisted sector observables. Let us examine this more closely.

Firstly, let us review the method devised in \cite{Hamidi} for computing interactions on orbifolds. Consider the bosonic field $\phi(\sigma)$ of the half-twisted sigma model on $X$. Let it be $g$-twisted such that  $\phi(\sigma + 2 \pi) = g \phi(\sigma)$.  A $g$-twisted field configuration such as $\phi(\sigma)$, inserted at a point $z$ on the worldsheet Riemann surface $\Sigma$, must result in a map $\Phi(z)$ from $\Sigma$ to the target space $X$, that satisfies $\Phi(e^{2 \pi i} z) = g \Phi(z)$. In other words, the twisted sectors involve multivalued maps $\Phi: \Sigma \to X$ that have specific monodromies around  points of insertion of twisted operators and states. However, as explained in \cite{Hamidi}, we can find an equivalent description involving $\it{single}$-$\it{valued}$ maps by choosing a cover $\tilde {\Sigma}$ of $\Sigma$ on which $G$ acts whilst preserving the metric and complex structure, that is, $\Sigma \cong \tilde{\Sigma} /G$. Then, the corresponding single-valued map with equivalent information will be given by $\tilde \Phi : \tilde {\Sigma} \to X$, and it must obey
\be
\tilde {\Phi} (g \tilde{z}) = g \tilde {\Phi}(\tilde z)
\label{newmap}
\ee
for any group element $g \in G$, so that $\phi(\sigma)$  will have the appropriate multi-valuedness about  each insertion point of a twisted operator or state. $\tilde z$ is the coordinate on $\tilde \Sigma$, and the relation in (\ref{newmap}) implies that the twisted field configurations are governed by single-valued, holomorphic equivariant maps $\tilde \Phi$ to $X$ when one considers the equivalent theory on $\tilde \Sigma$ instead of $\Sigma$.

An important point to note is the following. Consider a $g$-twisted observable that is inserted at a point $p_i$. Let the group action $g_i$ on this observable be such that a small loop around $p_i$ will lift to a line which connects $\tilde {p_i}$ to $g_i \tilde {p_i}$ in $\tilde \Sigma$. Now reduce the size of this loop gradually until it shrinks to the point $p_i$. The continuity of the $G$-action will mean that at $p_i$, we will have the condition $g_i\tilde {p_i} = \tilde {p_i}$ on $\tilde \Sigma$, that is, $p_i$ descends from a fixed point of $g_i$ on $\tilde \Sigma$. This applies for any general point $p$ on $\Sigma$ over which one can insert a $g$-twisted observable (that is, a $g$-twisted sector observable of the half-twisted model on $X$). Then, together with (\ref{newmap}), we will have ${\tilde \Phi}(g \tilde p) = {\tilde \Phi}( \tilde p) = g{\tilde \Phi}(\tilde p)$. In other words, ${\tilde \Phi}( \tilde p) \in X_g$. This observation has some non-trivial consequences as follows. Suppose that we consider ${\cal O}^g(p)$, a $g$-twisted operator observable inserted at a point $p$ on $\Sigma$. As explained above, it can be interpreted as some $(0,k)$-form with values in some tensor product bundle, which we will assume does not vanish on $X$ in all generality. Since a twisted sector observable at $p$ will always be evaluated at some ${\tilde \Phi(\tilde p}) \equiv x \in X_g$, it will mean that if the restriction of ${\cal O}^g(p)$ (as a $(0,k)$-form with values in some tensor product bundle)  to $X_g$ is zero, it's physical contribution to correlation functions will vanish. Thus, we find that the twisted sector observables of the half-twisted model on $X$, represented by ${\cal O}^g(z)$, are effectively supported on the fixed-point set $X_g$ of $X$. One can see that the argument will also hold for $G$-invariant operators ${\cal O}^g_G(p)$. Therefore, we can conclude that the twisted sector observables of the half-twisted orbifold model on $X/G$, represented by ${\cal O}^g_G(z)$, are effectively supported on the fixed-point set $X_g$ of $X$ as well.

\newsubsection{Sheaves of Chiral Algebras}

We shall now explain the idea of a ``sheaf of chiral algebras'' on $X$. To this end, note that both the local  operators in the $Q_R$-cohomology (that is, operators which are local on the Riemann surface $\Sigma$), and the fermionic symmetry generator $Q_R$, can be described locally on $X$. Hence, one is free to restrict the local operators to be well-defined not throughout $X$, but only on a given open set $U \subset X$. Since in perturbation theory, we are considering, in the untwisted and twisted sectors, trivial maps $\Phi :\Sigma \to X$ and $\tilde \Phi : \tilde \Sigma \to X$ with no multiplicities, any operator defined in an open set $U$ will have a sensible operator product expansion with another operator defined  in $U$. From here, one can naturally proceed to restrict the definition of the (untwisted and twisted) operators to smaller open sets, such that  a global definition of the (untwisted and twisted) operators can be obtained by gluing together the open sets on their unions and intersections. From this description, in which one associates a chiral algebra spanned by the local operators and their OPE's to every open set $U \subset X$, we get what is known mathematically as a ``sheaf of chiral algebras''. We shall call these sheaves of chiral algebras corresponding to observables in the untwisted and twisted sectors $\widehat {\cal A}$ and $\widehat{\cal A}^g$ respectively.

\vspace{0.0cm}\smallskip\noindent{\it Description of ${\cal A}_G$ and ${\cal A}^g_G$ via Cech Cohomology}

In perturbation theory, one can also describe the $Q_R$-cohomology classes of local operators by a form of Cech cohomology. This abstract description will take us to the mathematical point of view on the subject \cite{Frenkel}. In essence, we will show that the chiral algebras ${\cal A}_G$ and ${\cal A}^g_G$,  spanned by the $Q_R$-cohomology classes of $G$-invariant local operators in the untwisted and twisted sectors of the half-twisted orbifold sigma model on $X/G$, can be represented, in perturbation theory, by $G$-invariant classes of the Cech cohomology  of the sheaves $\widehat {\cal A}$ and $\widehat {\cal A}^g$ spanned by locally-defined chiral operators which are untwisted and twisted respectively. We shall demonstrate this for ${\cal A}_G$ first. The approach for ${\cal A}^g_G$ will be analogous.

To begin with, we shall demonstrate that the local operators $\cal F$ in the $Q_R$-cohomology of the half-twisted model on $X$, can be described in terms of a Cech cohomology. Thereafter, we will project onto the $G$-invariant subspace to obtain the corresponding operators ${\cal F}_G$ which span ${\cal A}_G$, thus providing a Cech cohomological description of ${\cal A}_G$ as claimed. To this end, let us start by considering an open set $U \subset X$ that is isomorphic to a contractible space such as an open ball in $\mathbb C^n$, where $n = {\textrm {dim}}_{\mathbb C} (X)$. Because $U$ is a contractible space, any bundle over $U$ will be trivial. By applying this statement on the tangent bundle over $U$, we find that the curvature of $U$ vanishes, i.e., it is flat. From the discussion in section 4.1, we find that $Q_R$ will then act as the $\bar \partial$ operator on any local operator $\cal F$ in $U$. In other words, $\cal F$ can be interpreted as a $\bar\partial$-closed $(0,k)$-form with values in a certain  tensor product bundle $\widehat F$ over $U$. Thus, in the absence of perturbative corrections at the classical level, any operator ${\cal F}$ on $U$ in the $Q_R$-cohomology will be classes of $H^{0,k}_{\bar \partial}(U, \widehat {F})$. As explained, $\widehat F$ must be a trivial bundle over $U$, which means that $\widehat {F}$ will always possess a global section, i.e., it corresponds to a soft sheaf.  Since the higher Cech cohomologies of a soft sheaf are trivial \cite{Wells}, we will have $ {H_{\textrm{Cech}}^{k} (U, {\widehat {F}} )} = 0$ for $k > 0$. Mapping this back to Dolbeault cohomology via the Cech-Dolbeault isomorphism, we find that $H^{0,k}_{\bar \partial}(U, \widehat {F}) = 0$ for $k > 0$.  Note that small quantum corrections in the perturbative limit can only annihilate cohomology classes and not create them. Hence, in perturbation theory, it follows that the local operators ${\cal F}$   on $U$ with positive values of $g_R$ must vanish in $Q_R$-cohomology.

Now consider a good cover of  $X$ by open sets $\{U_a \}$.  Since the intersection of open sets $\{U_a \}$ also give open sets (isomorphic to open balls in $\mathbb C^n$),  $\{U_a \}$ and all of their intersections have the same property as $U$    described above: $\bar\partial$-cohomology and hence $Q_R$-cohomology vanishes for positive values of $g_R$ on $\{U_a \}$ and their intersections.

Let the operator ${\cal F}_{1}$ on $X$ be a $Q_R$-cohomology class with $g_R = 1$. It is here that we shall demonstrate an isomorphism between the $Q_R$-cohomology and a Cech cohomology.  When restricted to an open set        $U_a$, the operator ${\cal F}_{1}$ must be trivial in $Q_R$-cohomology, i.e., ${\cal F}_{1} =\{Q_R,{\cal C}_a\}$, where $Q_R$ has $g_R=1$, and ${\cal C}_a$ is an operator of $g_R=0$ that is well-defined in $U_a$.

Now, since $Q_R$-cohomology classes such as ${\cal F}_{1}$ can be globally-defined on $X$, we have ${\cal F}_{1} =\{Q_R ,{\cal C}_a\}=\{Q_R,{{\cal C}_b}\}$ over the intersection  $U_a\cap U_b$, so $\{Q_R,{{\cal C}_a}- {{\cal C}_b}\}=0$. Let
${\cal C}_{ab}= {{\cal C}_a}- {{\cal C}_b}$.  For each $a$ and $b$, ${\cal C}_{ab}$ is defined in
$U_a\cap U_b$. Therefore, for all $a,b,c$, we have
\be
{\cal C}_{ab}=     -{\cal C}_{ba}, \quad {{\cal C}_{ab}}+ {{\cal C}_{bc}} + {{\cal C}_{ca}} =0.
\label{cab}
\ee
Moreover, for ($g_R =0$) operators ${\cal K}_a$ and ${\cal K}_b$, whereby $\{ Q_R, {{\cal K}_a} \} = \{ Q_R, {{\cal K}_b} \}= 0$, we have an equivalence relation
\be
{\cal C}_{ab} \sim  {{\cal C'}_{ab} = {{\cal C}_{ab} + {\cal K}_a - {\cal K}_b}}.
\label{cab1}
\ee
Note that the collection $\{C_{ab} \}$ are operators in the $Q_R$-cohomology with well-defined operator product expansions.

Since the local operators with positive values of $g_R$ vanish in $Q_R$-cohomology on an arbitrary open set $U$, the sheaf $\widehat {\cal A}$ of the chiral algebra of untwisted operators has for its local sections  the $\psi^{\bar i}$-independent (i.e. $g_R =0$) operators  ${\widehat {\cal F}} (\phi^i,\partial_z\phi^i,\dots; \phi^{\bar i},\partial_z\phi^{\bar i},\dots; \psi_{zi}, \partial_z \psi_{zi}, \dots; \newline \psi^i, \partial_z \psi^i, \dots)$ that are annihilated by $Q_R$. Each $C_{ab}$ with $g_R =0$ is thus a section of $\widehat{\cal A}$ over the intersection $U_a\cap U_b$. From (\ref{cab}) and (\ref{cab1}), we find that the collection $\{C_{ab} \}$ defines the elements  of the first Cech cohomology group $H_{{\rm Cech}}^1(X, \widehat{\cal A})$.

Next, note that the $Q_R$-cohomology classes are defined as those operators which are $Q_R$-closed, modulo those which can be globally written as $\{ Q_R, \dots \}$ on $X$. In other words, ${\cal F}_1$ vanishes in $Q_R$-cohomology if we can write it as ${\cal F}_1 = \{Q_R ,{\cal C}_a\}=\{Q_R,{{\cal C}_b}\} =  \{Q_R ,{\cal C}\}$, i.e., ${\cal C}_a = {\cal C}_b$ and hence ${\cal C}_{ab} = 0$. Therefore, a vanishing $Q_R$-cohomology with $g_R =1$ corresponds to a vanishing first Cech cohomology. Thus, we have obtained a map between the $Q_R$-cohomology with $g_R =1$ and a first Cech cohomology.

One can also run everything backwards and construct an inverse of this map. Suppose we are given a family $\{ {\cal C}_{ab} \}$ of sections of $\widehat {\cal A}$ over the corresponding intersections $\{U_a \cap U_b \}$, and they obey (\ref{cab}) and (\ref{cab1}) so that they define the elements of $H^1(X , \widehat{\cal A})$. We can then proceed as follows.  Let the set $\{f_a \}$ be partition of unity subordinates to the open cover of $X$ provided by $\{U_a\}$. This means that the elements of $\{f_a \}$ are continuous functions on $X$, and they  vanish outside the corresponding elements in  $\{U_a \}$ whilst obeying $\sum_a f_a=1$. Let ${\cal F}_{1,a}$ be a chiral operator defined in $U_a$ by ${\cal F}_{1,a}= \sum_c[ Q_R, f_c]  {\cal C}_{ac}$.\footnote{Normal ordering of the operator product of $[Q_R, f_c(\phi^i,\phi^{\bar i})]$ with ${\cal C}_{ac}$ is needed for regularisation purposes.} ${\cal F}_{1,a}$ is well-defined throughout $U_a$, since in $U_a$, $[Q_R, f_c]$ vanishes wherever ${\cal C}_{ac}$ is not defined. Clearly, ${\cal F}_{1,a}$ has $g_R =1$, since ${\cal C}_{ac}$ has $g_R =0$ and $Q_R$ has $g_R=1$. Moreover, since ${\cal F}_{1,a}$ is a chiral operator defined in $U_a$, it will mean that $\{  Q_R, {\cal F}_{1,a}\} = 0$  over $U_a$. For any $a$ and $b$, we have ${\cal F}_{1,a}- {\cal F}_{1,b} = \sum_c [Q_R , f_c]  ({\cal C}_{ac}- {\cal C}_{bc})$. Using (\ref{cab}), this is $\sum_c [Q_R , f_c] {\cal C}_{ab} = [Q_R, \sum_c f_c ] {\cal C}_{ab}$. This vanishes since $\sum_c f_c = 1$. Hence, ${\cal F}_{1,a} = {\cal F}_{1,b}$ on $U_a \cap U_b$, for $\it{all}$ $a$ and $b$. In other words, we have found a globally-defined $g_R =1$ operator ${\cal F}_1$ that obeys $\{Q_R, {\cal F}_1 \} = 0$ on $X$. Notice that ${\cal F}_{1,a}$ and thus ${\cal F}_1$ is not defined to be of the form $\{Q_R, \dots \}$. Therefore, we have obtained a map from the Cech cohomology group $H^1(X, \widehat {\cal A})$ to the $Q_R$-cohomology group with $g_R =1$, i.e., $Q_R$-closed, $g_R =1 $ operators modulo those that can be globally written as $\{\overline Q_+, \dots \}$. The fact that this map is an inverse of the first map can indeed be verified.

Since there is nothing unique about the $g_R =1$ case, we can repeat the above procedure for operators ${\cal F}_{g_R}$ with $g_R > 1$. In doing so, we find that  the $ Q_R$-cohomology of the half-twisted model on $X$ coincides with the Cech cohomology of $\widehat {\cal A}$ for all $g_R$. As mentioned, we will need to project onto the $G$-invariant subspace of local operators to get ${\cal F}_G$, the observables in the untwisted sector of the half-twisted orbifold model on $X/G$. In doing so, we find that the chiral algebra ${\cal A}_G$ will be given by  $\bigoplus_{g_R} H^{g_R}_{\textrm {Cech}} (X, {\widehat{\cal A}})^G$ as a vector space, where the superscript indicates the $G$-invariant subset of the Cech cohomology group.

By repeating the exact same arguments above but by considering ${\cal F}^g$ instead of $\cal F$, and $\widehat{\cal A}^g$ instead of $\widehat{\cal A}$, keeping in mind the shift in the value of $g_R$ in the twisted sectors as discussed in section 3.1, we find that for the twisted sectors of the half-twisted orbifold model on $X/G$, the chiral algebra ${\cal A}^g_G$ will be given by  $\bigoplus_{g_R -F_g} H^{g_R -F_g}_{\textrm {Cech}} (X, {\widehat{\cal A}^g})^G$ as a vector space, where again, the superscript indicates the $G$-invariant subset of the Cech cohomology group. As there will be no ambiguity, we shall henceforth omit the label ``Cech'' when referring to the cohomology of $\widehat {\cal A}$ and $\widehat {\cal A}^g$.

Note that in the mathematical literature, the sheaves $\widehat {\cal A}$ and $\widehat {\cal A}^g$ are also known as sheaves of vertex superalgebras. They are studied purely from the Cech viewpoint; the fields $\psi^{\bar i}$ and $\psi^{\bar i,g}$ are omitted and locally on $X$, one considers operators constructed only from $\phi^i$, $\phi^{\bar i}$, $\psi_{zi}$, $\psi^i$, and $\phi^{i,g}$, $\phi^{\bar i,g}$, $\psi^g_{zi}$, $\psi^{i,g}$, and their  $z$-derivatives respectively. The $Q_R$-cohomology classes spanning the chiral algebras ${\cal A}_G$ and ${\cal A}^g_G$ with positive $g_R$ and $k =(g_R - F_g)$ are correspondingly constructed as $G$-invariant Cech $g_R$- and $k$-cocycles respectively. However, in the physical description via a Lagrangian and $Q_R$ operator, the sheaves  $\widehat {\cal A}$ and $\widehat {\cal A}^g$,  and their  cohomologies,	 are given a $\bar \partial$-like description, where Cech $g_R$- and $k$-cocycles are represented by operators that are $g^{th}_R$ and $k^{th}$ order in the fields $\psi^{\bar i}$ and $\psi^{\bar i,g}$ respectively. Notice that the mathematical description does not involve any form of perturbation theory at all. Instead, it utilises the abstraction of Cech cohomology to define the spectrum of operators in the quantum model. It is in this sense that the study of the orbifold sigma model is given a rigorous foundation in the mathematical literature. In section 4.5, we will work out the specific type of vertex superalgebras that the sections of the sheaves $\widehat {\cal A}$ and $\widehat {\cal A}^g$ furnish.

\newsubsection{Relation to a Free $bc$-$\beta\gamma$ System}

Now, we shall express in a physical language a few key points that are made in the mathematical literature \cite{Frenkel} starting from a Cech viewpoint.  Let us start by providing a convenient description of the local structure of the sheaves $\widehat {\cal A}$ and $\widehat {\cal A}^g$ on $X$. To this end, we will describe in a new way the operators in the $Q_R$-cohomology that are regular in a small open set $U \subset X$, where we assume $U$ to be isomorphic to an open ball in $\mathbb C^n$ and is thus contractible. We shall first discuss the sheaf $\widehat {\cal A}$. The arguments involving $\widehat {\cal A}^g$ will be similar.

\vspace{0.4cm}\smallskip\noindent{\it The Sheaf $\widehat {\cal A}$ on $X$}

To describe the local structure, we can pick a hermitian metric that is flat when restricted to $U$. The action, in general, also contains deformation terms derived from an element of $H^1(X,\Omega^{2,cl}_X)$ as explained in section 2.3.  From (\ref{ST}) and the discussion thereafter, we see that these terms are also $Q_R$-exact locally, and therefore can be discarded in analysing the local structure in $U$. Thus, the local action (derived from the flat hermitian metric) of the half-twisted sigma model on $U$ is
\be
I = {1 \over 2 \pi} \int_{\Sigma} |d^2 z| \sum_{i, \bar j} \delta_{i \bar j} \left ( \partial_z \phi^{\bar j} \partial_{\bar z}\phi^i + \psi^{\bar j}_z \partial_{\bar z} \psi^i +  \psi ^i_{\bar z} \partial_z \psi^{\bar j}  \right).
\label{CDRlocalaction}
\ee
Now let us describe the $Q_R$-cohomology classes of untwisted operators regular in $U$.  From our previous discussions, these are operators of dimension $(n,0)$ that are independent of $\psi^{\bar i}$. In general, such operators are of the form ${\widehat {\cal F}} (\phi^i,\partial_z\phi^i,\dots; \phi^{\bar i},\partial_z \phi^{\bar i},\dots; \psi_{zi}, \partial_z \psi_{zi}, \dots ; \psi^i, \partial_z \psi^i, \dots)$. As explained earlier in section 4.1, on a flat target space such as $U$, there can be perturbative corrections to the action of $Q_R$ coming from classes in $H^1(X, \Omega^{2,cl}_X)$. However, as mentioned above, they are irrelevant when analysing the $Q_R$-cohomology on $U$. Hence, we can ignore the perturbative corrections to $Q_R$ for our present purposes. Therefore, on the classes of operators in $U$, $Q_R$ acts as ${\bar \partial} = \psi^{\bar i}\partial/\partial\phi^{\bar i}$, and the condition that $\widehat {\cal F}$ is annihilated by $Q_R$ is precisely that, as a function of $\phi^i$, $\phi^{\bar i}$, $\psi_{zi}$, $\psi^i$ and their $z$-derivatives, it is independent of $\phi^{\bar i}$ (as opposed to its derivatives), and depends only on the other variables, namely $\phi^i$, $\psi_{zi}$, $\psi^i$ and the derivatives of $\phi^i$, $\phi^{\bar i}$, $\psi_{zi}$ and $\psi^i$.\footnote{We can again ignore the action of $Q_R$ on $z$-derivatives of $\phi^{\bar i}$ because of the equation of motion $\partial_z\psi^{\bar i}=0$ and the symmetry transformation law $\delta \phi^{\bar i} = \psi^{\bar i}$.} Hence, the $Q_R$-invariant operators are of the form ${\widehat {\cal F}}(\phi^i,\partial_z\phi^i,\dots;\partial_z\phi^{\bar i},\partial_z^2\phi^{\bar i},\dots ; \psi_{zi}, \partial_z \psi_{zi}, \partial_z^2 \psi_{zi}, \dots; \psi^i, \partial_z \psi^i, \partial_z^2 \psi^i, \dots )$. In other words, the operators, in their dependence on the center of mass coordinate of the string whose worldsheet theory is the half-twisted sigma model, is holomorphic. The local sections of $\widehat {\cal A}$ are just given by the operators in the $Q_R$-cohomology of the local, half-twisted sigma model on $U$ with action (\ref{CDRlocalaction}).

\vspace{0.4cm}\smallskip\noindent{\it A Holomorphic, Twisted  $N=2$ Superconformal Structure}

Note that the local theory with action (\ref{CDRlocalaction}) has an underlying, holomorphic, (twisted) $N=2$ superconformal structure as follows. Firstly, the action is invariant under the following field transformations
\be
\delta\psi^i = \psi^i, \quad \delta \psi^{\bar i}_{z} = - \psi^{\bar i}_{z}, \qquad \textrm{and} \qquad \delta \phi^i = \psi^i, \quad \delta\psi^{\bar i}_{z} = - \partial_z \phi^{\bar i},
\label{CDRghostL}
\ee
where the corresponding conserved currents are given by the dimension one, bosonic and fermionic operators ${\widehat J}(z)$ and ${\widehat Q}(z)$ respectively. They can be written as
\be
{{\widehat J}(z)} = {\delta}_{i\bar j} \psi^{\bar j}_z \psi^i \qquad \textrm{and} \qquad {{\widehat Q}(z)} = {\delta}_{i \bar j} \psi^i \partial_z \phi^{\bar j}.
\label{CDRJ}
\ee
Note that we also have the relation $[{\widehat Q}, {\widehat J}(z)] = - {\widehat Q}(z)$, where $\widehat Q$ is the charge of the current ${\widehat Q}(z)$.  Secondly, the conserved, holomorphic stress tensor is given by
\be
{\widehat T}(z) = - \delta_{i \bar j} \partial_z \phi^i \partial_z \phi^{\bar j} - \delta_{i \bar j} \psi^{\bar j}_z \partial_z \psi^i,
\label{CDRT}
\ee
where one can derive another conserved, fermionic current ${\widehat G}(z)$, such that ${\widehat T}(z) = \{{\widehat Q}, {\widehat G}(z)\}$, and
\be
{\widehat G}(z) = {\delta}_{i \bar j} \psi^{\bar j}_z \partial_z \phi^i.
\label{CDRG}
\ee
One can verify that  the $G$-invariant operators ${\widehat J}(z)$, ${\widehat Q}(z)$, ${\widehat T}(z)$ and   ${\widehat G}(z)$ satisfy the same OPE relations as that satisfied by      $J(z)$, $Q(z)$, $T(z)$ and   $G(z)$ in (2.17). In other words, they furnish the same (twisted) $N=2$ superconformal algebra satisfied by $J(z)$, $Q(z)$, $T(z)$ and   $G(z)$  of  the global version of the classical half-twisted sigma model with action $S$. ${\widehat J}(z)$, ${\widehat Q}(z)$, ${\widehat T}(z)$ and   ${\widehat G}(z)$ are local versions of  $J(z)$, $Q(z)$, $T(z)$ and   $G(z)$ respectively. Hence, if there is no obstruction to a global definition of  ${\widehat J}(z)$, ${\widehat Q}(z)$, ${\widehat T}(z)$ and   ${\widehat G}(z)$ in the quantum theory, the symmetries associated with ${ J}(z)$, ${ Q}(z)$, ${ T}(z)$ and  ${ G}(z)$ will persist in the non-linear half-twisted sigma model at the quantum level.  Another way to see this is to first notice that ${J}(z)$, ${Q}(z)$, ${T}(z)$ and   ${G}(z)$ are $G$-invariant $\psi^{\bar i}$-independent operators and as such, will correspond to classes in $H^0(X, \widehat {\cal A})^G$ (from our $Q_R$-Cech cohomology dictionary). Hence, these operators will exist in the untwisted sector of the $Q_R$-cohomology of the half-twisted orbifold sigma model on $X/G$ if they exist as $G$-invariant global sections of  $\widehat {\cal A}$.

\vspace{0.4cm}\smallskip\noindent{\it The $bc$-$\beta\gamma$ System}

Now, let us set $\beta_i  =  \delta_{i \bar j} \partial_z \phi^{\bar j}$ and $\gamma^i = \phi^i$, whereby $\beta_i$ and $\gamma^i$ are bosonic operators of dimension $(1,0)$ and $(0,0)$ respectively. Next, let us set $\delta_{i\bar j} \psi^{\bar j}_{z} = b_i$ and $\psi^i = c^i$, whereby $b_i$ and $c^i$ are fermionic operators of dimension $(1,0)$ and $(0,0)$ accordingly. Then, the untwisted operators in the $Q_R$-cohomology that are regular in $U$ can be represented by arbitrary local functions of the form ${\widehat {\cal F}} (\gamma^i, \partial_z \gamma^i, \partial_z^2 \gamma^i, \dots, \beta^i, \partial_z \beta^i, \partial_z^2 \beta^i, \dots, b^i, \partial_z  b^i, \partial_z^2 b^i, \dots, c^i, \partial_z c^i, \partial_z^2 c^i, \dots)$ in the fields $\beta$, $\gamma$, $b$ and $c$. The operators $\beta$ and $\gamma$ have the operator products of a standard $\beta\gamma$ system.  The products $\beta\cdot\beta$ and
$\gamma\cdot\gamma$ are non-singular, while
\be
\beta_i(z)\gamma^j(z')=-{\delta_{ij}\over z-z'}+{\rm regular}.
\ee
Similarly, the operators $b$ and $c$ have the operator products of a standard $bc$ system. The products $b \cdot b$ and $c \cdot c$ are non-singular, while
\be
b_i(z) c^j(z')={\delta_{ij}\over z-z'}+{\rm regular}.
\ee
These statements can be deduced from the flat action (\ref{CDRlocalaction}) by standard methods. We can write down an action for the fields $\beta$, $\gamma$, $b$ and $c$, regarded as free elementary fields, which reproduces these OPE's.  It is simply the following action of a $bc$-$\beta\gamma$ system:
\be
I_{bc \textrm{-} \beta\gamma}= {1\over
2\pi} \int |d^2z|  \sum_i \left (\beta_i \partial_{\bar z}\gamma^i + b_i \partial_{\bar z} c^i \right).
\label{bcaction}
\ee
Hence, we find that the local (i.e. flat) $bc$-$\beta\gamma$ system above reproduces the $ Q_R$-cohomology of $\psi^{\bar i}$-independent operators of the half-twisted sigma model on $U$, that is, the local sections of the sheaf $\widehat{\cal A}$.

At this point, one can make some important observations about the relationship between the symmetries of the local half-twisted sigma model with action (\ref{CDRlocalaction}), and the symmetries of the local version of the $bc$-$\beta\gamma$ system above. Note that the free $bc$-$\beta\gamma$ action (\ref{bcaction}) is invariant under the following field variations
\be
\delta c^i = c^i, \quad \delta b_i = - b_i, \qquad \textrm{and} \qquad \delta\gamma^i = c^i, \quad \delta b_i = - \beta_i,
\ee
where the corresponding conserved, bosonic and fermionic currents will be given by ${\cal J}(z)$ and ${\cal Q}(z)$ respectively. They can be written as
\be
{\cal J}(z) = b_i c^i, \qquad \textrm{and} \qquad {\cal Q}(z) = \beta_i c^i.
\ee
In addition, we  have the relation $[ {\cal Q}, {\cal J}(z) ] = - {\cal Q}(z)$, where $\cal Q$ is the charge of the current ${\cal Q}(z)$. The action is also invariant under
\be
\delta c^i = \partial_z \gamma^i, \qquad \textrm{and} \qquad \delta \beta_i = \partial_z b_i,
\ee
where the corresponding conserved, fermionic current will be given by
\be
{\cal G}(z) = b_i \partial_z \gamma^i.
\ee
Finally, the stress tensor of the local $bc$-$\beta\gamma$ system is
\be
{\cal T}(z) = - \beta_i \partial_z \gamma^i - b_i \partial_z c^i,
\ee
where we also have the relation $\{ {\cal Q}, {\cal G}(z) \} = {\cal T}(z)$. (Note that we have omitted the normal-ordering symbol in writing the above conserved currents and tensor.) One can verify that just like the operators ${\widehat J}(z)$, ${\widehat Q}(z)$, ${\widehat G}(z)$ and ${\widehat T}(z)$, the  operators ${\cal J}(z)$, ${\cal Q}(z)$, ${\cal G}(z)$ and ${\cal T}(z)$ generate a holomorphic,  (twisted) $N=2$ superconformal algebra. In fact,  via the respective identification of the fields $\beta_i$ and $\gamma^i$ with $\delta_{i \bar j}\partial_z \phi^{\bar j}$ and $\phi^i$, $\psi_{zi}$ and $\psi^i$ with $b_i$ and $c^i$, we find that ${\widehat J}(z)$, ${\widehat Q}(z)$, ${\widehat T}(z)$ and ${\widehat G}(z)$ coincide with ${\cal J}(z)$, $ {\cal Q}(z)$, ${\cal T}(z)$ and ${\cal G}(z)$ respectively. This observation will be important in sections 5.1 and 5.2, when we consider explicit examples.

One may now ask the following question: does the $bc$-$\beta\gamma$ system reproduce the $Q_R$-cohomology of $\psi^{\bar i}$-independent operators globally on $X$, or only in a small open set $U$? Well, the $bc$-$\beta\gamma$ system will certainly reproduce the $Q_R$-cohomology of $\psi^{\bar i}$-independent operators globally on $X$ if there is no obstruction to defining the system globally on $X$, i.e., one finds, after making global sense of the action (\ref{bcaction}), that the corresponding theory remains anomaly-free. Let's look at this more closely.

First and foremost, the classical action (\ref{bcaction}) makes sense globally if we interpret the bosonic fields $\beta$, $\gamma$, and the fermionic fields $b$, $c$, correctly.  $\gamma^i$ defines a map $\gamma:\Sigma\to X$, and $\beta_i$ is a $(1,0)$-form on $\Sigma$ with values in the pull-back $\gamma^*(T^*X)$. The fermionic field $c^i$ is a scalar on $\Sigma$ with values in the pull-back $\gamma^*(TX)$, while the fermionic field $b_i$ is a $(1,0)$-form on $\Sigma$ with values in the pull-back $\gamma^* (T^*X)$. With this interpretation, the global version of (\ref{bcaction}) becomes the action of what one might call a non-linear $bc$-$\beta\gamma$ system. However, by choosing $\gamma^i$ to be local coordinates on a small open set $U\subset X$, and $c^i$ to be local sections of the pull-back $\gamma^* (TX)$ over $U$, one can make the action linear. In other words, a local version of (\ref{bcaction}) as considered earlier represents the action of a linear $bc$-$\beta \gamma$ system. To the best of the author's knowledge, the non-linear $bc$-$\beta\gamma$ system with action (\ref{bcaction}) does not seem to have been studied anywhere in the physics literature. Nevertheless, the results derived in this paper will definitely serve to provide additional insights into future  problems involving the application of this non-linear $bc$-$\beta\gamma$ system.

Now that we have made global sense of the action of the $bc$-$\beta\gamma$ system at the classical level, we move on to discuss what happens at the quantum level. The vanishing anomalies of the half-twisted sigma model can also be demonstrated in the nonlinear $bc$-$\beta\gamma$ system. Expand around a classical solution of the nonlinear $bc$-$\beta\gamma$ system, represented by a holomorphic map $\gamma_0:\Sigma \to X$, and a section $c_0$ of the pull-back $\gamma_0^*(TX)$. Setting ${\gamma} =\gamma_0 +\gamma'$, and $c = c_0 + c'$, the action, expanded to quadratic order about this solution, is $(1/2\pi) \left [ (\beta , {\overline D \gamma'}) + (b, \overline D c') \right]$. $\gamma'$, being a deformation of the coordinate $\gamma_0$ on $X$, is a section of the pull-back $\gamma_0^* (TX)$. Thus, the kinetic operator of the $\beta$ and $\gamma$ fields is the $\overline D$ operator on sections of $ \gamma_0^*(TX)$. Next, since $c'$ is a deformation of $c_0$, it will be a section of the pull-back $\gamma_0^*(TX)$. The kinetic operator of the $b$ and $c$ fields is therefore the $\overline D$ operator on sections of $\gamma_0^*(TX)$. Thus, the kinetic operator of the $\beta\gamma$ fields is the same as the kinetic operator of the $bc$ fields. However, there is a sign change in  the anomaly that is associated to the kinetic operator of the $\beta\gamma$ fields as these fields are bosonic rather than fermionic. Hence, the anomalies cancel out, and the non-linear $bc$-$\beta\gamma$ system has vanishing anomalies, consistent with the underlying half-twisted sigma model. Thus, the $bc$-$\beta\gamma$ system will reproduce the $Q_R$-cohomology of $\psi^{\bar i}$-independent operators globally on $X$. In other words, one can always find a global section of $\widehat {\cal A}$.

Via the identification of the various fields mentioned above, and the ghost symmetry of the local action (\ref{CDRlocalaction}), we see that the left-moving fields $b_i$ and $c^i$ will have ghost numbers $g_L = -1$ and $g_L = 1$ respectively. However, note that the $bc$-$\beta\gamma$ system lacks the presence of right-moving fermions and thus, the right-moving ghost number $g_R$ carried by the fields $\psi^i_{\bar z}$ and $\psi^{\bar i}$ of the underlying half-twisted sigma model. Locally, the $Q_R$-cohomology of the sigma model is non-vanishing only for $g_R =0$. Globally however, there can generically be cohomology in higher degrees. Since the chiral algebra of operators furnished by the linear $bc$-$\beta\gamma$ system gives the correct description of the $Q_R$-cohomology of $\psi^{\bar i}$-independent operators on $U$, one can then expect the globally-defined chiral algebra of operators furnished by the non-linear $bc$-$\beta\gamma$ system to correctly describe the $ Q_R$-cohomology classes of zero degree (i.e. $g_R =0$) on $X$. How then can one use the non-linear $bc$-$\beta\gamma$ system to describe the higher cohomology? The answer lies in the analysis carried out in section 4.2. In the $bc$-$\beta\gamma$ description, we do not have a close analog of $\bar \partial$ cohomology at our convenience. Nevertheless, we can use the more abstract notion  of Cech cohomology. As before, we begin with a good cover of $X$ by small open sets $\{U_a \}$, and, as explained in section 4.2, we can then describe the         $Q_R$-cohomology classes of positive degree (i.e. $g_R > 0$) by Cech $g_R$-cocycles, i.e., they can be described by the $g^{th}_R$ Cech cohomology of the sheaf $\widehat {\cal A}$ of the chiral algebra of the linear $bc$-$\beta\gamma$ system with action being a linearised version of (\ref{bcaction}). Thus, these operators in the untwisted sector of the $Q_R$-cohomology of the half-twisted orbifold model on $X/G$, correspond to $G$-invariant classes in the $g^{th}_R$ Cech cohomology group of the sheaf $\widehat {\cal A}$. Although unusual from a physicist's perspective, this Cech cohomology approach has been taken as a starting point for the present subject in the mathematical literature \cite{Frenkel}.

\vspace{0.4cm}\smallskip\noindent{\it The Sheaf $\widehat {\cal A}^g$ on $X$}

The discussion involving the sheaf $\widehat {\cal A}^g$ is similar. One just needs to consider the twisted fields $\phi^{i,g}$, $\phi^{\bar i, g}$, $\psi^{i, g}$, $\psi^g_{z i}$, $\psi^{\bar i,g}$ and $\psi^g_{\bar z \bar i}$ instead, and apply the same  arguments above. In doing so, we find that the $Q_R$-invariant, twisted operators are of the form ${\widehat {\cal F}}^g(\phi^{i,g},\partial_z\phi^{i,g},\dots;\partial_z\phi^{\bar i,g},\partial_z^2\phi^{\bar i ,g},\dots ; \psi^g_{zi}, \partial_z \psi^g_{zi}, \partial_z^2 \psi^g_{zi}, \dots; \psi^{i,g}, \partial_z \psi^{i,g}, \partial_z^2 \psi^{i,g}, \dots )$. Thus, the local sections of $\widehat {\cal A}^g$ are just given by the twisted operators in the $Q_R$-cohomology of the local, half-twisted sigma model on $U$ with action (\ref{CDRlocalaction}).

Since the twist commutes with the worldsheet supersymmetries, one can also obtain, as in the previous discussion on  $\widehat{\cal A}$ involving untwisted fields, the following $G$-invariant conserved currents and tensors in terms of twisted fields:
\begin{eqnarray}
{{\widehat J}^g(z)} & = & {\delta}_{i\bar j} \psi^{\bar j,g}_z \psi^{i,g} + F_g z^{-1}, \\
{{\widehat Q}^g(z)}& = & {\delta}_{i \bar j} \psi^{i,g} \partial_z \phi^{\bar j,g},\\
{\widehat T}^g(z) & = & - \delta_{i \bar j} \partial_z \phi^{i,g} \partial_z \phi^{\bar j,g} - \delta_{i \bar j} \psi^{\bar j,g}_z \partial_z \psi^{i,g}, \\
{\widehat G}^g(z) & = & {\delta}_{i \bar j} \psi^{\bar j,g}_z \partial_z \phi^{i,g},
\end{eqnarray}
where the additional term of $F_g z^{-1}$ in ${\widehat J}^g(z)$ is to account for the shift in the value of $g_L$ in the twisted sector. One can verify that  ${\widehat J}^g(z)$, ${\widehat Q}^g(z)$, ${\widehat T}^g(z)$ and   ${\widehat G}^g(z)$ satisfy the same (twisted) $N=2$ superconformal OPE algebra relations as that satisfied by $J^g(z)$, $Q^g(z)$, $T^g(z)$ and   $G^g(z)$ of the global version of the classical half-twisted sigma model. ${\widehat J}^g(z)$, ${\widehat Q}^g(z)$, ${\widehat T}^g(z)$ and   ${\widehat G}^g(z)$ are local versions of  $J^g(z)$, $Q^g(z)$, $T^g(z)$ and   $G^g(z)$ respectively. Hence, if there is no obstruction to a global definition of  ${\widehat J}^g(z)$, ${\widehat Q}^g(z)$, ${\widehat T}^g(z)$ and   ${\widehat G}^g(z)$ in the quantum theory, the twisted $N=2$ superconformal structure associated with ${ J}^g(z)$, ${ Q}^g(z)$, ${ T}^g(z)$ and  ${ G}^g(z)$ will persist in the non-linear half-twisted sigma model at the quantum level.  Since $J^g(z)$, ${Q}^g(z)$, ${T}^g(z)$ and   ${G}^g(z)$ are $G$-invariant $\psi^{\bar i,g}$-independent operators, they will correspond to classes in $H^0(X, \widehat {\cal A}^g)^G$ (from our $Q_R$-Cech cohomology dictionary). Hence, these operators will exist in the twisted sectors of the $Q_R$-cohomology of the half-twisted orbifold sigma model on $X/G$ if they exist as $G$-invariant global sections of  $\widehat {\cal A}^g$.

Next, let us set $\beta^g_i  =  \delta_{i \bar j} \partial_z \phi^{\bar j,g} = \partial_z \phi^{g}_i$ and $\gamma^{i,g} = \phi^{i,g}$, whereby $\beta^g_i$ and $\gamma^{i,g}$ are $\it{twisted}$ versions of the bosonic operators $\beta_i$ and $\gamma^i$ of dimension $(1,0)$ and $(0,0)$ respectively. Next, let us set $\delta_{i\bar j} \psi^{\bar j,g}_{z} =\psi^{g}_{z i}  = b^g_i$ and $\psi^{i,g} = c^{i,g}$, whereby $b^g_i$ and $c^{i,g}$ are $\it{twisted}$ versions of the fermionic operators $b_i$ and $c^i$ of dimension $(1,0)$ and $(0,0)$ accordingly. Then, the twisted operators in the $Q_R$-cohomology that are regular in $U$ can be represented by arbitrary local functions in the fields $\beta^g$, $\gamma^g$, $b^g$ and $c^g$ of the form ${\widehat {\cal F}}^g (\gamma^{i,g}, \partial_z \gamma^{i,g}, \partial_z^2 \gamma^{i,g}, \dots, \beta^{i,g}, \partial_z \beta^{i,g}, \partial_z^2 \beta^{i,g}, \dots, b^{i,g}, \partial_z  b^{i,g}, \newline\partial_z^2 b^{i,g}, \dots, c^{i,g}, \partial_z c^{i,g}, \partial_z^2 c^{i,g}, \dots)$. The twist condition and mode expansion in (\ref{twistphi}) and (\ref{phiexpand}) tells us that
\begin{eqnarray}
\label{twistgamma}
\gamma^{j,g} (e^{2 \pi i}z) = e^{- 2 \pi i \theta_j} \gamma^{j,g} (z), & \qquad & \beta^g_j (e^{2 \pi i}z) = e^{ 2 \pi i \theta_j} \beta^g_j (z),
\end{eqnarray}
and that
\begin{eqnarray}
\label{gammaexpand}
\gamma^{j,g} (z) = \sum_{m \in \, \theta_j + {\mathbb Z}} {\gamma^{j}_m \over z^{m}}, & \qquad & \beta^g_j (z) = \sum_{m \in \, -\theta_j + {\mathbb Z}} {\beta_{jm} \over z^{m+1}}.
\end{eqnarray}
Note that the $\it{twisted}$ mode expansions in (\ref{gammaexpand}) have also been derived via a purely mathematical approach in section 4.3 of \cite{Frenkel} using Li's results in \cite{Li}. In addition, one can see from (\ref{twistgamma}) that $\beta^g_i$ and $\gamma^{i,g}$ have opposite twists. The $\beta^g$ and $\gamma^g$ twisted fields will therefore have the operator products of a standard $\beta\gamma$ system:
\be
\beta^g_i(z)\gamma^{j,g}(z')=-{\delta_{ij}\over z-z'}+{\rm regular}.
\ee
Similarly, from the twist conditions and mode expansions in (\ref{twistpsi}) and (\ref{psiexpand}), we find that
\begin{eqnarray}
\label{twistc}
c^{j,g} ( e^{2 \pi i} z) =  e^{- 2\pi i \theta_j} c^{j,g} (z), & \qquad  & b^g_{j}(e^{2 \pi i} z) = e^{ 2\pi i \theta_j} b^g_{j}(z),
\end{eqnarray}
and that
\begin{eqnarray}
\label{cexpand}
c^{j,g} (z) = \sum_{m \in \, \theta_j + {\mathbb Z}} {c^{j}_m \over z^{m}},  & \qquad  & b^g_{j}(z) = \sum_{m \in \, - \theta_j + {\mathbb Z}} {b_{jm} \over z^{m+1}}.
\end{eqnarray}
Note that the $\it{twisted}$ mode expansions in (\ref{cexpand}) have also been derived via a purely mathematical approach in section 4.3 of \cite{Frenkel} using Li's results in \cite{Li}. In addition, one can see from (\ref{twistc}) that $b^g_i$ and $c^{i,g}$ have opposite twists. The $b^g$ and $c^g$ twisted fields will therefore have the operator products of a standard $bc$ system:
\be
b^g_i(z) c^{j,g}(z')={\delta_{ij}\over z-z'}+{\rm regular},
\ee
while the OPE's $\beta^g_i(z) \cdot \beta^g_i (z')$, $\gamma^{i,g}(z) \cdot \gamma^{i,g} (z')$, $b^g_i(z) \cdot b^g_i (z')$ and $c^{i,g}(z) \cdot c^{i,g}(z')$ are non-singular.\footnote{As long as $z'$ is not at the origin where the twist fields are, the OPE's will be regular as $z \to z'$. The author wishes to thank L. Dixon for clarifying this point.} Hence, we find that the twisted sector of the linear $bc$-$\beta\gamma$ system    reproduces the $ Q_R$-cohomology of $\psi^{\bar i ,g}$-independent operators of the half-twisted sigma model on $U$, that is, the local sections of the sheaf $\widehat{\cal A}^g$.

One can also construct the following conserved currents and tensors from the twisted fields $\beta^g$, $\gamma^g$, $b^g$ and $c^g$. They can be written as follows:
\begin{eqnarray}
{\cal J}^g(z) & = & b^g_i c^{i,g} + F_g z^{-1}, \\
{\cal Q}^g(z)  & = &  \beta^g_i c^{i,g}, \\
{\cal T}^g(z) & = & - \beta^g_i \partial_z \gamma^{i,g} - b^g_i \partial_z c^{i,g},\\
{\cal G}^g(z)& = & b^g_i \partial_z \gamma^{i,g}.
\end{eqnarray}
(Note that we have again omitted the normal-ordering symbol in writing the above conserved currents and tensor for notational convenience.) The additional term of $F_g z^{-1}$ in ${\cal J}^g(z)$ is to account for the shift in the fermion number of the vacuum of the twisted sector in the $bc$-$\beta\gamma$ system.  Via the respective identification of the fields $\beta^g_i$ and $\gamma^{i,g}$ with $\delta_{i \bar j}\partial_z \phi^{\bar j,g}$ and $\phi^{i,g}$, $\psi^g_{zi}$ and $\psi^{i,g}$ with $b^g_i$ and $c^{i,g}$, we find that ${\widehat J}^g(z)$, ${\widehat Q}^g(z)$, ${\widehat T}^g(z)$ and ${\widehat G}^g(z)$ coincide with ${\cal J}^g(z)$, $ {\cal Q}^g(z)$, ${\cal T}^g(z)$ and ${\cal G}^g(z)$ respectively, and thus furnish a holomorphic,  (twisted) $N=2$ superconformal structure. This observation will be important in sections 5.1 and 5.2, when we consider explicit examples.

Note that the operators ${\cal J}^g(z)$, ${\cal Q}^g(z)$, ${\cal T}^g(z)$ and ${\cal G}^g(z)$ have also been obtained using an entirely different approach in the mathematical literature \cite{Frenkel} using Li's twisted iterate formula \cite{Li}. Moreover, it has also been argued in \cite{Frenkel} that these operators satisfy the OPE's of a (twisted) $N=2$ superconformal algebra.

As explained earlier, the global, non-linear $bc$-$\beta\gamma$ system is anomaly-free. This means that one can always find a global section of the sheaf $\widehat{\cal A}^g$. In addition, based on a similar explanation regarding the untwisted operators in the $Q_R$-cohomology of the half-twisted model on $X$,  we find that the twisted operators with $g_R = k + F_g$ (i.e., those which correspond to $(0,k)$-forms with values in a tensor product bundle over $X$) can be described by Cech $k$-cocycles, that is, they can be described by the $k^{th}$ Cech cohomology of the sheaf $\widehat {\cal A}^g$ of the $\it{twisted}$ chiral algebra of the linear $bc$-$\beta\gamma$ system with action being a linearised version of (\ref{bcaction}). Thus, these operators in the twisted sector of the $Q_R$-cohomology of the half-twisted orbifold model on $X/G$, correspond to $G$-invariant classes in the  $k^{th}$ Cech cohomology group of the sheaf $\widehat {\cal A}^g$. This Cech cohomology approach for the twisted sector has also been taken as a starting point for the present subject in the mathematical literature \cite{Frenkel}.

\newsubsection{Local Symmetries}

So far, we have obtained an understanding of the local structure of the sheaves $\widehat {\cal A}$ and $\widehat{\cal A}^g$ via the free, linear $bc$-$\beta\gamma$ system on an open set $U \subset X$. We shall now proceed towards our real objective of obtaining an understanding of its global structure. In order to do, we will need to glue the local descriptions that we have studied above together.

To this end, we must first cover $X$ by small open sets $\{U_a\}$.  Recall here that in each $U_a$, the $Q_R$-cohomology of the half-twisted model is described by the local operators in the chiral algebra of the free, linear $bc$-$\beta\gamma$ system on $U_a$. Next, we will need to glue these local descriptions together over the intersections $\{ U_a \cap U_b \}$, so as to describe the global structure of this $Q_R$-cohomology in terms of a globally-defined sheaf of chiral algebras over the entire manifold $X$.

Note that the gluing has to be carried out using the automorphisms of the free, linear $bc$-$\beta\gamma$ system. Thus, one must first ascertain the underlying symmetries of the system, which are in turn divided  into  geometrical and non-geometrical symmetries. The geometrical symmetries are used in gluing together the local sets $\{ ({TX})_f \times {U_a} \}$ into the entire tangent bundle ${TX}$, where $(TX)_f$ just denotes the fibre of the tangent bundle over $U_a$.  The non-geometrical symmetries on the other hand, are used in gluing the local descriptions at the algebraic level. However, in the case of the half-twisted A-model where one has vanishing anomalies, we only need to consider the geometrical symmetries in gluing the local descriptions together. This has been explained in \cite{MC} and shown to be consistent with the mathematical results of \cite{MSV1}.

As usual, the generators of these geometrical symmetries will be given by the charges of the conserved currents of the free $bc$-$\beta\gamma$ system. In turn, these generators will furnish the Lie algebra $\mathfrak h$ of the geometrical symmetry group. Let the elements of $\mathfrak h$  be   written as ${\mathfrak h} = ({\mathfrak v} , \mathfrak f)$, where  $\mathfrak v$ generates the geometrical symmetries of $U$, while $\mathfrak f$ generates the fibre space symmetries of the tangent bundle over $U$. Since the conserved charges must also be conformally-invariant, it will mean that an element of $\mathfrak h$ must be given by an integral of a dimension one current, modulo total derivatives.

\vspace{0.4cm}\smallskip\noindent{\it Gluing the Local Descriptions of the Untwisted Sector}

With the above considerations in mind, let us construct the dimension one currents of the free $bc$-$\beta \gamma$ system in the untwisted sector.  Firstly, if we have a holomorphic vector field $V$ on $X$ where $V = V^i (\gamma) {\partial \over {\partial \gamma^i}}$, we can construct a dimension one current $J_V=-V^i \beta_i$. The corresponding conserved charge is then given by $K_V=\oint J_V dz  $. A computation of the operator product expansion with the elementary fields $\gamma$ gives
\be
J_V(z)\gamma^k(z')\sim {V^k(z')\over z-z'}.
\label{jv}
\ee
Under the symmetry transformation generated by $K_V$, we have $\delta \gamma^k = i \epsilon [ K_V, \gamma^k ]$, where $\epsilon$ is a infinitesinal transformation parameter. Thus, we see from (\ref{jv}) that $K_V$ generates the infinitesimal diffeomorphism $\delta\gamma^k=i \epsilon V^k$ of $U$. In other words, $K_V$ generates the holomorphic diffeomorphisms of the target space $X$. Therefore, $K_V$ spans the $\mathfrak v$ subset
of $\mathfrak h$. For finite diffeomorphisms, we will have a coordinate transformation ${\tilde \gamma}^k = g^k (\gamma)$, where each $g^k (\gamma)$ is a holomorphic function in the $\gamma^k$s. Since we are using the symmetries of the $bc$-$\beta \gamma$ system to glue the local descriptions over the intersections $\{U_a \cap U_b\}$, on an arbitrary intersection $U_a \cap U_b$, $\gamma^k$ and ${\tilde \gamma}^k$ must be defined in $U_a$ and $U_b$ respectively.

Next, let $[t(\gamma)]$ be an arbitrary $N \times N$ matrix over $X$ (where $N= \textrm{dim}_{\mathbb C}X$) whose components are holomorphic functions in $\gamma$. One can then construct a  dimension one current involving the fermionic fields $b$ and $c$ as $J_F = c^m [t(\gamma)] _m{}^n b_n$, where the indices $m$ and $n$ on the matrix $[t(\gamma)]$ denote its $(m,n)$ component, and $m,n = 1, 2, \dots, \textrm{dim}_{\mathbb C}X$. The corresponding conserved charge is thus given by $K_F = \oint J_F dz$. A computation of the operator product expansion with the elementary fields $c$ gives
\be
J_F(z) c^n(z') \sim {  {c^m (z') t_m{}^n  }  \over z-z'},
\label{jf1}
\ee
while a computation of the operator product expansion with the elementary fields $b$ gives
\be
J_F(z) b_n(z') \sim - {  { t_n{}^m b_m (z')  }  \over z-z'}.
\label{jf2}
\ee
Under the symmetry transformation generated by $K_F$, we have $\delta c^n = i \epsilon [ K_F, c^n]$ and $\delta b_n = i \epsilon [ K_F, b_n]$. Hence, we see from (\ref{jf1}) and (\ref{jf2}) that $K_F$ generates the infinitesimal transformations $\delta c^n=i \epsilon c^m t_m{}^n$ and $\delta b_n= - i \epsilon t_n{}^m b_m$. For finite transformations, we will have ${\tilde c}^n = c^m A_m{}^n$ and ${\tilde b}_n = (A^{-1})_n{}^mb_m$, where $A$ is an $N \times N$ matrix holomorphic in $\gamma$ and is given by $[A(\gamma)] = e^{i \alpha [t(\gamma)]}$, where $\alpha$ is a finite transformation parameter.  As before, since we are using the symmetries of the free, linear $bc$-$\beta \gamma$ system to glue the local descriptions over the intersections $\{U_a \cap U_b\}$, on an arbitrary intersection $U_a \cap U_b$, $(c^n, b_n)$ and $({\tilde c}^n, {\tilde b}_n)$ must be defined in $U_a$ and $U_b$ respectively. Recall at this point that the $c^n$'s transform as holomorphic sections of the pull-back $\gamma^*(TX)$, while the $b_n$'s transform as holomorphic sections of the pull-back $\gamma^* (T^*X)$. Moreover, note that the transition function matrix of a dual bundle is simply the inverse of the transition function matrix of the original bundle. This means that we can consistently identify $[A(\gamma)]$ as the holomorphic transition matrix of the tangent bundle $TX$, i.e., $[A(\gamma)]_m{}^n = \partial {\tilde \gamma^n} / \partial \gamma^m$ and $[A^{-1}(\gamma)]_m{}^n = \partial {\gamma^n} / \partial {\tilde \gamma^m}$,  and that $K_F$ spans the $\mathfrak f$ subset   of $\mathfrak h$. It is thus clear from the discussion so far how one can use the geometrical symmetries generated by $K_V$ and $K_F$ to glue the local sets $\{ (TX)_f \times U_a \}$ together on  intersections of small open sets to form the entire bundle $TX$.

Let us now describe how the different fields of the  free, linear $bc$-$\beta \gamma$ system on $U$ transform under the geometrical symmetries generated by $K_H= K_V + K_F$ of $\mathfrak h$. Firstly, note that the symmetries generated by $K_F$ act trivially on the $\gamma$ fields, i.e., the $\gamma$ fields have non-singular OPE's with $J_F$. Secondly, note that the symmetries generated by $K_V$ act trivially on both the $b$ and $c$ fields, i.e., the $b$ and $c$ fields have non-singular OPEs with $J_V$. As for the $\beta$ fields,    they transform non-trivially under $\it{all}$ the symmetries, i.e., the OPE's of the $\beta$ fields with $J_V$ and $J_F$ all contain simple poles. In summary, via a computation of the relevant OPE's, we find that the fields transform under the geometrical symmetries of the free, linear $bc$-$\beta\gamma$ system on $U$ as follows:
\begin{eqnarray}
\label{autoCDRgamma}
{\tilde \gamma}^i & = & g^i (\gamma) ,\\
\label{autoCDRbeta}
{\tilde \beta}_i  & = &   {\partial \gamma^k \over {\partial{\widetilde \gamma^i}}}\beta_k   +  {\partial^2 \gamma^l \over{\partial {\widetilde \gamma^i} \partial{\widetilde \gamma^j}}} {\partial {\widetilde \gamma^j}\over{\partial \gamma^k}} b_l c^k,  \\
\label{autoCDRc}
{\tilde c}^i & = & {\partial{\widetilde \gamma^i} \over{\partial \gamma^k}}c^k, \\
\label{autoCDRb}
{\tilde b}_i & = & {\partial \gamma^k \over{\partial{\widetilde \gamma^i}}}b_k ,
\end{eqnarray}
where $i,j,k, l = 1, 2, \dots, {\textrm{dim}_{\mathbb C} X}$. We thus conclude that in the untwisted sector, the untwisted fields must undergo the above transformations (\ref{autoCDRgamma})-(\ref{autoCDRb}) when we glue a local description of the untwisted sector (in a small open set) to another local description of the untwisted sector (in another small open set) on the mutual intersection of open sets using the automorphisms of the free, linear $bc$-$\beta\gamma$ system.

\vspace{0.4cm}\smallskip\noindent{\it Gluing the Local Descriptions of the Twisted Sectors}

Likewise, one can also construct dimension one currents in the twisted sectors of the free $bc$-$\beta \gamma$ system by using twisted fields in place of the untwisted ones.   Therefore, if we have a holomorphic vector field $V^g$ on $X$, where $V^g = V^{i,g} (\gamma^g) {\partial \over {\partial \gamma^{i,g}}}$, we can construct a dimension one current $J^g_V=-V^{i,g} \beta^g_i$. The corresponding conserved charge is then given by $K^g_V=\oint J^g_V dz  $. A computation of the operator product expansion with the elementary twisted fields $\gamma^g$ gives
\be
J^g_V(z)\gamma^{k,g}(z')\sim {V^{k,g}(z')\over z-z'}.
\label{jvg}
\ee
Under the symmetry transformation generated by $K^g_V$, we have $\delta \gamma^{k,g} = i \epsilon [ K^g_V, \gamma^{k,g} ]$, where $\epsilon$ is a infinitesinal transformation parameter. Thus, we see from (\ref{jvg}) that $K^g_V$ generates the infinitesimal diffeomorphism $\delta\gamma^k=i \epsilon V^k$ on $U$. For finite diffeomorphisms, we will have a coordinate transformation ${\tilde \gamma}^{k,g} = g^{k,g} (\gamma^g)$, where each $g^{k,g} (\gamma^g)$ is a holomorphic function in the $\gamma^{k,g}$s. Since we are using the symmetries of the $bc$-$\beta \gamma$ system to glue the local descriptions over the intersections $\{U_a \cap U_b\}$, on an arbitrary intersection $U_a \cap U_b$, $\gamma^{k,g}$ and ${\tilde \gamma}^{k,g}$ must be defined in $U_a$ and $U_b$ respectively.

Analogous to the untwisted case, let $[\bar t(\gamma^g)]$ be an arbitrary $N \times N$ matrix over $X$  whose components are holomorphic functions in $\gamma^g$. One can then construct a dimension one current involving the twisted fermionic fields $b^g$ and $c^g$ as $J^g_F = c^{m,g} [{\bar t}(\gamma^g)] _m{}^n b^g_n$, where the indices $m$ and $n$ on the matrix $[{\bar t}(\gamma^g)]$ denote its $(m,n)$ component, and $m,n = 1, 2, \dots, \textrm{dim}_{\mathbb C}X$. The corresponding conserved charge is then given by $K^g_F = \oint J^g_F dz$. A computation of the operator product expansion with the elementary twisted fields $c^g$ gives
\be
J^g_F(z) c^{n,g}(z') \sim {  {c^{m,g} (z') {\bar t}_m{}^n  }  \over z-z'},
\label{jf1g}
\ee
while a computation of the operator product expansion with the elementary twisted fields $b^g$ gives
\be
J^g_F(z) b^g_n(z') \sim - {  { {\bar t}_n{}^m b^g_m (z')  }  \over z-z'}.
\label{jf2g}
\ee
Under the symmetry transformation generated by $K^g_F$, we have $\delta c^{n,g} = i \epsilon [ K^g_F, c^{n,g}]$ and $\delta b^g_n = i \epsilon [ K^g_F, b^g_n]$. Hence, we see from (\ref{jf1g}) and (\ref{jf2g}) that $K^g_F$ generates the infinitesimal transformations $\delta c^{n,g}=i \epsilon c^{m,g} {\bar t}_m{}^n$ and $\delta b^g_n= - i \epsilon {\bar t}_n{}^m b^g_m$. For finite transformations, we will have ${\tilde c}^{n,g} = c^{m,g} {\bar A}_m{}^n$ and ${\tilde b}^g_n = ({\bar A}^{-1})_n{}^m b^g_m$, where $\bar A$ is an $N \times N$ matrix holomorphic in $\gamma^g$ and is given by $[{\bar A}(\gamma)] = e^{i \alpha [{\bar t}(\gamma)]}$, where $\alpha$ is a finite transformation parameter.  As before, since we are using the symmetries of the free, linear $bc$-$\beta \gamma$ system to glue the local descriptions over the intersections $\{U_a \cap U_b\}$, on an arbitrary intersection $U_a \cap U_b$, $(c^{n,g}, b^g_n)$ and $({\tilde c}^{n,g}, {\tilde b}^g_n)$ must be defined in $U_a$ and $U_b$ respectively. Similar to the untwisted case, since the $c^{n,g}$'s transform as holomorphic sections of the pull-back $\gamma^*(TX)$, while the $b^g_n$'s transform as holomorphic sections of the pull-back $\gamma^* (T^*X)$, we can consistently identify $[{\bar A}(\gamma^g)]$ as the holomorphic transition matrix (in the twisted $\gamma^{i,g}$ coordinates) of the tangent bundle $TX$, i.e., $[{\bar A}(\gamma^g)]_m{}^n = \partial {\tilde \gamma^{n,g}} / \partial {\gamma^{m,g}}$ and $[{\bar A}^{-1}(\gamma^g)]_m{}^n = \partial { \gamma^{n,g}} / \partial {\tilde \gamma^{m,g}}$ .

Let us now describe how the different twisted fields of the  free, linear $bc$-$\beta \gamma$ system on $U$ transform under the geometrical symmetries generated by $K^g_H= K^g_V + K^g_F$. Firstly, note that the symmetries generated by $K^g_F$ act trivially on the $\gamma^g$ fields, i.e., the $\gamma^g$ fields have non-singular OPE's with $J^g_F$. Secondly, note that the symmetries generated by $K^g_V$ act trivially on both the $b^g$ and $c^g$ fields, i.e., the $b^g$ and $c^g$ fields have non-singular OPE's with $J^g_V$. As for the $\beta^g$ fields,    they transform non-trivially under the symmetries generated by both $K^g_V$ and $K^g_F$, i.e., the OPE's of the $\beta$ fields with $J^g_V$ and $J^g_F$ all contain simple poles. In summary, via a computation of the relevant OPE's, we find that the twisted fields transform under the geometrical symmetries (generated by the twisted charges $K^g_H$) of the free, linear $bc$-$\beta\gamma$ system on $U$ as follows:
\begin{eqnarray}
\label{gautoCDRgamma}
{\tilde \gamma}^{i,g} & = & g^i (\gamma^g) ,\\
\label{gautoCDRbeta}
{\tilde \beta}^g_i  & = &   {\partial \gamma^{k,g} \over {\partial{\widetilde \gamma^{i,g}}}}\beta^g_k   +  {\partial^2 \gamma^{l,g} \over{\partial {\widetilde \gamma^{i,g}} \partial{\widetilde \gamma^{j,g}}}} {\partial {\widetilde \gamma^{j,g}}\over{\partial \gamma^{k,g}}} b^g_l c^{k,g},  \\
\label{gautoCDRc}
{\tilde c}^{i,g} & = & {\partial{\widetilde \gamma^{i,g}} \over{\partial \gamma^{k,g}}}c^{k,g}, \\
\label{gautoCDRb}
{\tilde b}^g_i & = & {\partial \gamma^{k,g} \over{\partial{\widetilde \gamma^{i,g}}}}b^g_k ,
\end{eqnarray}
where $i,j,k, l = 1, 2, \dots, {\textrm{dim}_{\mathbb C} X}$. We thus conclude that in the twisted sectors, the twisted fields must undergo the above transformations (\ref{gautoCDRgamma})-(\ref{gautoCDRb}) when we glue a local description of the $g$-twisted sector (in a small open set) to another local description of the $g$-twisted sector (in another small open set) on the mutual intersection of open sets using the automorphisms of the free, linear $bc$-$\beta\gamma$ system.

\newsubsection{The Sheaves $\widehat{\Omega}^{ch}_X$ and $\widehat{\Omega}^{ch,g}_X$ on $X$}

Note that in computing (\ref{autoCDRgamma})-(\ref{autoCDRb}), we have just rederived, from a purely physical perspective, the set of field transformations (3.17a)-(3.17d) of \cite{MSV1}, which define the admissible automorphisms of a sheaf of conformal vertex superalgebras mathematically known as the chiral de Rham complex! Hence, we learn that $\widehat {\cal A}$ is the sheaf $\widehat\Omega^{ch}_X$ of chiral de Rham complex on $X$.

In addition, this also means that the set of twisted field transformations (\ref{gautoCDRgamma})-(\ref{gautoCDRb}) will define the admissible automorphisms of a twisted version of the sheaf of chiral de Rham complex on $X$. Hence, we learn that $\widehat {\cal A}^g$ is the sheaf $\widehat\Omega^{ch,g}_X$ of the $g$-twisted chiral de Rham complex on $X$ defined in \cite{Frenkel}. In other words, the sheaf $\widehat\Omega^{ch,g}_X$ over $X$ is given by $H^0(X, \widehat{\cal A}^g)$.

Recall from our $Q_R$-Cech cohomology dictionary that the global sections $H^0(X, \widehat{\cal A}^g)$ correspond to $g$-twisted observables in the $Q_R$-cohomology of the half-twisted model on $X$. Thus, from the discussion in section 4.1, we find that $H^0(X, \widehat{\cal A}^g)$ is supported on $X_g$, the fixed-point set of $X/G$. This physical observation is consistent with the mathematical definition of $\widehat\Omega^{ch,g}_X$ in \cite{Frenkel} as a sheaf supported on $X_g$. There is no condition of this sort on untwisted observables, and the sheaf $\widehat\Omega^{ch}_X$ continues to be supported on all of $X$.

Another observation that one can make, based on the fact that $\widehat\Omega^{ch,g}_X$ is a sheaf supported on $X_g$, is that $V^g$ (of section 4.4) is actually a holomorphic vector field along $X_g$. Since the $G$-action maps all of $X_g$ to itself, it will mean that the components of $V^g$ will be invariant under the action of $G$,  i.e., $V^g$ is a $g$-invariant, holomorphic vector field along $X_g$. This observation is consistent with the mathematical construction in section 4.6 of \cite{Frenkel}.

Last but not least, note that if $c_1(X) =0$, one has a state-operator isomorphism of the half-twisted A-model on $X$. This means that the Hilbert space of $g$-twisted states in the sigma model on $X$, i.e., ${\cal H}_g$, can be represented by its space of local, $g$-twisted, $Q_R$-closed operators. Therefore, the Hilbert space of states in the $g$-twisted sector of the orbifold sigma model on $X/G$, i.e., ${\cal H}^{G}_g$, can be represented by the $G$-invariant subspace of $g$-twisted, $Q_R$-closed, local  operators in the sigma model on $X$. This $G$-invariant subspace corresponds to the set of physical operators in the chiral algebra ${\cal A}^g_G$ (elaborated in section 4.2). From the vector space expansion of ${\cal A}^g_G$ in section 4.2, the identification of $\widehat {\cal A}^g$ with $\widehat \Omega^{ch,g}_X$, and the fact that $G = C(g)$ for an abelian orbifold considered in this paper, the isomorphism relation in (\ref{isomorphism}) translates to the statement that
\be
\bigoplus_{k=0}^{\textrm{dim}_{\mathbb C}X}H^k(X, \Omega^{ch,g}_X)^{C(g)} \cong \bigoplus_{k=0}^{\textrm{dim}_{\mathbb C}X}H^k(X, \Omega^{ch,g'}_X)^{C(g')}.
\label{iso}
\ee
The above relation has also been proven from a purely mathematical approach in Theorem 4.3 of \cite{Frenkel}.

\newsection{Examples of Sheaves of CDR}

In this section, we study in detail, examples of sheaves of CDR and their cohomologies  by considering the half-twisted model on two different orbifolds. Our main objective is to illustrate the rather abstract discussion in section 4. In the process, we will obtain an interesting and novel understanding of the relevant physics in terms of pure mathematical data.

\newsubsection{The Sheaves of CDR and the  Half-Twisted A-Model on $\mathbb{CP}^1/ {\mathbb Z}_K$}

For our first example, we take $X =\mathbb{CP}^1$.  In other words, we will be analysing the ($G$-invariant) untwisted and twisted local operators in the $Q_R$-cohomology  which respectively span the chiral algebras ${\cal A}_G$ and ${\cal A}^g_G$ of the half-twisted A-model on the orbifold  $X/G =\mathbb {CP}^1/{\mathbb Z}_K$. To this end, we will work locally on the worldsheet $\Sigma$, choosing a local complex parameter $z$.

Before we proceed further, recall that as vector spaces, ${\cal A}_G$ and ${\cal A}^g_G$ can be expressed as ${\cal A}_G = \bigoplus_{g_R} H^{g_R} (X, \widehat {\cal A})^G$ and ${\cal A}^g_G = \bigoplus_{g_R -F_g} H^{g_R-F_g} (X, \widehat {\cal A}^g)^G$. In addition, as explained in section 4.5, $\widehat{\cal A}$ and $\widehat {\cal A}^g$ are given by the sheaves ${\Omega}^{ch}_X$ and ${\Omega}^{ch,g}_X $ respectively. Hence, in order to study ${\cal A}_G$ and ${\cal A}^g_G$, one simply needs to study the $G$-invariant Cech cohomologies of the sheaves $\widehat{\Omega}^{ch}_X$ and $\widehat{\Omega}^{ch,g}_X $ on $X$.

To this end, first note that $X= \mathbb{CP}^1$ can be regarded as the complex $\gamma$-plane plus a point at infinity.  Thus, we can cover it by two open sets, $U_1$ and $U_2$, where $U_1$ is the complex $\gamma$-plane, and $U_2$ is the complex $\tilde \gamma$-plane, where $\tilde\gamma=1/\gamma$.

Since $U_1$ is isomorphic to $\mathbb{C}$, the sheaves of CDR in $U_1$ can be described by a single, free $bc$-$\beta\gamma$ system with action
\be
I={1\over 2\pi}\int|d^2z| \ \beta \partial_{\bar z} \gamma + b \partial_{\bar z} c.
\label{actionU1}
\ee
Here  $\beta$, $b$, and $c$, $\gamma$, are  fields of dimension $(1,0)$ and $(0,0)$ respectively. They obey the usual free-field OPE's; there are no singularities in the operator products $\beta(z)\cdot \beta(z')$, $b(z) \cdot b(z')$, $\gamma(z)\cdot\gamma(z')$ and $c(z) \cdot c(z')$,  while
\be
\beta(z) \gamma(z')  \sim  -{1\over z-z'} \quad  \textrm{and} \quad  b(z) c(z')  \sim  {1\over z-z'}.
\ee
In a $g$-twisted sector, one must consider the $g$-twisted counterpart of the untwisted fields above, i.e., $\beta^g$, $b^g$, and $c^g$, $\gamma^g$, which are also of dimension $(1,0)$ and $(0,0)$ respectively. They obey the usual free-field OPE's; there are no singularities in the operator products $\beta^g(z)\cdot \beta^g(z')$, $b^g(z) \cdot b^g(z')$, $\gamma^g(z)\cdot\gamma^g(z')$ and $c^g(z) \cdot c^g(z')$,  while
\be
\beta^g(z) \gamma^g(z')  \sim  -{1\over z-z'} \quad  \textrm{and} \quad  b^g(z) c^g(z')  \sim  {1\over z-z'}.
\ee

Similarly, the sheaves of CDR in $U_2$ can be described by a single, free ${\tilde b} {\tilde c}$-$\tilde\beta\tilde\gamma$ system with action
\be
I= {1\over 2\pi}\int|d^2z| \ \tilde
\beta \partial_{\bar z} \tilde\gamma +  \tilde b \partial_{\bar z} \tilde c,
\label{actionU2}
\ee
where the fields $\tilde \beta$, $\tilde b$, $\tilde \gamma$ and $\tilde c$ obey the same OPE's as $\beta$, $b$, $\gamma$ and $c$. In other words, the non-trivial OPE's are given by
\be
\tilde \beta(z) \tilde \gamma(z')  \sim  -{1\over z-z'} \quad  \textrm{and} \quad  \tilde b(z) \tilde c(z')  \sim  {1\over z-z'}.
\ee
Likewise, in a $g$-twisted sector, one must consider the $g$-twisted counterpart of the above untwisted fields, i.e., $\tilde \beta^g$, $\tilde b^g$, $\tilde \gamma^g$ and $\tilde c^g$. They obey the same OPE's as $\beta^g$, $b^g$, and $c^g$, $\gamma^g$. In other words, the non-trivial OPE's are given by
\be
\tilde \beta^g(z) \tilde \gamma^g(z')  \sim  -{1\over z-z'} \quad  \textrm{and} \quad  \tilde b^g(z) \tilde c^g(z')  \sim  {1\over z-z'}.
\ee

In order to describe the globally-defined sheaves $\widehat\Omega^{ch}_{\mathbb P^1}$ and $\widehat\Omega^{ch,g}_{\mathbb P^1}$ of CDR on $\mathbb {CP}^1$, one will need to glue the free conformal field theories  with actions (\ref{actionU1}) and (\ref{actionU2})  in the overlap region $U_1 \cap U_2$. To do so,  one must use the admissible automorphisms of the free conformal field theories defined in (\ref{autoCDRgamma})-(\ref{autoCDRb}) and (\ref{gautoCDRgamma})-(\ref{gautoCDRb}) to glue respectively,  the free-fields in the untwisted and twisted sectors together. In the case of $X = \mathbb {CP}^1$, the automorphisms of the untwisted sector will be given by
\begin{eqnarray}
\label{autoCP1gamma}
{\tilde \gamma} & = & {1 \over \gamma},\\
\label{autoCP1beta}
{\tilde \beta} & = &   - \gamma^2 \beta   - 2 \gamma b c,  \\
\label{autoCP1c}
{\tilde c} & = & - {c \over {\gamma^2}}, \\
\label{autoCP1b}
{\tilde b} & = & -\gamma^2 b,
\end{eqnarray}
while the automorphisms of a $g$-twisted sector will be given by
\begin{eqnarray}
\label{gautoCP1gamma}
{\tilde \gamma^g} & = & {1 \over \gamma^g},\\
\label{gautoCP1beta}
{\tilde \beta^g} & = &   - (\gamma^g)^2 \beta^g   - 2 \gamma^g b^g c^g,  \\
\label{gautoCP1c}
{\tilde c^g} & = & - {c^g \over {(\gamma^g)^2}}, \\
\label{gautoCP1b}
{\tilde b^g} & = & -(\gamma^g)^2 b^g.
\end{eqnarray}
As explained in \cite{MC}, since the half-twisted A-model on $\it{any}$ smooth manifold $X$ has vanishing anomalies, it will mean that there is no obstruction to the above gluing, and the sheaves of CDR can be globally-defined on the target space $\mathbb {CP}^1$ (but only locally-defined on the worldsheet $\Sigma$ of the conformal field theory, because we are using a local complex parameter $z$ to define it).

\bigskip\noindent{\it $\mathbb Z_K$-Invariant Global Sections of  $\widehat\Omega^{ch}_{\mathbb P^1}$}

Since $X= \mathbb {CP}^1$ is of complex dimension $1$, the chiral algebra ${\cal A}_{\mathbb Z_K}$ will be given by ${\cal A}_{\mathbb Z_K} = \bigoplus_{g_R = 0}^{g_R = 1} H^{g_R}({\mathbb {CP}^1}, {\widehat \Omega}^{ch}_{\mathbb P^1})^{\mathbb Z_K}$ as a vector space. Thus, in order to understand ${\cal A}_{\mathbb Z_K}$, the chiral algebra in the untwisted sector of the half-twisted orbifold model on $\mathbb {CP}^1/{\mathbb Z_K}$, one needs to study $H^{0}({\mathbb {CP}^1}, {\widehat \Omega}^{ch}_{\mathbb P^1})^{\mathbb Z_K}$, the ${\mathbb Z_K}$-invariant global sections of the sheaf ${\widehat \Omega}^{ch}_{\mathbb P^1}$, and $H^1( \mathbb {CP}^1, {\widehat \Omega}^{ch}_{\mathbb P^1})^{\mathbb Z_K}$, the ${\mathbb Z_K}$-invariant first Cech cohomology of ${\widehat \Omega}^{ch}_{\mathbb P^1}$.

 It will be useful to ascertain the action of $\mathbb Z_K$ on the various fields before we proceed any further. From (\ref{1})-(\ref{3}), and the identifications $\beta = \partial_z \phi$, $\gamma = \phi$, $b = \psi_{z}$ and $c = \psi$, we find that these fields will transform under the action of $\mathbb Z_K$ as $\beta\to e^{-2 \pi i \theta} \beta$, $\gamma \to e^{2 \pi i \theta}\gamma$, $b \to e^{-2 \pi i \theta}b$ and $c \to e^{2 \pi i \theta}c$, where $\theta = m /K$, and $m = 0,1, 2, \dots, K-1$.

Next, note  that $H^{0}({\mathbb {CP}^1}, {\widehat \Omega}^{ch}_{\mathbb P^1})$ has already been thoroughly analysed in \cite{MC}. For brevity, we shall just quote the relevant results from \cite{MC}. To this end, let us denote $\Omega^{ch}_{\mathbb P^1; n}$ as the sheaf of chiral de Rham complex on $\mathbb {CP}^1$ at dimension $n$, i.e., the corresponding chiral algebra $\widehat A$ consists of dimension $(n,0)$ operators only.

In short, one finds that at dimension 0, $H^{0}({\mathbb {CP}^1}, {\widehat \Omega}^{ch}_{\mathbb P^1;0})$ is one-dimensional and generated by 1. While at dimension 1, i.e., $H^{0}({\mathbb {CP}^1}, {\widehat \Omega}^{ch}_{\mathbb P^1;1})$, one has $J_-  = - \gamma^2 \beta - 2 \gamma b c$, $J_+ =\beta$, $J_3 = \gamma\beta + bc$, $j_- = - \gamma^2 b$, $j_+ = b$ and $j_3 = \gamma b$. It can be shown \cite{MC} that $\{J_-, J_+, J_3, j_-, j_+, j_3\}$ generate a super-affine algebra of $SL(2)$ at level 0 in the Wakimoto free-field representation. However, notice that only the generator 1, $J_3$ and $j_3$ are $\mathbb Z_K$-invariant.  Hence, $\{1, J_3, j_3\} \in  H^{0}({\mathbb {CP}^1}, {\widehat \Omega}^{ch}_{\mathbb P^1})^{\mathbb Z_K}$, but $\{J_-, J_+,  j_-, j_+, \} \notin H^{0}({\mathbb {CP}^1}, {\widehat \Omega}^{ch}_{\mathbb P^1})^{\mathbb Z_K}$. Therefore, in contrast to the situation observed in \cite{MC} of the half-twisted A-model on $\mathbb {CP}^1$, the subset of the infinite-dimensional space of physical operators in the half-twisted A-model on the orbifold $\mathbb {CP}^1 / \mathbb Z_K$ (represented by the $G$-invariant global sections of the sheaf $\Omega^{ch}_{\mathbb P^1}$) $\it{do}$ $\it{not}$ furnish a super-affine algebra of $SL(2)$. Note that the space of operators (in the untwisted sector) spanned by $\{1,J_3, j_3\}$ has a structure of a chiral algebra in the full physical sense; it obeys all the physical axioms of a chiral algebra, including reparameterisation invariance on the $z$-plane or worldsheet $\Sigma$. We will substantiate this last statement momentarily by showing that the holomorphic stress tensor exists in the $Q_R$-cohomology of the half-twisted A-model on $\mathbb {CP}^1 / \mathbb Z_K$.

Still on the subject of global sections, recall from section 4.3 and our $Q_R$-Cech cohomology dictionary that in the untwisted sector, there will be $\mathbb Z_K$-invariant, $\psi^{\bar i}$-independent operators ${J}(z)$, ${Q}(z)$, ${T}(z)$ and ${G}(z)$ in the $Q_R$-cohomology of the underlying half-twisted A-model on $\mathbb {CP}^1/\mathbb Z_K$ if and only if the corresponding $\mathbb Z_K$-invariant operators ${\widehat J}(z)$, ${\widehat Q}(z)$, ${\widehat T}(z)$ and ${\widehat G}(z)$ can be globally-defined, i.e., the $\mathbb Z_K$-invariant operators ${\cal J}(z)$, ${\cal Q}(z)$, ${\cal T}(z)$ and ${\cal G}(z)$  of the free $bc$-$\beta\gamma$ system belong in $H^0(\mathbb {CP}^1, {\widehat \Omega}^{ch}_{\mathbb P^1})^{\mathbb Z_K}$ - the space of $\mathbb Z_K$-invariant global sections of  ${\widehat \Omega}^{ch}_{\mathbb P^1}$. Let's look at this more closely.

For $X= \mathbb {CP}^1$, we have
\begin{eqnarray}
{\cal J}(z)& =& :bc: (z), \\
{\cal Q}(z) &= &:\beta c: (z), \\
{\cal T}(z) & = &  -: \beta \partial_z \gamma: (z) - : b\partial_z c: (z), \\
{\cal G}(z) & = & :b \partial_z \gamma:(z),
\end{eqnarray}
where the above operators are defined and regular in $U_1$. Similarly, we also have
\begin{eqnarray}
\label{A1}
{\widetilde{\cal J}(z)}& =& :\tilde b \tilde c: (z), \\
{\widetilde{\cal Q}(z)} &= &:\tilde\beta \tilde c: (z), \\
{\widetilde{\cal T}(z)} & = &  -: \tilde\beta \partial_z \tilde\gamma: (z) - : \tilde b\partial_z \tilde c: (z), \\
\label{A4}
{\widetilde{\cal G}(z) }& = & :\tilde b \partial_z \tilde\gamma:(z),
\end{eqnarray}
where the above operators are defined and regular in $U_2$. By substituting the automorphism relations (\ref{autoCP1gamma})-(\ref{autoCP1b}) into (\ref{A1})-(\ref{A4}), a small computation shows that in $U_1 \cap U_2$, we have
\be
{\widetilde{\cal T}(z)}  =  {\cal T}(z),
\label{B1}
\ee
\be
{\widetilde{\cal Q}(z)} - {\cal Q}(z)  =  {2  \partial_z \left ( {c \over \gamma} \right )(z)},
\label{B2}
\ee
\be
{\widetilde{\cal G}(z)}  =  {\cal G}(z),
\label{B3}
\ee
\be
{\widetilde{\cal J}(z)} - {\cal J}(z) = -{2  \left ({\partial_z \gamma \over \gamma} \right )(z)},
\label{B4}
\ee
where an operator that is a ($\mathbb Z_K$-invariant) global section of ${\widehat \Omega}^{ch}_{\mathbb P^1}$ must agree in $U_1 \cap U_2$. Notice that in $U_1 \cap U_2$, we have ${\widetilde {\cal J}} \neq {\cal J}$ and ${\widetilde {\cal Q}} \neq {\cal Q}$. One can argue that there is no consistent way to modify $\cal J$ and $\widetilde {\cal J}$, or $\cal Q$ and $\widetilde {\cal Q}$, so as to agree on $U_1\cap U_2$.\footnote{The only way to consistently modify $\cal J$ and $\widetilde {\cal J} $ so as to agree on $U_1\cap U_2$, is to shift them by a multiple of the term ${(\partial_z \gamma) / \gamma} = -{(\partial_z \tilde \gamma) / \tilde \gamma}$. However, this term has a pole at both $\gamma=0$ and $\tilde\gamma=0$. Thus, it cannot be used to redefine $\cal J$ or $\widetilde {\cal J} $ (which has to be regular in $U_1$ or $U_2$ respectively). The only way to consistently modify $\cal Q$ and $\widetilde {\cal Q} $ so as to agree on $U_1\cap U_2$, is to shift them by a linear combination  of the terms $({\partial_z c}) / \gamma = - \tilde \gamma \partial_z (\tilde c / {\tilde \gamma}^2)$, and $({c \partial_z \gamma}) / \gamma^2 = ({\tilde c \partial_z \tilde \gamma}) / {\tilde \gamma}^2$. Similarly, these terms have poles at both $\gamma = 0$ and $\tilde\gamma = 0$, and hence, cannot be used to redefine $\cal Q$ or $\widetilde {\cal Q}$ (which also has to be regular in $U_1$ or $U_2$ respectively). This has previously been discussed in   \cite{MC}.} Therefore,    we conclude that ${\cal T}(z)$ and ${\cal G}(z)$ belong in $H^0(\mathbb {CP}^1, {\widehat \Omega}^{ch}_{\mathbb P^1})^{\mathbb Z_K}$, while ${\cal J}(z)$ and ${\cal Q}(z)$ $\it{do}$ $\it{not}$ belong in $H^0(\mathbb {CP}^1, {\widehat \Omega}^{ch}_{\mathbb P^1})^{\mathbb Z_K}$. This means that $T(z)$ and $G(z)$ are in the $Q_R$-cohomology of the underlying half-twisted A-model on $\mathbb {CP}^1/\mathbb Z_K$, while $J(z)$ and $Q(z)$ are not. This last statement is in perfect agreement with the physical picture presented in section 3.3, which in this case states that since $\mathbb {CP}^1$ is not Calabi-Yau, i.e., $c_1(\mathbb {CP}^1) \neq 0$, the symmetries associated with $J(z)$ and $Q(z)$ ought to be broken so that  $J(z)$ and $Q(z)$ cease to exist in the $Q_R$-cohomology at the quantum level. Moreover, it is also explained in section 3.3, that the symmetries associated with $T(z)$ and $G(z)$ are exact in quantum perturbation theory, and that these operators will remain in the $Q_R$-cohomology at the quantum level. This just corresponds to the mathematical fact that the sheaf ${\widehat \Omega}^{ch}_{X}$ on $\it{any}$ $X$ has the structure of a conformal vertex superalgebra, such that  we will have ${\widetilde{\cal T}} = {\cal T}$ and ${\widetilde {\cal G}} = {\cal G}$ always, regardless of whether $c_1(X)$ vanishes or not. Via (\ref{B1})-(\ref{B4}), we have obtained a purely mathematical interpretation of a physical result concerning the holomorphic structure of the underlying, `massive' half-twisted A-model on $\mathbb {CP}^1/\mathbb Z_K$; the reduction  from an $N=2$ to an $N=1$ algebra in the holomorphic structure of the half-twisted A-model on $\mathbb {CP}^1/\mathbb Z_K$, is due to an obstruction in gluing, on overlaps, the ($\mathbb Z_K$-invariant) ${\cal J}(z)$'s and ${\cal Q}(z)$'s as ($\mathbb Z_K$-invariant) global sections of the sheaf ${\widehat \Omega}^{ch}_{\mathbb P^1}$.

One can go further to ascertain the relationship between the obstructing terms on the RHS of (\ref{B2}) and (\ref{B4}), and the  first Chern class $c_1(\mathbb {CP}^1)$. One can then check to see if there is any correlation between a non-vanishing obstruction and a non-zero first Chern class of $\mathbb {CP}^1$, and vice-versa. To this end, one may  substitute the automorphism relations (\ref{autoCDRgamma})-(\ref{autoCDRb}) into ${\widetilde{\cal J}(z)}$, ${\widetilde{\cal Q}(z)}$, ${\widetilde{\cal T}(z)}$ and ${\widetilde{\cal G}(z)}$, and compute that for any $X$ \cite{MSV1}
\be
{\widetilde{\cal T}(z)}  =  {\cal T}(z),
\label{C1}
\ee
\be
{\widetilde{\cal Q}(z)} - {\cal Q}(z)  = \partial_z \left( {\partial \over {\partial \tilde \gamma^k}} [\mathrm{Tr} \ \mathrm{ln} ( \partial \gamma^i / \partial \tilde \gamma^j)] \tilde c^k (z) \right),
\label{C2}
\ee
\be
{\widetilde{\cal G}(z)}  =  {\cal G}(z),
\label{C3}
\ee
\be
{\widetilde{\cal J}(z)} - {\cal J}(z) = \partial_z \left( \mathrm{Tr} \ \mathrm{ln} ( \partial \tilde \gamma^i / \partial \gamma^j) \right),
\label{C4}
\ee
where $i,j,k = 1, \dots, \textrm{dim}_{\mathbb C} X$. It has been shown in \cite{MSV1} that the terms on the RHS of (\ref{C2}) and (\ref{C4}) vanish if and only if $c_1(X) = 0$, whence  the structure of the sheaf ${\widehat \Omega}^{ch}_X$ is promoted to that of a topological vertex superalgebra, with $G$-invariant global sections ${{\cal T}(z)}$, ${{\cal G}(z)}$, ${{\cal J}(z)}$ and ${{\cal Q}(z)}$ obeying the OPE's in  (2.17) of a holomorphic, (twisted) $N=2$ superconformal algebra. Thus, the terms on the RHS of (\ref{B2}) and (\ref{B4}) indeed appear because $c_1(\mathbb {CP}^1) \neq 0$. This observation provides a purely mathematical perspective on the presence or absence of a holomorphic, (twisted) $N=2$ superconformal structure in the half-twisted A-model on an orbifold $X/G$, when $X/G$ is Calabi-Yau or otherwise.

\bigskip\noindent{\it The $\mathbb Z_K$-Invariant First Cohomology of  $\widehat\Omega^{ch}_{\mathbb P^1}$}

We shall now proceed to make a few comments about the $\mathbb Z_K$-invariant first cohomology group $H^1(\mathbb{CP}^1, {\widehat \Omega}^{ch}_{\mathbb P^1})^{\mathbb Z_K}$. Once again, the first cohomology $H^1(\mathbb{CP}^1, {\widehat \Omega}^{ch}_{\mathbb P^1})$ has already been analysed in \cite{MC}. Thus, for brevity, we shall just state the observations in \cite{MC} which will be relevant to our present analysis.

In short, we find that $H^1(\mathbb {CP}^1, {\widehat \Omega}^{ch}_{\mathbb P^1 ; 0})$, the first cohomology group at dimension 0, must be one-dimensional and generated by $c$. However, since $c$ is not $\mathbb Z_K$-invariant, it will mean that $c \notin H^1(\mathbb {CP}^1, {\widehat \Omega}^{ch}_{\mathbb P^1 ; 0})^{\mathbb Z_K}$, and therefore, $H^1(\mathbb {CP}^1, {\widehat \Omega}^{ch }_{\mathbb P^1 ; 0})^{\mathbb Z_K}$ vanishes. This is in contrast to the first cohomology of the half-twisted A-model on $\mathbb {CP}^1$.

In dimension 1, we learn that because of (\ref{B4}),  we have ${\partial_z \gamma / {\gamma}} \notin H^1(\mathbb {CP}^1, {\widehat \Omega}^{ch}_{\mathbb {P}^1 ; 1})$ and therefore, ${\partial_z \gamma / {\gamma}} \notin H^1(\mathbb {CP}^1, {\widehat \Omega}^{ch}_{\mathbb {P}^1 ; 1})^{\mathbb Z_K}$. Similarly, we  learn that because of (\ref{B2}), we have  $\partial_z ( {c / \gamma}) \notin H^1(\mathbb {CP}^1, {\widehat \Omega}^{ch}_{\mathbb {P}^1 ; 1})$ and therefore, $\partial_z ( {c / \gamma}) \notin H^1(\mathbb {CP}^1, {\widehat \Omega}^{ch}_{\mathbb {P}^1 ; 1})^{\mathbb Z_K}$. From a purely physical perspective, one can view these observations as due to quantum effects in perturbation theory.

It can be explained, using chiral Poincar\'e duality \cite{CPD}, that since $\{J_+, J_-, J_3, j_+, j_-, j_3 \} \newline \in H^0(\mathbb {CP}^1, {\widehat \Omega}^{ch}_{\mathbb P^1})$, the space $H^1(\mathbb {CP}^1, {\widehat \Omega}^{ch}_{\mathbb P^1})$ is also a module for a super-affine algebra of $SL(2)$ at level 0 \cite{MC}. However, recall that we only have $\{1, J_3, j_3 \} \in H^0(\mathbb {CP}^1, {\widehat \Omega}^{ch}_{\mathbb P^1})^{\mathbb Z_K}$. Hence, in contrast to the half-twisted A-model on $\mathbb {CP}^1$,  the space $H^1(\mathbb {CP}^1, {\widehat \Omega}^{ch}_{\mathbb P^1})^{\mathbb Z_K}$ is $\it{not}$ a module for a super-affine algebra of $SL(2)$.

In dimension 2 and higher, we do not have relations that are analogous to (\ref{B4}) and (\ref{B2}) in dimension 1. Thus, we could very well borrow the results from standard algebraic geometry to ascertain the relevant operators of dimension 2 and higher in the first cohomology, and project onto $\mathbb Z_K$-invariant operators. We will omit the computation of  these operators  for brevity.

\bigskip\noindent{\it $\mathbb Z_K$-Invariant Global Sections of $\widehat\Omega^{ch,g}_{\mathbb P^1}$}

Note that since $X= \mathbb {CP}^1$ is of complex dimension $1$, the chiral algebra ${\cal A}^g_{\mathbb Z_K}$ will be given by ${\cal A}^g_{\mathbb Z_K} = \bigoplus_{l= 0}^{k= 1} H^{l}({\mathbb {CP}^1}, {\widehat \Omega}^{ch,g}_{\mathbb P^1})^{\mathbb Z_K}$ as a vector space, where $l =g_R- F_g$. Thus, in order to understand ${\cal A}^g_{\mathbb Z_K}$, the chiral algebra in the $g$-twisted sector of the half-twisted orbifold model on $\mathbb {CP}^1/{\mathbb Z_K}$, one needs to study $H^{0}({\mathbb {CP}^1}, {\widehat \Omega}^{ch,g}_{\mathbb P^1})^{\mathbb Z_K}$, the ${\mathbb Z_K}$-invariant global sections of the sheaf ${\widehat \Omega}^{ch,g}_{\mathbb P^1}$, and $H^1( \mathbb {CP}^1, {\widehat \Omega}^{ch,g}_{\mathbb P^1})^{\mathbb Z_K}$, the ${\mathbb Z_K}$-invariant first Cech cohomology of ${\widehat \Omega}^{ch,g}_{\mathbb P^1}$.

Note that the action of $\mathbb Z_K$ on the various twisted fields $\beta^g$, $\gamma^g$, $b^g$ and $c^g$, is the same as its action on the untwisted fields $\beta$, $\gamma$, $b$ and $c$. Thus, we find that these twisted fields will transform under the action of $\mathbb Z_K$ as $\beta^g\to e^{-2 \pi i \theta} \beta^g$, $\gamma^g \to e^{2 \pi i \theta}\gamma^g$, $b^g \to e^{-2 \pi i \theta}b^g$ and $c^g \to e^{2 \pi i \theta}c^g$, where $\theta = m /K$, and $m = 0,1, 2, \dots, K-1$.

Next, note  that the analysis of $H^{0}({\mathbb {CP}^1}, {\widehat \Omega}^{ch}_{\mathbb P^1})$ in \cite{MC} depends purely on the automorphism relations in (\ref{autoCP1gamma})-(\ref{autoCP1b}), and the target space interpretation of the fields $\beta$, $\gamma$, $b$ and $c$, i.e., the role of $\beta$, $\gamma$, $b$ and $c$ as worldsheet fields is irrelevant in determining the result that  $\{1, J_-, J_+, J_3, j_-, j_+, j_3\} \in H^{0}({\mathbb {CP}^1}, {\widehat \Omega}^{ch}_{\mathbb P^1})$ and therefore, only $\{1, J_3, j_3\} \in H^{0}({\mathbb {CP}^1}, {\widehat \Omega}^{ch}_{\mathbb P^1})^{\mathbb Z_K}$ while $\{J_-, J_+,  j_-, j_+, \} \notin H^{0}({\mathbb {CP}^1}, {\widehat \Omega}^{ch}_{\mathbb P^1})^{\mathbb Z_K}$. This means that even though the twisted fields $\beta^g$, $\gamma^g$, $b^g$ and $c^g$ have non-trivial monodromy on the worldsheet, they can be analysed in just the same way as the untwisted fields. Thus, by applying the same arguments, noting the fact that the automorphism relations in (\ref{gautoCP1gamma})-(\ref{gautoCP1b}) involving the twisted fields are the same as the automorphism relations in (\ref{autoCP1gamma})-(\ref{autoCP1b}) involving the untwisted fields, and the fact that the action of $\mathbb Z_K$ on the twisted fields is the same, we find that even though $\{1, J^g_-, J^g_+, J^g_3, j^g_-, j^g_+, j^g_3\} \in H^{0}({\mathbb {CP}^1}, {\widehat \Omega}^{ch,g}_{\mathbb P^1})$, one only has $\{1, J^g_3, j^g_3\} \in H^{0}({\mathbb {CP}^1}, {\widehat \Omega}^{ch,g}_{\mathbb P^1})^{\mathbb Z_K}$ while $\{J^g_-, J^g_+,  j^g_-, j^g_+, \} \notin H^{0}({\mathbb {CP}^1}, {\widehat \Omega}^{ch,g}_{\mathbb P^1})^{\mathbb Z_K}$, where $J^g_-  = - (\gamma^g)^2 \beta^g - 2 \gamma^g b^g c^g$, $J_+ =\beta^g$, $J^g_3 = \gamma^g\beta^g + b^gc^g$, $j^g_- = - (\gamma^g)^2 b^g$, $j^g_+ = b^g$ and $j^g_3 = \gamma^g b^g$. Note that in the twisted sector, the space of operators spanned by $\{1,J^g_3, j^g_3\}$ has a structure of a chiral algebra in the full physical sense; it obeys all the physical axioms of a chiral algebra, including reparameterisation invariance on the $z$-plane or worldsheet $\Sigma$. We will substantiate this last statement shortly by showing that the holomorphic stress tensor $T^g(z)$ exists in the $Q_R$-cohomology of the half-twisted A-model on $\mathbb {CP}^1 / \mathbb Z_K$.

Besides the above operators, one can also find other global sections. Recall from section 4.3 and our $Q_R$-Cech cohomology dictionary that in the twisted sector, there will be $\mathbb Z_K$-invariant, $\psi^{\bar i,g}$-independent operators ${J}^g(z)$, ${Q}^g(z)$, ${T}^g(z)$ and ${G}^g(z)$ in the $Q_R$-cohomology of the underlying half-twisted A-model on $\mathbb {CP}^1/\mathbb Z_K$ if and only if the corresponding $\mathbb Z_K$-invariant operators ${\widehat J}^g(z)$, ${\widehat Q}^g(z)$, ${\widehat T}^g(z)$ and ${\widehat G}^g(z)$ can be globally-defined, i.e., the $\mathbb Z_K$-invariant operators ${\cal J}^g(z)$, ${\cal Q}^g(z)$, ${\cal T}^g(z)$ and ${\cal G}^g(z)$  of the free $bc$-$\beta\gamma$ system belong in $H^0(\mathbb {CP}^1, {\widehat \Omega}^{ch,g}_{\mathbb P^1})^{\mathbb Z_K}$ - the space of $\mathbb Z_K$-invariant global sections of  ${\widehat \Omega}^{ch,g}_{\mathbb P^1}$. Let's look at this in greater detail.

Note that  for $X= \mathbb {CP}^1$, we have
\begin{eqnarray}
{\cal J}^g(z)& =& :b^gc^g: (z) + F_g z^{-1}, \\
{\cal Q}^g(z) &= &:\beta^g c^g: (z), \\
{\cal T}^g(z) & = &  -: \beta^g \partial_z \gamma^g: (z) - : b^g \partial_z c^g: (z), \\
{\cal G}^g (z) & = & :b^g \partial_z \gamma^g:(z),
\end{eqnarray}
where the above operators are defined and regular in $U_1$. Similarly, we also have
\begin{eqnarray}
\label{gA1}
{\widetilde{\cal J}^g (z)}& =& :\tilde b^g \tilde c^g: (z) + F_g z^{-1}, \\
{\widetilde{\cal Q}^g (z)} &= &:\tilde\beta^g \tilde c^g: (z), \\
{\widetilde{\cal T}^g (z)} & = &  -: \tilde\beta^g \partial_z \tilde\gamma^g: (z) - : \tilde b^g \partial_z \tilde c^g: (z), \\
\label{gA4}
{\widetilde{\cal G}^g (z) }& = & :\tilde b^g \partial_z \tilde\gamma^g:(z),
\end{eqnarray}
where the above operators are defined and regular in $U_2$. By substituting the automorphism relations (\ref{gautoCP1gamma})-(\ref{gautoCP1b}) into (\ref{gA1})-(\ref{gA4}), a small computation shows that in $U_1 \cap U_2$, we have
\be
{\widetilde{\cal T}^g (z)}  =  {\cal T}^g (z),
\label{gB1}
\ee
\be
{\widetilde{\cal Q}^g (z)} - {\cal Q}^g (z)  =  {2  \partial_z \left ( {c^g \over \gamma^g } \right )(z)},
\label{gB2}
\ee
\be
{\widetilde{\cal G}^g (z)}  =  {\cal G}^g (z),
\label{gB3}
\ee
\be
{\widetilde{\cal J}^g(z)} - {\cal J}^g(z) = -{2  \left ({\partial_z \gamma^g \over \gamma^g } \right )(z)},
\label{gB4}
\ee
where an operator that is a ($\mathbb Z_K$-invariant) global section of ${\widehat \Omega}^{ch,g}_{\mathbb P^1}$ must agree in $U_1 \cap U_2$. Notice that in $U_1 \cap U_2$, we have ${\widetilde {\cal J}^g } \neq {\cal J}^g $ and ${\widetilde {\cal Q}^g } \neq {\cal Q}^g$. One can again, one can argue that there is no consistent way to modify ${\cal J}^g$ and $\widetilde {\cal J}^g$, or ${\cal Q}^g$ and $\widetilde {\cal Q}^g$, so as to agree on $U_1\cap U_2$. Therefore, we conclude that ${\cal T}^g(z)$ and ${\cal G}^g (z)$ belong in $H^0(\mathbb {CP}^1, {\widehat \Omega}^{ch,g}_{\mathbb P^1})^{\mathbb Z_K}$, while ${\cal J}^g (z)$ and ${\cal Q}^g (z)$ $\it{do}$ $\it{not}$ belong in $H^0(\mathbb {CP}^1, {\widehat \Omega}^{ch,g}_{\mathbb P^1})^{\mathbb Z_K}$. This means that $T^g(z)$ and $G^g(z)$ are twisted sector operators in the $Q_R$-cohomology of the underlying half-twisted A-model on $\mathbb {CP}^1/\mathbb Z_K$, while $J^g(z)$ and $Q(z)$ are not. This last statement is in perfect agreement with the physical picture presented in section 3.3, which in this case states that since $\mathbb {CP}^1$ is not Calabi-Yau, i.e., $c_1(\mathbb {CP}^1) \neq 0$, the symmetries of the twisted sector, associated with $J^g(z)$ and $Q^g(z)$, will be broken such that  $J^g (z)$ and $Q^g (z)$ will cease to exist in the $Q_R$-cohomology at the quantum level. Moreover, it is also explained in section 3.3, that the symmetries associated with $T^g (z)$ and $G^g (z)$ are exact in quantum perturbation theory, and that these operators will remain in the $Q_R$-cohomology at the quantum level. This just corresponds to the mathematical fact that the sheaf ${\widehat \Omega}^{ch,g}_{X}$ on $\it{any}$ $X$ has a conformal vertex superalgebraic structure, such that  we will have ${\widetilde{\cal T}^g } = {\cal T}^g$ and ${\widetilde {\cal G}^g} = {\cal G}^g$ always, regardless of whether $c_1(X)$ is zero or not. Via (\ref{gB1})-(\ref{gB4}), we have obtained a purely mathematical interpretation of a physical result concerning the holomorphic structure of the twisted sector in the underlying, `massive' half-twisted A-model on $\mathbb {CP}^1/\mathbb Z_K$; the reduction  from a holomorphic $N=2$ to a holomorphic $N=1$  structure in the twisted sectors of the half-twisted A-model on $\mathbb {CP}^1/\mathbb Z_K$, is due to an obstruction in gluing, on overlaps, the ($\mathbb Z_K$-invariant) ${\cal J}^g (z)$'s and ${\cal Q}^g (z)$'s as ($\mathbb Z_K$-invariant) global sections of the sheaf ${\widehat \Omega}^{ch,g}_{\mathbb P^1}$, which is a twisted version of the chiral de Rham complex on $\mathbb {CP}^1$.

The above observation also provides a purely mathematical perspective on the presence or absence of a holomorphic, twisted $N=2$ superconformal structure in the twisted sectors of the half-twisted A-model on an orbifold $X/G$, when $X/G$ is Calabi-Yau or otherwise.

\bigskip\noindent{\it The $\mathbb Z_K$-Invariant First Cohomology of $\widehat\Omega^{ch,g}_{\mathbb P^1}$}

Let us now proceed to make a few comments about the $\mathbb Z_K$-invariant first cohomology group $H^1(\mathbb{CP}^1, {\widehat \Omega}^{ch,g}_{\mathbb P^1})^{\mathbb Z_K}$. Once again, the role of $\beta$, $\gamma$, $b$ and $c$ as worldsheet fields is irrelevant to the analysis of the first cohomology $H^1(\mathbb{CP}^1, {\widehat \Omega}^{ch}_{\mathbb P^1})$ in \cite{MC}; the analysis depends on their target space interpretation only. Thus, even though the twisted fields $\beta^g$, $\gamma^g$, $b^g$ and $c^g$ have non-trivial monodromy on the worldsheet, they can be analysed in just the same way as the untwisted fields where the first cohomology is concerned. Hence, the observations made in \cite{MC} of $H^1(\mathbb{CP}^1, {\widehat \Omega}^{ch}_{\mathbb P^1})$ will apply equally to $H^1(\mathbb{CP}^1, {\widehat \Omega}^{ch,g}_{\mathbb P^1})$.

In summary, we find that in the twisted sector, $H^1(\mathbb {CP}^1, {\widehat \Omega}^{ch,g}_{\mathbb P^1 ; 0})$, the first cohomology group at dimension 0, must be one-dimensional and generated by $c^g$. However, since $c^g$ is not $\mathbb Z_K$-invariant, it will mean that $c^g \notin H^1(\mathbb {CP}^1, {\widehat \Omega}^{ch,g}_{\mathbb P^1 ; 0})^{\mathbb Z_K}$, and therefore, $H^1(\mathbb {CP}^1, {\widehat \Omega}^{ch,g}_{\mathbb P^1 ; 0})^{\mathbb Z_K}$ vanishes.

In dimension 1, we learn that because of (\ref{gB4}),  we have ${\partial_z \gamma^g / {\gamma^g}} \notin H^1(\mathbb {CP}^1, {\widehat \Omega}^{ch,g}_{\mathbb {P}^1 ; 1})$ and therefore, ${\partial_z \gamma^g / {\gamma^g}} \notin H^1(\mathbb {CP}^1, {\widehat \Omega}^{ch,g}_{\mathbb {P}^1 ; 1})^{\mathbb Z_K}$. Similarly, we  learn that because of (\ref{gB2}), we have  $\partial_z ( {c^g / \gamma^g}) \notin H^1(\mathbb {CP}^1, {\widehat \Omega}^{ch,g}_{\mathbb {P}^1 ; 1})$ and therefore, $\partial_z ( {c^g / \gamma^g}) \notin H^1(\mathbb {CP}^1, {\widehat \Omega}^{ch,g}_{\mathbb {P}^1 ; 1})^{\mathbb Z_K}$. Once again, from a purely physical perspective, one can view these observations as due to quantum effects in perturbation theory.

In dimension 2 and higher, we do not have relations that are analogous to (\ref{gB4}) and (\ref{gB2}) in dimension 1. Therefore, we could very well borrow the results from standard algebraic geometry to ascertain the relevant operators of dimension 2 and higher in the first cohomology, and project onto $\mathbb Z_K$-invariant operators. We will omit the computation of  these operators  for brevity.

\newsubsection{The Sheaves of CDR and the  Model on $(\bf{S}^3 \times \bf{S}^1) / \mathbb Z_K$}

In our second and last example, we shall take $X/G =({\bf{S}^3 \times \bf{S}^1})/ \mathbb Z_K$, where $\bf{S}^3 \times \bf{S}^1$ is a parellelisable manifold with vanishing first Chern class. In other words, the $\mathbb Z_K$-invariant, global sections of the sheaves of CDR on $\bf{S}^3 \times \bf{S}^1$ which we will be constructing, must correspond to operators in the chiral algebra (or alternatively, $Q_R$-cohomology) of an underlying half-twisted A-model on $({\bf{S}^3 \times \bf{S}^1})/ \mathbb Z_K$ that is ``$\it{non}$-$\it{massive}$''. Thus, as per our discussion in section 3.3, one can expect to find a holomorphic (twisted) $N=2$ superconformal structure that persists in the quantum theory.

Let us begin by noting that $\bf{S}^3 \times \bf{S}^1$ can be expressed as $(\Bbb{C}^2-\{0\})/\Bbb{Z}$, where
$\Bbb{C}^2$ has coordinates $v^1,$ $v^2$, and $\{0\}$ is the origin in $\Bbb{C}^2$ (the point $v^1=v^2=0$) which should be
removed before dividing by $\Bbb{Z}$. Also, $\Bbb{Z}$ acts by $v^i \to \lambda^n v^i$, where $\lambda$ is a nonzero complex number
of modulus less than 1, and $n$ is any integer.  $\lambda$ is a modulus of $\bf{S}^3 \times \bf{S}^1$ that we shall keep fixed.

To construct the most basic sheaves of CDR with target $X=\bf{S}^3 \times \bf{S}^1$ that correspond to operators in the untwisted and twisted sectors, one simply defines the scalar coordinate variables $v^i$ as free bosonic fields of spin 0, with conjugate spin 1 fields $V_i$. One will also need to introduce fermionic fields $w^i$ of spin 0, with conjugate spin 1 fields $W_i$. Since $\bf{S}^3 \times \bf{S}^1$ has complex dimension 2, the index $i$ in all fields will run from 1 to 2. Therefore, the free field action that one must consider is given by

\be
{I={1\over
2\pi}\int|d^2z| \left(\ V_1\bar\partial v^1+ V_2 \bar \partial
v^2 +  W_1\bar\partial w^1+ W_2 \bar \partial
w^2   \right).  }
\label{lagrangian}
\ee
Notice that the above $Vv$-$Ww$ system is just the usual $\beta\gamma$-$bc$ system with nontrivial OPE's $V_i(z)v^j(z')\sim -\delta^i_j/(z-z')$ and $W_i(z)w^j(z')\sim \delta^i_j/(z-z')$. In a $g$-twisted sector, one needs to consider the $g$-twisted fields $v^{i,g}$, $V^g_i$, $w^{i,g}$ and $W^g_i$, where the fields $v^{i,g}$, $w^{i,g}$ and $V^g_i$, $W^g_i$ have opposite twists. The non-trivial OPE's are then given by $V^g_i(z)v^{j,g}(z')\sim -\delta^i_j/(z-z')$ and $W^g_i(z)w^{j,g}(z')\sim \delta^i_j/(z-z')$.

In the above representation of $\bf{S}^3 \times \bf{S}^1$, the action of $\mathbb Z$ represents a geometrical symmetry of the system. Thus, the only allowable operators spanning the space of global sections of the sheaves $\widehat\Omega^{ch}_X$ and $\widehat\Omega^{ch,g}_X$ on $X={\bf{S}^3 \times \bf{S}^1}$, are those which are invariant under the finite action of $\mathbb Z$.  Under this symmetry, $v^i$ and $v^{i,g}$ transform as $v^i \to {\tilde v}^i = \lambda v^i$ and $v^{i,g} \to {\tilde v}^{i,g} = \lambda v^{i,g}$.  In order to ascertain how the rest of the fields ought to transform under this symmetry, we simply substitute $v^i$ and ${\tilde v}^i$ (noting that it is equivalent to $\gamma^i$ and ${\tilde \gamma}^i$ respectively) into (\ref{autoCDRgamma})-(\ref{autoCDRb}), and substitute $v^{i,g}$ and ${\tilde v}^{i,g}$ (noting that it is equivalent to $\gamma^{i,g}$ and ${\tilde \gamma}^{i,g}$ respectively) into (\ref{gautoCDRgamma})-(\ref{gautoCDRb}).   In short, the operators in the untwisted sector which correspond to global sections of $\widehat\Omega^{ch}_X$ are those which are invariant under $v^i \to \lambda v^i$, $V_i \to \lambda^{-1} V_i$, $w^i \to \lambda w^i$ and $W^i \to \lambda^{-1} W^i$, while the operators in the twisted sectors which correspond to global sections of $\widehat\Omega^{ch,g}_X$ are those which are invariant under $v^{i,g} \to \lambda v^{i,g}$, $V^g_i \to \lambda^{-1} V^g_i$, $w^{i,g} \to \lambda w^{i,g}$ and $W^{i,g} \to \lambda^{-1} W^{i,g}$.

However, since we are really looking for $\mathbb Z_K$-invariant global sections $H^0(X, \widehat\Omega^{ch}_X)^{\mathbb Z_K}$ and $H^0(X, \widehat\Omega^{ch,g}_X)^{\mathbb Z_K}$ which correspond to operators in the chiral algebra of the half-twisted A-model on the orbifold $({\bf{S}^3 \times \bf{S}^1})/ \mathbb Z_K$, we must further project onto the  $\mathbb Z_K$-invariant subspace of operators. Via the correspondence between the $Vv$-$Ww$ and the $\beta\gamma$-$bc$ fields, together with the action of $\mathbb Z_K$ on the $\beta\gamma$-$bc$ fields as elucidated in section 5.1, we find that the action of $\mathbb Z_K$ on the fields will be given by $V_j \to e^{-2 \pi i \theta_j} V_j$, $v^j \to e^{2 \pi i \theta_j}v^j$, $W_j \to e^{-2 \pi i \theta_j}W_j$ and $w^j \to e^{2 \pi i \theta_j}w^j$, where $\theta_j = m_j /K$, and $m_j = 0,1, 2, \dots, K-1$. Likewise, we will have $V^g_j \to e^{-2 \pi i \theta_j} V^g_j$, $v^{j,g} \to e^{2 \pi i \theta_j}v^{j,g}$, $W^g_j \to e^{-2 \pi i \theta_j}W^g_j$ and $w^{j,g} \to e^{2 \pi i \theta_j}w^{j,g}$.

One operator that possesses the above stated invariances is the stress-energy tensor:
\be
{{\cal  T}_{zz} \sim \sum_i (V_i\partial v^i + W_i\partial w^i).}
\ee
Other operators that also possess the above stated invariances include:
\be
{\cal G}_{zz} \sim \sum_i W_i \partial v^i, \quad  {\cal J}_z \sim \sum_i W_i w^i,  \quad  {\cal Q}_z \sim \sum_i V_i v^i.
\ee
Thus, ${\cal T}_{zz}$, ${\cal G}_ {zz}$, ${\cal J}_z$ and ${\cal Q}_z$ belong in $H^0(X, \widehat \Omega^{ch}_X)^{\mathbb Z_K}$, and therefore correspond to operators in the untwisted sector of the chiral algebra of the half-twisted A-model on $({\bf{S}^3 \times \bf{S}^1})/ \mathbb Z_K$.

Note that the corresponding operators in the twisted fields $v^{i,g}$, $w^{i,g}$, $V^g_i$ and $W^g_i$ given by
\be
{\cal T}^g_{zz} \sim \sum_i (V^g_i \partial v^{i,g} + W^g_i \partial w^{i,g}), \quad  {\cal G}^g_{zz} \sim \sum_i W^g_i \partial v^{i,g},
\ee
and
\be
{\cal J}^g_z \sim \sum_i W^g_i w^{i,g} + F_g z^{-1},  \qquad  {\cal Q}^g_z \sim \sum_i V^g_i v^{i,g},
\ee
also possess the above stated invariances. (Once again, the term $F_g z^{-1}$ is added to ${\cal J}_z$ to account for the shift in the fermionic charge of the twisted sector vacuum.) Hence, ${\cal T}^g_{zz}$, ${\cal G}^g_ {zz}$, ${\cal J}^g_z$ and ${\cal Q}^g_z$ belong in $H^0(X, \widehat \Omega^{ch,g}_X)^{\mathbb Z_K}$, and therefore correspond to operators in the twisted sector of the chiral algebra of the half-twisted A-model on $({\bf{S}^3 \times \bf{S}^1})/ \mathbb Z_K$.

One can verify that the two sets of operators $\{{\cal T}_{zz}, {\cal G}_{zz}, {\cal J}_z, {\cal Q}_z \}$ and $\{{\cal T}^g_{zz}, {\cal G}^g_{zz}, {\cal J}^g_z, {\cal Q}^g_z \}$ both generate a holomorphic (twisted) $N=2$ superconformal OPE algebra, thereby reflecting the anticipated (quantum) superconformal invariance of the underlying ``non-massive'' model.

One can continue to make the following observation. In \cite{MC}, it was shown that the chiral algebra of the underlying half-twisted A-model on $\bf{S}^3 \times \bf{S}^1$ also contains the dimension 1 currents $J^i_j= -(V_jv^i + W_j w^i)$, where $i\neq j$. In addition, these operators furnish a $GL(2)$ current algebra at level $0$. However, note that the $J^i_j$'s are not $\mathbb Z_K$-invariant. Hence, $J^i_j \notin H^0(X, \widehat \Omega^{ch}_X)^{\mathbb Z_K}$. Therefore, the $J^i_j$'s are not operators in the chiral algebra of the half-twisted A-model on the orbifold $(\bf{S}^3 \times \bf{S}^1)/ {\mathbb Z}_K$. Thus, in contrast to the non-orbifold model, the space of physical operators in the orbifold model $\it{does}$ $\it{not}$ furnish a $GL(2)$ current algebra at level 0, and therefore, according to the analysis of \cite{CDO}\cite{MC}, the symmetry at the level of its $Q_R$-cohomology $\it{will}$ $\it{not}$ be given by $U(2)$.\footnote{Note that the symmetries of the underlying sigma model are readily complexified in the sheaf of CDR. This means that if we have a $GL(2)$ current algebra furnished by the sheaf of CDR, the symmetry at the level of the $Q_R$-cohomology will be given by $U(2)$.} On the other hand, it is easy to see that the operator $K=-{1\over 2}\left(V_1v^1+ W_1w^1 + V_2v^2 + W_2w^2 \right)$ satisfies the requisite invariances stated above. Hence, it belongs in the chiral algebra of the orbifold model. This operator can be easily shown to generate the current algebra, at level 0, of $GL(1)$, the centre (at the Lie algebra level) of $GL(2)$ \cite{MC}. This implies that the symmetry at the level of the $Q_R$-cohomology of the orbifold sigma model is abelian and given by $U(1)$ instead. By replacing the untwisted fields in the $J^i_j$ and $K$ operators by twisted ones, one can make the exact same observation concerning the symmetry of the $Q_R$-cohomology in the twisted sectors.

Another pertinent observation that will allow us to make contact with the results in \cite{Frenkel} is the following. First, notice that $c_1(X) = 0$ for $X = \bf{S}^3 \times \bf{S}^1$. Thus, via the state-operator isomorphism, we find that the Hilbert space of untwisted and twisted states in the sigma model on $\bf{S}^3 \times \bf{S}^1$,  can be represented by the sum of all its $Q_R$-closed, local operators $\cal F$ and ${\cal F}^g$. Therefore, the Hilbert space of all states in the orbifold sigma model on $(\bf{S}^3 \times \bf{S}^1) / {\mathbb Z_K}$, can be represented by the $G$-invariant subspace of all $Q_R$-closed, local operators, i.e., ${\cal F}_G$ and ${\cal F}^g_G$. Next, note that the $G$-invariant subspace just corresponds to the set of physical operators in the chiral algebra $\bigoplus_{g\in G}{\cal A}_G^g$, where ${\cal A}^1_G = {\cal A}_G$. Finally, note that for an abelian group such as $\mathbb Z_K$, we have $G =C(g)$. Hence, from the vector space expansion of ${\cal A}_G$ and ${\cal A}^g_G$ in section 4.2, the Hilbert space expression for an orbifold sigma model in (\ref{hilbert space}) and (\ref{twisted sectors}), and the identification of $\widehat{\cal A}$ and $\widehat {\cal A}^g$ with ${\widehat \Omega}^{ch}_X$ and ${\widehat \Omega}^{ch,g}_X$ respectively, we find that the Hilbert space of all states in the half-twisted A-model on the orbifold $(\bf{S}^3 \times \bf{S}^1) / {\mathbb Z_K}$, can be expressed as
\be
{\cal H} = \bigoplus_{[g] \in T} \bigoplus _{g \in [g]} \bigoplus_{k =0}^{\textrm{dim}_{\mathbb C}X}H^k (X, {\widehat \Omega}^{ch,g}_X)^{C(g)},
\label{space}
\ee
where $T$ is the set of conjugacy classes in $G$, $C(1) =G$, and ${\widehat \Omega}^{ch,1}_X ={\widehat \Omega}^{ch}_X$.  In this abelian case, $T$ is given by a single conjugacy class, which is the group $\mathbb Z_K$ itself.  Note that (\ref{space}) coincides with the conjectural space of states of an orbifold model as defined in the mathematical literature \cite{Frenkel}.

Lastly, note  that the $Q_R$-cohomology of the model on $\bf{S}^3 \times \bf{S}^1$ does not receive worldsheet instanton corrections. For any
target space $X$, such corrections (because they are local on the
Riemann surface $\Sigma$, albeit global in $X$) come only from
holomorphic curves in $X$ of genus zero. There is no such curve in $\bf{S}^3 \times \bf{S}^1$.\footnote{The author wishes to thank Ed Witten for a detailed explanation of this point.} This means that our above analysis of the chiral algebra or $Q_R$-cohomology of the half-twisted model on $(\bf{S}^3 \times \bf{S}^1)/ {\mathbb Z}_K$ is exact in the full theory.

\vspace{0.5cm}
\hspace{-1.0cm}{\large \bf Acknowledgements:}\\
I would like to take this opportunity to thank E. Frenkel, M. Szczesny, E. Witten and in particular L. Dixon, for providing their expert opinion on various issues over our email correspondences.
\newline

\end{document}